\documentclass{JHEP}
\usepackage{amsmath}
\input epsf
\usepackage{epsfig}
\usepackage{amssymb}
\usepackage{graphicx}
\usepackage{colordvi}
\usepackage[active]{srcltx}
\usepackage{subfigure}
\usepackage{subfig}
\usepackage{cancel}

\newcommand{\bea}{\begin{eqnarray}}
\newcommand{\eea}{\end{eqnarray}}
\newcommand{\bean}{\begin{eqnarray*}}
\newcommand{\eean}{\end{eqnarray*}}
\newcommand{\nn}{\nonumber \\}

\def\O #1{\overline{#1}}

\def\W #1{\widetilde{#1}}
\def\WH #1{\widehat{#1}}

\def\braket#1{\left\langle #1 \right\rangle}
\def\bra#1{\left\langle #1\right|}
\def\ket#1{\left| #1\right\rangle}
\def\bbra#1{\left[ #1\right|}
\def\bket#1{\left| #1\right]}
\def\gb #1{ \left\langle #1 \right]}
\def\tgb #1{ \left[ #1 \right\rangle}
\newcommand \widebar [1] {\overline{#1}}

\def\Tr{\mathop{\rm Tr}}
\def\det{\mathop{\rm det}}

\def\wh{\widehat}

\def\a{{\alpha}}

\def\b{{\beta}}

\def\la{\lambda}
\def\eps{\epsilon}

\def\vev{\braket}
\def\tgb #1{ \left[ #1 \right\rangle}
\def\bket#1{\left| #1\right]}
\def\bvev#1{\left[ #1 \right]}
\def\Spaa{\vev}
\def\Spbb{\bvev}
\def\Spab{\gb}
\def\Spba{\tgb}

\def\l<{\langle}\def\r>{\rangle}

\def\Label#1{\label{#1}%
  \smash{\hbox to0pt{\raise1ex\hbox{\tiny[#1]}\hss}}}

\title{An Introduction to On-shell Recursion Relations}
\author{Bo Feng${}^{a,b}$,  Mingxing Luo${}^{a}$,  \\
$^a$\small Zhejiang Institute of Modern Physics, Zhejiang
University, Hangzhou, 310027, P. R. China\\$^b$\small Center of
Mathematical Science, Zhejiang University, Hangzhou, China }
\date{\today}
\abstract{This article provides an introduction to on-shell
recursion relations for calculations of tree-level amplitudes.
Starting with the basics, such as spinor notations and color
decompositions, we expose analytic properties of gauge-boson
amplitudes, BCFW-deformations, the large $z$-behavior of amplitudes,
and on-shell recursion relations of gluons. We discuss further
developments of on-shell recursion relations, including
generalization to other quantum field theories, supersymmetric
theories in particular, recursion  relations for off-shell currents,
recursion relation with nonzero boundary contributions, bonus
relations, relations for rational parts of one-loop amplitudes,
recursion relations in 3D and a proof of CSW rules. Finally, we
present samples of applications, including solutions of split
helicity amplitudes and of ${\cal N} = 4$ SYM theories, consequences
of consistent conditions under recursion relation, Kleiss-Kuijf (KK)
and Bern-Carrasco-Johansson (BCJ) relations for color-ordered gluon
tree amplitudes, Kawai-Lewellen-Tye (KLT) relations. }

\begin{document}

\section{Introduction}

It might be hard to over-state the importance of scattering
amplitudes in particle physics and quantum field theories. They are
basic building blocks of quantum field theories. Without them,
quantum field theories would have lost most of their physics
contents. They are indispensable in explaining almost all phenomena
in particle physics. In particle physics experiments, such as those
at the on-going Large Hadron Collider, scattering amplitudes are
needed to make sense out of the huge mountain of experimental data.
With the increasing precision of experiments and due to large
contributions from gluon-loop processes, one has to calculate
scattering amplitudes of many particles and to go beyond the
leading-order. Leading-order calculation may tell little if anything
about these experiments, neither affirmative nor negative. For many
physics process, next-leading-order calculations are mandatory to
understand the standard model and to delineate new
physics\footnote{See the Table 2 of \cite{Bern:2008ef} where
$K$-factors (i.e., the ratio of NLO/LO) are presented for several
processes. For example, $K$-factor of process with Higgs plus one
jet is about $2.02$.}. In the Les Houches accord \cite{Bern:2008ef},
there is the so-called wish-list for LHC experiments. The list of
scattering processes are deemed particularly important and precise
calculations of their amplitudes are highly desirable.

Traditionally, one relies on Feynman diagrams to calculate
scattering amplitudes. Feynman diagrams provide a clear picture of
physics and a systematic procedure of calculations. They are in
textbooks and widely used. Calculations in quantum field theory must
have been extremely difficult before their appearance. Julius
Schwinger  commented:\footnote{ J. Schwinger, "Quantum
Electrodynamics-An Individual View," J. Physique 43, Colloque C-8,
Supplement au no. 12, 409 (1982) and "Renormalization Theory of
Quantum Electrodynamics: An Individual View," in The Birth of
Particle Physics, Cambridge University Press, 1983, p. 329.} ``Like
the silicon chips of more recent years, the Feynman diagram was
bringing computation to the masses." Regardless the undertone, it
conveyed the feeling then as well as the historical significance of
Feynman diagrams.

But Feynman diagrams are not efficient in complicated calculations
for high energy physics. Increasing the number of particles in a
scattering, the number of Feynman diagrams increase exponentially.
If gauge fields are involved, one easily encounters thousands of
diagrams. For example, for pure non-Abelian gauge theory, the number
of Feynman diagrams for $n$-gluons at tree-level is given by
 \bea
\begin{tabular}{|c|c|c|c|c|c|c|c|} \hline
 $n=$ & $4$ & $5$ & $6$ & $7$ & $8$ & $9$ & $10$ \\\hline
& $4$ & $25$ &  $220$ & $2485$ & $34300$ & $559,405$ & $10,525,900$
\\ \hline
\end{tabular}~~\label{numcount}\eea
Similarly, for seven gluons at one loop, there are $227,585$ Feynman
diagrams. Not only with huge number of diagrams, the expression for
a single Feynman diagram can also be very complicated. For example,
the three-graviton vertex  has almost 100 terms. It is almost
unimaginable to calculate scattering amplitudes of gravitons
directly from Feynman diagrams. For gauge theories, single Feynman
diagram usually depends on the gauge. Many terms cancel with each
other in the process of calculation. In practice, one does not even
know where to start most times. They are way beyond the ability of
present-day computers. However,  final results are usually simple
and tidy. They are gauge invariant, as physical quantities  should
be.

Simply put, it is not economical and sometimes unrealistic to
calculate scattering amplitudes in gauge theories via Feynman
diagrams. The wish-list of LHC was thought to be a big challenge to
the community. Their calculations are particularly hard and almost
impossible via conventional methods. New ways have to be devised to
calculate them efficiently and accurately. In last several years,
with the development of various new methods, there has been
tremendous progress. It has made many difficult calculations
possible. In fact, most processes in the wish-list have now been
calculated with the help of these methods.\footnote{A review of
recent developments of one-loop calculations can be found in
\cite{Britto:2010xq}.}

As anything in scientific advancement, there are always numerous
efforts before and after, sometimes heroic. It is difficult to pick
up  threads. One (not too arbitrary) starting point might be the
formalism introduced by Xu, Zhang and Chang \cite{XZC} (sometimes it
is called "Chinese magic"). Among other things, it provided a set of
symbols inherited by most of late developments and proposed a new
way to represent polarization vectors of massless gauge bosons.
Then, neat formulas were conjectured for the so-called
maximally-helicity-violating (MHV) tree-level amplitudes for gluons
\cite{Parke:1986gb}. They were proved a few years later
\cite{Berends:1987me}. Despite the elegance of these formulas, no
deep understanding was achieved, though a connection with string
theories was pointed out.

In a seminal work \cite{Witten:2003nn}, Witten re-expressed known
scattering amplitudes in ${\cal N}=4$ super-Yang-Mills (SYM)
theories in the language of twistors. The importance of MHV
amplitudes\footnote{The ${\cal N}=4$-version of MHV amplitude is
given in \cite{Nair:1988bq}.} was emphasized. After elaborated
analysis and considerations of symmetries, a duality was pointed out
between ${\cal N}=4$ SYM theories and type-B topological string
theories. Unfortunately, this duality does not survive beyond the
leading order, as gauge bosons and gravitons do not decouple from
each other in loops.
Along the way,  an intuitive geometrical picture was provided for
scattering amplitudes in ${\cal N}=4$ SYM theories, by taking MHV
amplitudes as localized on straight lines in twistor space.

Starting from the deep insight thus gained and detailed expositions
of known amplitudes, a novel method (CSW) was suggested to calculate
scattering amplitudes by taking MHV amplitudes as vertices, linked
by scalar propagators \cite{Cachazo:2004kj}. The CSW method
reproduces known results easily and generates new ones. Incidently,
consistency checks of CSW rules helped to find holomorphic anomalies
\cite{Cachazo:2004by}, which played an important role in later
developments, such as the completion of unitarity cut method
\cite{Landau:1959fi,Bern:1994cg,Bern:1994zx,Britto:2004nc,Britto:2005ha,Anastasiou:2006jv}
and direct evaluation of coefficients of one-loop amplitudes
\cite{Cachazo:2004dr, Britto:2004nj}. It was extended to contain
more physical contents and beyond the tree-level
\cite{hep-th/0407214,hep-th/0510253, Bullimore:2010pj}. CSW rules
were proved via several avenues\cite{Britto:2005fq, Risager:2005vk},
where one of proofs will be given in late section. Compared with
ordinary Feynman diagrams, CSW method has obvious advantages, with
all inputs being gauge invariant on-shell amplitudes, less diagrams
and much more compact expressions. However, when the number of
external lines increases, expressions in intermediate calculations
become extremely complex.

By analysis of relations between one-loop and tree-level amplitudes
in ${\cal N}=4$ SYM theories under the infrared limit and insights
from holomorphic anomalies,  on-shell recursion relations was
conjectured to calculate scattering amplitudes \cite{Britto:2004ap}.
Soon after, the conjecture was  proved  by using fundamental
properties of Feynman diagrams and the correctness of CSW method was
shown \cite{Britto:2005fq}. This breakthrough provided an extremely
powerful tool to calculate scattering amplitudes, generating new
ways in calculations of multi-particle process at tree and loop
levels. Starting from here, tremendous progresses have been made.

The on-shell recursion method starts with an important property of
tree-level amplitudes, as they are meromorphic functions of momenta,
a property due to basic physical principles such as causality. A
meromorphic function is completely determined by its pole locations
and corresponding residues, or cuts, which are closely related to
physical quantities. For tree-amplitudes, there are only poles but
no cuts. Residues are in factorized form, as products of two
sub-amplitudes. Compared with Feynman diagrams or even the CSW
method, on-shell recursion relations handle much less terms. All
intermediate results are on-shell and gauge symmetries are
automatically satisfied. Final results are very compact.

As a simple way to gauge the progress, one notices that the
lowest-order of $W$ production with three jets at hadron collider
could not be calculated in the early 1980's. Recently, processes of
$W,Z$ production with four jets have been computed at the 1-loop
level \cite{arXiv:1009.2338,arXiv:1108.2229}.
 Such computations, if done by Feynman
diagrams, would have required thousands of Feynman diagrams, which
is obviously unrealistic. On the other hand, there are now four loop
calculations of amplitudes in ${\cal N}=4$ SYM theory
\cite{arXiv:1008.3327} and in  ${\cal N}=8$ super-gravity
\cite{arXiv:0905.2326}, which was unthinkable just a few years back.

There are subtle links between on-shell recursion and the S-matrix
program \cite{S-matrix}. Both uses complexified momenta. In the
S-matrix program, all components of external momenta are
complexified, generating complicated complex functions of
multi-variables. In on-shell recursion relations, only two external
momenta are complexified in a controlled way, by introducing a
complex variable. One get rather simple meromorphic functions, for
which the apparatus of single-variable complex analysis is
available.

This paper intends to introduce some of these progresses and ideas
behind them. In particular, we intend to provided an introduction,
starting from the simplest on-shell recursion relation to its
various ramifications as well as sample applications. In detail, we
will start with the basics in section 2, including a simple tour on
spinor notations, color decompositions, and the concept of partial
amplitudes in gauge theories. In section 3, we expose analytic
properties of gauge-boson amplitudes, the so-called
BCFW-deformation, the large $z$-behavior of amplitudes under
BCFW-deformations, and finally on-shell recursion relations of
gluons. In section 4, we discuss further developments of on-shell
recursion relations, including generalization to other quantum field
theories, supersymmetric theories in particular, on-shell recursion
relations for off-shell currents, recursion relations with nonzero
boundary contributions, bonus relations, recursion relations for
rational parts of one-loop amplitudes, recursion relations in 3D and
a proof of CSW rules via on-shell recursion relations. In section 5,
we present sample applications of on-shell recursion relations,
including solutions of split helicity amplitudes  and of ${\cal N} =
4$ SYM theories, consequences of consistent conditions from on-shell
recursion relations, Kleiss-Kuijf (KK) and  Bern-Carrasco-Johansson
(BCJ) relations, Kawai-Lewellen-Tye (KLT) relations and their
proofs.  Due to limits of space and also, limits of our abilities,
many subjects are not covered. Remarks and a partial list of
omissions are presented in section 6.

Now we begin our exposition.

\section{Basics}

In this section, we will introduce two concepts useful for
simplifying calculations of scattering amplitudes, especially when
they involve massless particles and fields with gauge symmetry. The
first concept is spinor notation and the second one, the color
decomposition.

\subsection{Spinor notations}

We start with a short review of spinor notations, which may not be
familiar to readers. Detailed expositions can be found in
\cite{Dixon:1996wi,Maitre:2007jq}, where a Mathematica package ${\rm
S@M}$ has also been developed.

Given a null  momentum $k_\mu$ in four space-time dimensions, one
may define a two-dimensional Weyl spinor $\la$ and an anti-spinor
$\W \la$ through Dirac equations
\bea k_{\dot a a} \la^a(k)=0,~~~~\W \la^{\dot a}(k) k_{\dot a
a}=0~~~\label{Spinor-def}\eea
where we have transformed the vector representation 
of the Lorentz group to bi-spinor notation through the
$\sigma$-matrices:
\bea k_{\dot{a} a}\equiv
k_\mu (\sigma^\mu)_{\dot a a},~~~~\sigma^{\mu}=(1,
\vec{\sigma}),~~~\dot{a},a=1,2~\label{P-bispinor}\eea
It is easy to show that because $k_\mu \cdot k^\mu= \det(k_{\dot{a}
a})$,  a null momentum means the matrix $k_{\dot{a} a}$ is
degenerate and can be decomposed as
\bea k_{\dot{a} a}=\W \la_{\dot a} \la_a,~~~\label{P-la}\eea
Spinor indices can be raised or lowered by anti-symmetric matrices
$\epsilon^{ab}$  and $\epsilon_{ab}$ via\footnote {We have
$\epsilon^{12}=1, \epsilon_{12}=-1$. See appendices A and B of
\cite{Wess} for details.}
\bea
\la^a=\epsilon^{ab}\la_b,~~~~\la_a=\epsilon_{ab}\la^b~,~~~\label{up-down}\eea
and similarly for dotted indices. Using $\epsilon$ we can also
define Lorentz invariant inner products  of two spinors or
anti-spinors
\bea \Spaa{i|j} \equiv \la^a_i \la_{ja},~~~~~\Spbb{i|j}\equiv
\W\la_{i\dot{a}} \W \la_j^{\dot a}~.\eea

In standard textbooks of quantum field theories, free fermions are
usually described  by the Dirac equation $(\cancel{k}- m) u(p)=0$
for positive frequencies and $(\cancel{k}+ m) v(p)=0$ for negative
frequencies, where the $4\times 4$ $\gamma$-matrices are taken in
chiral representation. For $m\neq 0$, there is no direct relation
between these two types of solutions. When $m=0$, positive and
negative frequency solutions are identical up to normalization
conventions. Solutions of definite helicity can be identified with
each other as the following:
\bea u_{\pm}(k) &= & {1\pm \gamma_5\over 2} u(k),~~~v_{\mp}(k)={1\pm
\gamma_5\over 2} u(k), \\
\O{u_{\pm}(k)} &= & \O {u(k)}{1\mp \gamma_5\over
2},~~~\O{v_{\mp}(k)} = \O {v(k)}{1\mp \gamma_5\over 2}\eea
Depending on the representation of $\gamma$-matrices, whether Weyl
or Majorana, explicit expressions of $u,v$ by $\la, \W \la$ may be
different. In our calculations, we can take the following
identification\footnote{In the literature, there are two conventions
to make the identification: the QCD convention and the twistor
convention. We will use the QCD convention, same as the one used in
 the Mathematica package ${\rm S@M}$.  The translation between
these two conventions is simply $\Spbb{~}_{\rm QCD}=-\Spbb{~}_{\rm
twistor}$.}:
\bea \ket{i} &\equiv&  \ket{k_i^+}=u_+(k_i)=v_-(k_i),~~~~~~\bket{i}
\equiv
\ket{k_i^-}=u_-(k_i)=v_+(k_i),\\
\bra{i} &\equiv&
\bra{k_i^-}=\O{u_-}(k_i)=\O{v_+}(k_i),~~~~~~\bbra{i} \equiv
\bra{k_i^+}=\O{u_+}(k_i)=\O{v_-}(k_i)\eea

Within these conventions, we can translate familiar expressions in
Feynman diagrams to  spinor notations, such as
\bea \O{ u_+(k_i)} \cancel{k}_j u_+(k_l) \equiv \Spba{i| k_j|l},~~~ \O{
u_+(k_i)} \cancel{k}_j \cancel{k}_m u_-(k_l) \equiv \Spbb{i| k_j k_m|l}\eea
For simplification and without confusion, we have written $k$ instead of $\cancel{k}$ at
the right-handed sides of these equations.
One can also demonstrate following properties of spinor variables straightforwardly:
\begin{itemize}
\item {\bf Antisymmetries}:
\bea \Spaa{i|j}=-\Spaa{j|i},~~~~\Spbb{i|j}=-\Spbb{j|i}, \label{anti}
\eea
Because the anti-symmetry we have $\Spaa{i|i}=\Spbb{i|i}=0$. Also
$\Spab{i|j}=\Spba{i|j}=0$, i.e., the inner product of a spinor and
an anti-spinor vanishes.

\item {\bf Schouten identity}:
\bea \ket{i} \Spaa{j|k}+ \ket{j} \Spaa{k|i}+ \ket{k}
\Spaa{i|j}=0,~~~\label{Schouten} \eea
and similar ones for anti-spinor $\W\la$ with $\Spaa{~}\to
\Spbb{~}$. Notice that there are only two independent components in
each spinor, so spinor $\ket{i}$ can always be expressed linearly in
terms of other two spinors $\ket{j},\ket{k}$. The identity then
follows trivially.
\item {\bf Projection operator}:
\bea \ket{i}\bbra{i}={1+\gamma_5\over
2}\cancel{k}_i,~~~~\bket{i}\bra{i}={1-\gamma_5\over 2}\cancel{k}_i,
~~~\cancel{k}_i=\ket{i}\bbra{i}+\bket{i}\bra{i},~~~\label{Project}
\eea
Using this we can calculate
\bea \Spaa{i|j}\Spbb{j|i} & = & \Spab{i|\cancel{k}_j|i}= {\rm
Tr}({1-\gamma_5\over 2}\cancel{k}_i\cancel{k}_j)= 2k_i\cdot k_j=
(k_i+k_j)^2 \equiv s_{ij}~~~\label{sij}\eea
from which follows the {\bf Gordon identity}
\bea \Spab{i|\gamma^\mu|i}=\Spba{i|\gamma^\mu|i}=2
k_i^\mu,~~~\label{Gordon}\eea
as well as {\bf Fierz rearrangement}
\bea \Spba{i|\gamma^\mu|j}\Spba{k|\gamma_\mu|l}=2
\Spbb{i|k}\Spaa{l|j}~~~\label{Fierz}\eea
where we have used (\ref{Gordon}) and (\ref{sij}) after identifying
$\ket{i}\to \ket{j}$.
\end{itemize}

Further identities can be derived by using above results. The
following are particularly useful in practical manipulations
\bea & & \Spaa{i| p q|j}+\Spaa{i|q p|j}=(2 p\cdot
q)\Spaa{i|j},~~~\Spaa{i|p p|j}=p^2 \Spaa{i|j}\\
& & \Spab{i|p|j}\Spab{j|p|i} =(2 k_i\cdot p)(2k_j\cdot p)-p^2
(2k_i\cdot k_j)\eea
as well as
\bea & & \Spaa{i|j}\Spbb{j|\ell}\Spaa{\ell|m}\Spbb{m|i}={\rm
tr}\left( {1-\gamma_5\over 2}\cancel{k}_i \cancel{k}_j\cancel{k}_\ell\cancel{k}_m\right)\\
& = & {1\over 2}\left[(2k_i\cdot k_j)(2k_\ell\cdot k_m)+(2k_i\cdot
k_m)(2k_\ell\cdot k_j)-(2k_i\cdot k_\ell)(2k_j\cdot
k_m)-4i\eps(i,j,\ell,m)\right]\eea
where $\eps(i,j,\ell,m)\equiv \epsilon_{\mu\nu\rho\sigma} p_i^\mu
p_j^\nu p_{\ell}^\rho p_m^\sigma$.

In standard Feynman rules, scattering amplitudes are expressed as functions
of momenta and wave functions of external particles.
For a scalar, the wave function is just one. For a fermion, we have identified $u, v$ with spinor and anti-spinor.
For a vector, the wave function is a polarization vector.
By (\ref{P-la}) we have transformed null momenta to
spinors and we need to do similar things to polarization vectors of gauge bosons.
Using the Gordon identity (\ref{Gordon}) we can write down
\bea \eps^+_{\nu}(k|\mu)= {\Spab{\mu|\gamma_\nu|k}\over
\sqrt{2}\Spaa{\mu|k}},~~~~~~\eps^-_{\nu}(k|\mu)=
{\Spba{\mu|\gamma_\nu|k}\over
\sqrt{2}\Spbb{\mu|k}},~~~~~~\label{QCD-eps}\eea
where $\mu$ is an arbitrary null momentum not parallel to $k$. The
choice of $\mu$ corresponds to a choice of gauge and $\sqrt{2}$
is a normalization factor.
(\ref{QCD-eps}) can be converted into spinor notations
\bea  \cancel\eps^+ (k|\mu)= {\la_\mu \W\la_k\over \sqrt{2}
\Spaa{\mu|k}},~~~~~~\cancel\eps^-(k|\mu)= {-\la_k \W\la_\mu\over
\sqrt{2}\Spbb{\mu|k}},~~~~~~\label{Spinor-eps}\eea
Polarization vectors defined above have following properties:
\bea k\cdot \eps^{\pm}(k|\mu) &= & \mu\cdot
\eps^{\pm}(k|\mu)=0,~~~~~\eps^{\pm}(\W k|\mu)\cdot \eps^{\pm}(k|\mu)=0\\
0 & = & \eps^{+}(\W k|k)\cdot \eps^{-}(k|\mu)=\eps^{+}(\W
k|\mu)\cdot
\eps^{-}(k|\W k) \\
0 & = & \cancel\eps^+(k|\mu)\ket{\mu}=\bra{\mu}
\cancel{\eps}^+(k|\mu)= \cancel\eps^-(k|\mu)\bket{\mu}=
\bbra{\mu}\cancel\eps^-(k|\mu)\eea
%

\subsubsection{Spinor notations for massive particles}

Spinor formalism was first developed for massless particles, and
helped to simplify calculations. However, spinor formalism can also
be defined for massive particles \cite{Dittmaier:1998nn,
Ozeren:2006ft, Schwinn:2007ee, Schwinn:2005pi, Rodrigo:2005eu}. In
contrast with massless particles, the helicity of massive particles
do not transform co-variantly under Lorentz transformations.
Helicity eigenstates are frame dependent. This makes the formalism
complicated. There are two approaches  for massive spinor formalism.
The first is to use massive Dirac equations to solve wave functions.
The second is to decompose massive momentum $p_\mu$ with the help of
an auxiliary null-momentum $q_\mu$. These two methods are equivalent
and we now present the second approach.

Using a reference null-momentum $q$,
a non-null momentum $p^2=-m^2$ can be decomposed as (using the QCD convention)
\bea p_\mu= p^\flat_\mu-{m^2\over 2p\cdot q} q_\mu=
p^\flat_\mu-{m^2\over \Spab{q|p|q}}
q_\mu~.~~~~\label{Massive-p-qdec} \eea
Since $(p^\flat)_\mu$ is massless (i.e., $(p^\flat)^2=0$), we can
find its spinor $\ket{p^\flat}$ and anti-spinor $\bket{p^\flat}$,
which are given by
\bea p^\flat= \ket{p^\flat}\bbra{p^\flat}+
\bket{p^\flat}\bra{p^\flat},~~~~\bket{p^\flat}={ \ket{p|q}\over
\sqrt{\Spab{q|p|q}}},~~\ket{p^\flat}={ \bket{p|q}\over
-\sqrt{\Spab{q|p|q}}}~.~~~\label{Massive-p-q-spinor}\eea

In the formula (\ref{Massive-p-q-spinor}), the denominator
$\sqrt{\Spab{q|p|q}}$ is for normalization. In the massless limit
$m\to 0$, spinor components return to standard ones only by
rescaling with factor $t$, i.e., $ \ket{p}\to t\ket{p}$, and
$\bket{p}\to t^{-1}\bket{p}$ with $t=\sqrt{\Spbb{q|p}/\Spaa{p|q}}$.
We can take another normalization by using $2p\cdot
q=\Spaa{p^\flat|q}\Spbb{q|p^\flat}$, so
\bea \bket{p^\flat}={ \ket{p|q}\over
\Spaa{p^\flat|q}},~~\ket{p^\flat}={ \bket{p|q}\over
\Spbb{p^\flat|q}}~,\eea
which will reduce to standard notation under the limit $m\to 0$.
With this new normalization,  wave functions of fermion and
anti-fermion of massive particles can be defined as follows
\cite{Schwinn:2005pi, Schwinn:2007ee}
\bea  u_+(p) & = &  \ket{p^\flat}+{m\over
\Spbb{p^\flat|q}}\bket{q},~~~~~~ u_-(p)=\bket{p^\flat}+{m\over
\Spaa{p^\flat|q}}\ket{q}\nn
 v_-(p) & = &  \ket{p^\flat}-{m\over
\Spbb{p^\flat|q}}\bket{q} ,~~~~~~ v_+(p)=\bket{p^\flat}-{m\over
\Spaa{p^\flat|q}}\ket{q}\nn
 \O u_{-}(p) &= & \bra{p^\flat}+{m\over
\Spbb{q|p^\flat}}\bbra{q},~~~~~~ \O u_{+}(p)=\bbra{p^\flat}+{m\over
\Spaa{q|p^\flat}}\bra{q},\nn
\O v_{+}(p)& = &\bra{p^\flat}-{m\over
\Spbb{q|p^\flat}}\bbra{q},~~~~~~ \O v_{-}(p)=\bbra{p^\flat}-{m\over
\Spaa{q|p^\flat}}\bra{q}~~~\label{SW-off-massive-def}\eea
%

\subsection{Color decompositions}

The complication of Feynman diagrams increases dramatically when
interactions with gauge fields are added. Feynman rules now contain
two kinds of information: one is dynamical and the another one is
group algebra carried by representations of fields. Color
decompositions \cite{Mangano:1990by,Dixon:1996wi} separate the
information  so one can deal with one thing once upon a time. We
take care of the group structure first, then concentrate on the
dynamical part. To see how it works, we take pure  $SU(N_c)$ gauge
theory as an example.

Gluons carry adjoint representation with color indices $a = 1, 2, .
. . ,N^2_c -1$, while quarks and antiquarks carry fundamental
representation $N_c$ or anti-fundamental representation $\O N_c$
indices, $i,\O j= 1, . . . ,N_c$. Generators of $SU(N_c)$ group in
the fundamental representation are traceless hermitian $N_c\times
N_c$ matrices, $(T^a)_i^{\O j}$, normalized according to $\Tr(T^a
T^b) = \delta_{ab}$. With this normalization we have
\bea f^{abc}={-i\over \sqrt{2}}{\rm Tr}(T^a[T^b,
T^c]),~~~~\label{f-abc} \eea
thus we can use the trace structure at the right-handed side to
replace $f^{abc}$ in Feynman rules. When we glue two vertices
together by a propagator, we need to sum over  colors
\bea \sum_{a=1}^{N_c^2-1} (T^a)_{i_1}^{\O j_1} (T^a)_{i_2}^{\O j_2}
=\delta_{i_1}^{\O j_2} \delta_{i_2}^{\O j_1}-{1\over N_c}
\delta_{i_1}^{\O j_1} \delta_{i_2}^{\O j_2}~~~\label{color-sum} \eea
which, when putting into the trace structure, is given by "Fierz"
identity (or completeness relations)
\bea & & \sum_a {\rm Tr}( X T^a) {\rm Tr}(T^a Y)= {\rm Tr}( XY)
-{1\over N_c} {\rm Tr}( X){\rm Tr}( Y)\nn
& & \sum_a {\rm Tr} ( X T^a Y T^a) = {\rm Tr}(X) {\rm Tr}(Y)-{1\over
N_c} {\rm Tr}(XY)~.~~~~~\label{color-sum-trace}\eea
Applying (\ref{f-abc}) and (\ref{color-sum-trace}) to all Feynman
diagrams, with simple algebraic manipulations (or more intuitive
double line notations \cite{Dixon:1996wi}), tree-level amplitudes of
gluons can be decomposed into the following structure\footnote{The
${1/N_c}$ parts  in (\ref{color-sum-trace})  are always canceled.}
\bea {\cal A}_{\rm tot}(\{k_i,\epsilon_i, a_i\}) & = &
\sum_{\sigma\in S_n/Z_n} {\rm Tr}( T^{a_{\sigma(1)}}
T^{a_{\sigma(2)}}...T^{a_{\sigma(n)}}) A_n^{\rm tree} (\sigma(1),
\sigma(2),...,\sigma(n))~~~\label{color-decom}\eea
where group information is separated from dynamical ones. The sum is
over all permutations of $n$-particle up to cyclic ordering. The
decomposition (\ref{color-decom}) is usually referred to as {\sl
color decomposition}, while $A_n^{\rm tree}$ are called {\sl partial
amplitudes} which contain all kinematic information.

The advantages of color decomposition are following:
\begin{itemize}
\item (1) Group information and kinematic information are separated.
When we change the gauge group from $SU(N_1)$ to $SU(N_2)$, no
calculations are needed, except those changing the fundamental
representation matrix of $SU(N_1)$ to $SU(N_2)$ in the trace part.
All partial amplitudes $A_{n}^{\rm tree}$ are same.
\item (2) The partial amplitude $A_{n}^{\rm tree}$ is the minimal gauge
invariant object. It is cyclic in the sense that
$A_n(\sigma_1,\sigma_2,...,\sigma_n)=A_n(\sigma_n,\sigma_1,...,\sigma_{n-1})$.
Partial amplitudes are simpler than full amplitudes and can be
calculated by using simpler color-ordered Feynman rules (see
\cite{Dixon:1996wi}). Due to color ordering,  they only receive
contributions from diagrams of a particular cyclic ordering of
gluons. Singularities of partial amplitudes, poles and cuts (in
loops), can only occur in a limited set of kinematic channels, those
made out of sums of cyclically adjacent momenta. These characters
make it easier to analyze partial amplitudes than full amplitudes.
\end{itemize}

Partial amplitudes have very interesting relations among themselves:
\begin{itemize}
\item {\sl Color-order reversed identity}
\bea  A_n(1,2,...,n-1,n)=(-)^n
A_n(n,n-1,...,2,1),~~~\label{color-reverse}\eea
\item {\sl $U(1)$-decoupling identity}
\bea \sum_{\sigma~ \rm cyclic}
A(1,\sigma(2),...,\sigma(n))=0,~~~\label{U1-decouple} \eea

\item {\sl Kleiss-Kuijf relations} were conjectured in \cite{Kleiss:1988ne}
and proved in \cite{DelDuca:1999rs}.
They are
\bea  A_n(1,\{\a\}, n,\{\b\}) = (-1)^{n_\b}\sum_{\sigma\in
OP(\{\a\},\{\b^T\})} A_n(1,\sigma, n)~.~~~~\label{KK-rel}\eea
The order-preserved (OP) sum is over all permutations of the set $\a \bigcup \b^T$,
where relative orderings in $\a$  and $\b^T$ (the reversed ordering of set $\b$) are preserved.
$n_\b$ is the number of $\b$ elements.
For example, six gluon amplitude are related as
\bea A(1,2,3,6,4,5) & = & A(1,2,3,5,4,6)+ A(1,2,5,3,4,6)+
A(1,2,5,4,3,6)\nn & & +
A(1,5,4,2,3,6)+A(1,5,2,4,3,6)+A(1,5,2,3,4,6)~.
~~~\label{KK-6-point}\eea
\item {\sl Bern-Carrasco-Johansson Relations} were conjectured in \cite{Bern:2008qj},
proved in string theory by
\cite{BjerrumBohr:2009rd,Stieberger:2009hq} and  in field theory by
\cite{Feng:2010my,Chen:2011jx}.
\end{itemize}

The cyclic property reduces the number of independent partial
amplitudes $A_n$ from $n!$ to $(n-1)!$ while the KK-relation reduces
it to $(n-2)!$ by fixing, for example, the first and last particles
to be $1$ and $n$. The newly discovered BCJ relation enable us to
reduce the number further down to $(n-3)!$. That is, we can fix
three particles at three fixed positions, for example, the first one
and last two.\footnote{In open string theory, with conformal
invariance, we can fix locations of three vertexes along the
boundary of disk diagram. This explains intuitively why we can fix
three positions in basis. } General expressions of other partial
amplitudes in this basis are quite complicated and given in
\cite{Bern:2008qj}. However, there are very simple relations, termed
as ``fundamental BCJ relations", which can be used to derive all
other expressions. Examples of fundamental BCJ relations are
\bea 0 & = & I_4=A(2,4,3,1)(s_{43}+s_{41})+A(2,3,4,1) s_{41}\nn
0 & = & I_5=A(2,4,3,5,1)(s_{43}+s_{45}+s_{41})\nn & &
+A(2,3,4,5,1)(s_{45}+ s_{41})+A(2,3,5,4,1)s_{41}\nn
0 & = & I_6=A(2,4,3,5,6,1)(s_{43}+s_{45}+s_{46}+s_{41}) \nn & &
+A(2,3,4,5,6,1)(s_{45}+s_{46}+s_{41})\nn & &
+A(2,3,5,4,6,1)(s_{46}+s_{41})+
A(2,3,5,6,4,1)s_{41}~~~\label{BCJ-funda}\eea
where we observe the pattern of how the particle $4$ is moving from
the second position to the $(n-1)$-th position with proper kinematic
factors $s_{ij}=(p_i+p_j)^2$.

In the above, we have only discussed color decompositions of
tree-level amplitudes of pure gluons. Similar decompositions can be
performed when we include matter fields like fermions or scalars, in different representations.
Loop-level decompositions also exist although things will become more complicated.

\subsection{Partial amplitudes in gauge theories}

Having introduced notions of color decompositions and partial amplitudes in gauge theory,
we now give a few examples for illustration.

On-shell gluons can have two choices of helicities (negative and
positive)\footnote{We will always define the helicity respect to
outgoing momenta.}, so partial amplitudes can be fixed by
color-ordering, helicity configuration and their momenta. The
simplest example is the partial amplitudes with  at most one of
particle to be negative helicity and their values are
\bea A_n^{\rm tree}(1^\pm, 2^+,..., n^+)=0~. \eea
The vanishing of these partial amplitudes can be explained by using
Feynman rules or supersymmetric Ward identities \cite{Dixon:1996wi}.
A less trivial example is the celebrated MHV (maximal helicity
violating) amplitude, where two particles are of negative  helicity
and all others are of positive,
\bea A_n(1^+,..,i^-, ..., j^-,...,n^+)= { \Spaa{i|j}^4\over
\Spaa{1|2}\Spaa{2|3}...\Spaa{n-1|n}\Spaa{n|1}}.~~~~~~\label{MHV}\eea
It was conjectured in \cite{Parke:1986gb} and proven in
\cite{Berends:1987me}. In particular, the three-point partial
amplitudes are
\bea A_3(1^-, 2^-, 3^+) = {\Spaa{1|2}^3\over \Spaa{2|3}
\Spaa{3|1}},~~~A_3(1^+, 2^+, 3^-) = {\Spbb{1|2}^3\over \Spbb{2|3}
\Spbb{3|1}}.~~~\label{3-point}\eea
Taking the conjugation of \eqref{MHV} we can obtain amplitudes
of two positive helicities only.

For an on-shell massless particle in spinor notation, there is an
operator counting its helicity $h$
\bea \left( \W\la^a {\partial \over \partial \W \la^a}-\la^a
{\partial \over \partial \la^a}\right) A = 2h_a A,~~{\rm given}
~a~~~\label{helicity-counting} \eea
Explicitly, each $\W\la$ is assigned charge one and $\la$ charge
minus one. In (\ref{MHV}), we see that for particle $i$ of negative
helicity $h_i=-1$, there are four $\la_i$ in the numerator and two
$\la_i$ in the denominator, thus we have $4(-1)-2(-1)=-2$. This
counting is simple, but very useful as a consistent check in
practical calculations.

MHV or its conjugation $\overline {\rm MHV}$ are the simplest non-zero partial amplitudes of gluons.
Calculations of other helicity configurations are not trivial, especially when the number of
gluons increases. Traditional methods relying on Feynman diagrams lose their
power because (1) there are too many diagrams;  (2) there are too many terms in each
diagram; (3) one diagram is usually not gauge invariant and many diagrams are
related by gauge invariance; (4) we get gauge invariant result
only when summing over certain  subsets. Consequently,
intermediate expressions tend to be vastly more complicated than
final results, when the latter are represented properly.
Feynman diagram method is simply not efficient to do such calculations.

To deal with the complexity in these calculations, many methods have
been developed over the years. One of the early methods is the
recursion relation of off-shell currents proposed in
\cite{Berends:1987me}. A recent method is the CSW method
\cite{Cachazo:2004kj} and another one is to use on-shell recursion
relations \cite{Britto:2004ap,Britto:2005fq}. Both methods were
trigged by Witten's twistor program \cite{Witten:2004cp}. A recent
review of the CSW method can be found in \cite{Brandhuber:2011ke}.
Our focus will be on-shell recursion relations.

\section{On-shell recursion relations}

In this section, we show how to use general analytic properties
of gluon amplitudes to derive on-shell recursion relations.
Originally, they were discovered \cite{Britto:2004ap}
by comparing infrared divergences of one-loop
amplitudes with tree-level amplitudes
\cite{Kunszt:1994mc,Bern:1994zx,Britto:2004nc,Roiban:2004ix}.

\subsection{Analytic properties of scattering amplitudes}

Naively, scattering amplitudes are functions of real variables
$p_\mu$. However, scattering amplitudes are fundamentally
meromorphic functions of complexified momenta, as consequences of
unitarity and causality. One obvious place to see these properties
is in the familiar $p^2+i\epsilon$  prescription of propagators.

The importance of analyticity was realized long ago. The so-called
S-matrix program\cite{S-matrix}, was proposed to understand
scattering amplitudes (especially in strong interactions) based only
on some general principles, such as Lorentz invariance, locality,
causality, gauge symmetry as well as analytic properties. Different
from the Lagrangian paradigm, the S-matrix program has generality
as its most distinguished feature: results so obtained do not rely
on any details of the theory.
 However, exactly because of its generality and with so little
assumptions, there are not many tools available and its study is
very challenging.
In this article, we will see that when we combine
the idea of S-matrix program and on-shell recursion
relations, many important results can be derived.


One analytic property most easily seen is in the pole structure.
Dictated by Feynman rules, amplitudes are constructed by connecting
interaction vertices through propagators. Because of locality,
obtained expressions by Feynman rules are always  {\sl meromorphic
functions of momenta and polarization vectors}, which have pole
structures when propagators are {\sl on-shell}. Branch cuts and
other singularities appear when these expressions are integrated
over. The singularity structure of tree-level amplitudes is very
simple: there are only poles.

Another well studied property is the {\sl soft limit}, i.e., when
all components of a momentum $k_\mu$ go to zero. For massive
particles, on-shell conditions prevent this to happen. For massless
particles, nothing prevents this. In QCD amplitudes, soft limits
have universal behaviors. For example, for tree-level partial
amplitudes one find
\bea A_n^{\rm tree}(...,a,s^+,b,...)\to {\rm
Soft}(a,s^+,b)A_{n-1}^{\rm tree}(...,a,b,...),~~~~k_s\to
0,~~~\label{Soft-QCD} \eea
where the soft or ``eikonal" factor is
\bea  {\rm Soft}(a,s^+,b)= {\Spaa{a|b}\over
\Spaa{a|s}\Spaa{s|b}}~.~~~\label{soft-factor}\eea

When momenta of external particles take certain particular values,
such that one inner propagator becomes on-shell, the amplitude will {\sl factorize}.
When this happens, all Feynman diagrams are divided into two categories,
those with and those without this particular propagator.
The leading contribution comes from the first category.
This can be expressed rigorously as the factorization property.
A simple example is the tree-level partial amplitude of gluons when $p_{1m}^2\to 0$
\bea A_n^{\rm tree}(1,2,...,n)\sim \sum_{h} A_{m+1}^{\rm
tree}(1,2,..,m,-p_{1,m}^h){1\over p_{1m}^2} A_{n-m+1}^{\rm
tree}(p_{1,m}^{-h}, m+1,..,n),~~~\label{Factorization}\eea
where the sum is over two physical helicities of the immediate
particle $p_{1,m}$. Under the limit $p_{1m}^2\to 0$, the amplitudes
at the left- and right-handed sides of the propagator become
on-shell amplitudes. The leading contribution depends on products of
two on-shell amplitudes.

If the number of particles involved in the factorization is equal to
or more than three, we will call it a multi-particle channel.
However, there is a special case called {\sl collinear channel},
where only two particles are involved in the factorization. For a
collinear channel, one on-shell amplitude is the three-point
amplitude. As we will see presently, by quite general reasonings,
the form of three-point amplitude can be uniquely fixed. Similar to
the soft limit, we have an universal behavior under collinear limit,
captured by the {\sl splitting function} \cite{Berends:1988zn,
Dixon:1996wi}. One handy way (up to signs) to derive these functions
is to use (\ref{3-point}). Assuming $p_a=-z p_c, p_b=-(1-z)p_c$,
the splitting function is
\bean {\rm Split}_{-}^{\rm tree} (a^+, b^+) = A_3(a^+, b^+, c^-) {1\over
s_{ab}}\sim { \Spbb{a|b}^3\over \Spbb{b|c}\Spbb{c|a}}{1\over
\Spaa{a|b}\Spbb{b|a}}= { -\Spbb{a|b}^2\over
\Spbb{b|c}\Spbb{c|a}}{1\over \Spaa{a|b}}\eean
Following from the real condition of momentum, the
complex conjugation of $\la$ is equal to $\W \la$.
So,
$|\Spaa{a|b}|=|\Spbb{a|b}|\sim \sqrt{ s_{ab}}=\sqrt{p_c^2}$ and
$|\Spbb{a|c}|\sim \sqrt{(p_a+p_c)^2}= \sqrt{(1-z) p_c^2}$,
$|\Spbb{b|c}|\sim \sqrt{(p_b+p_c)^2}= \sqrt{z p_c^2}$.
Plugging in all expressions, we have
\bea {\rm Split}_{-}^{\rm tree} (a^+, b^+)= {1\over \sqrt{z(1-z)}}
{1\over \Spaa{a|b}}\eea
Using similar method we have\footnote{To fix the sign, see \cite{Dixon:1996wi} for a more rigorous derivation.}
\bea {\rm Split}_{+}^{\rm tree} (a^+, b^+)=0,~~{\rm Split}_{+}^{\rm
tree} (a^+, b^-)={ (1-z)^2\over \sqrt{z(1-z)}}{1\over
\Spaa{a|b}},~~~{\rm Split}_{-}^{\rm tree} (a^+, b^-)={-z^2\over
\sqrt{z(1-z)}}{1\over \Spbb{a|b}}~.\eea
%

\subsection{BCFW-deformation}

As scattering amplitudes are understood to be meromorphic functions,
they are defined over complexified momenta. Starting from a momentum
configuration $(p_1,...,p_n)$ we can deform each of them in complex
planes. However, arbitrary deformation will loose momentum
conservation and bring on-shell momenta to off-shell. We are
interesting in some complex deformations, such that on-shell
conditions and the momentum conservation are kept. One of such
deformations is the {\sl BCFW deformation}.\footnote{As we shall see
in later sections, there can be other deformations. One example was
introduced in \cite{Risager:2005vk} for the proof of CSW rule. It is
very useful  in the proof of on-shell recursion relations for
gravitons \cite{Benincasa:2007qj} and has been applied in
supersymmetric field theories (see, for example,
\cite{Bianchi:2008pu}).}

In BCFW deformation, we pick two special momenta, for example, $p_i,
p_j$ as reference and deform them as
\bea p_i(z)= p_i+z q,~~~~p_j(z)=p_j-z q,~~~~\label{BCFW-deform-mom}
\eea
so momentum conservation is kept. To insist on-shell
conditions for arbitrary $z$, we require
\bea q^2= q\cdot p_i=q\cdot p_j=0~.~~\label{BCFW-con-mom}\eea
These equations can be solved for $q$ when the
dimension of space-time is four or above and the solution is in
general complex.  For example, we can take $q=\la_i \W \la_j$. In
this case, we have
\bea \la_i\to \la_i,~~~ \W\la_i\to \W\la_i+z\W\la_j,~~~ \W\la_j\to
\W\la_j,~~~\la_j\to \la_j-z\la_i~. \eea
This will be the mostly used deformation in this article and in the
literature. It will be referred to as the $\Spba{i|j}$-deformation
(it is worth to mention that $\Spba{i|j}$-deformation is different
from $\Spba{j|i}$-deformation). We will also use $(i|j)$-deformation
to denote either $\Spba{i|j}$- or $\Spba{j|i}$-deformations. In the
literature, the BCFW deformation is usually written in the above
spinor form. However, the momentum form (\ref{BCFW-deform-mom})
provides another perspective.\footnote{As we will discuss later,
BCFW deformation can be applied to massive theory too. In that case,
the momentum form is more natural.} Under this deformation, the
original partial amplitude\footnote{In this review, most times we
use $A$ to denote amplitudes of gauge theory while $M$ to,
amplitudes of other theories.} $A(p_i,p_j,...)$ becomes a
meromorphic function $A(z)$ of a single variable $z$, for which
powerful mathematical tools and claims are available. One of the
claims is following: {\sl for an meromorphic function with only pole
structures in complex plane, the knowledge of its pole locations and
corresponding residues uniquely determine the function}. We will see
consequences of this idea shortly.

Before ending this subsection, we mention one more thing about on-shell three-point amplitudes.
On-shell conditions tell that
\bea 2p_i\cdot p_i=\Spaa{i|j}\Spbb{j|i}=0,~~~~\forall i,j=1,2,3\eea
so either $\Spaa{i|j}$ or $\Spbb{i|j}$ must be zero.
There are two possible solutions:
\bea {\rm Solution}~(A): &~~ & \la_1\sim \la_2\sim
\la_3,~~~~\label{Sol-A}
\\ {\rm Solution}~(B): &~~ & \W\la_1\sim \W\la_2\sim \W\la_3,~~~~\label{Sol-B}
\eea
If we insist momenta to be real, $\W\la_i$ will be the complex
conjugation of $\la_i$. Thus we would have (\ref{Sol-A}) and
(\ref{Sol-B}) at same time and expression (\ref{3-point}) is zero.
That is, three-point on-shell amplitudes are zero for real momenta.
However, if momenta are complex, spinors $\la$ are independent of
anti-spinors $\W\la$, so (\ref{Sol-A}) and (\ref{Sol-B}) have not to
be true at same time. Three-point on-shell MHV-amplitudes are well
defined for the case (\ref{Sol-B}) and $\O{\rm MHV}$-amplitudes well
defined for the case (\ref{Sol-A}). This fact was first pointed in
\cite{Witten:2003nn} and is crucial for on-shell recursion
relations.

\subsection{Large $z$-behavior of amplitudes under BCFW-deformations}

Now we apply deformation (\ref{BCFW-deform-mom}) to MHV amplitudes
(\ref{MHV}) with various choices of reference momenta (remembering
that $i,j$ are negative helicities) and check its behavior when
$z\to \infty$:
\bea p_i(z) &= & p_i+z \la_i \W\la_j,~~~p_j(z)=p_j-z\la_i \W\la_j,
\Longrightarrow \left\{\begin{array}{l} A(z)\to  z^{-2},~~~i,j~{\rm not~nearby}\\
A(z)\to z^{-1},~~~i,j~{\rm nearby}
\end{array}\right.\nn
p_i(z) &= & p_i+z \la_i \W\la_k (k\neq j),~~~p_k(z)=p_k-z\la_i
\W\la_k, \Longrightarrow \left\{\begin{array}{l} A(z)\to z^{-2},~~~i,k~{\rm not~nearby}\\
A(z)\to z^{-1},~~~i,k~{\rm nearby}
\end{array}\right.\nn
p_i(z) &= & p_i+z \la_k \W\la_i (k\neq j),~~~p_k(z)=p_k-z\la_k
\W\la_i, \Longrightarrow \left\{\begin{array}{l} A(z)\to {z^2},~~~i,k~{\rm not~nearby}\\
A(z)\to {z^3},~~~i,k~{\rm nearby}
\end{array}\right.\nn
p_k(z) &= & p_k+z \la_l \W\la_k ,~~~p_l(z)=p_l-z\la_l
\W\la_k,~~k,l\neq i,j \Longrightarrow \left\{\begin{array}{l}
A(z)\to z^{-2},~~~l,k~{\rm not~nearby}\\
A(z)\to  z^{-1},~~~l,k~{\rm nearby}
\end{array}\right.~~~\label{Gluon-Bound}\eea
Although these behaviors are observed in MHV amplitudes, they are
true for all helicity configurations of gluons, as we will explain
shortly \cite{ArkaniHamed:2008yf}. They are nontrivial since
individual Feynman diagram does not vanish when $z\to \infty$ in
general. For gluons, no matter what helicity configuration one has,
there is at least a BCFW deformation available such that when $z\to
\infty$, $A(z)\to0$. It is not generally true for other theories.
One obvious example is the $\la \phi^4$ theory and we will come back
to these theories later.

Occasionally, the large $z$-behavior of amplitudes under BCFW-deformation
can be understood by direct inspections of Feynman diagrams.
In many circumstances, naive analysis of Feynman diagrams leads to wrong results, so we need a better way to
deal with the problem.

We start with a simple example, where the momentum of particle $i$
of positive helicity is shifted as $p_i(z)=(\la_i+z\la_j) \W\la_i$
and the momentum particle $j$ of negative helicity is shifted as
$p_j(z)=\la_j(\W\la_j-z\W\la_i)$, i.e., the
$\Spba{j|i}$-deformation. Under these, we see $\epsilon^+_i\sim
z^{-1}$ and $\epsilon_j^-\sim z^{-1}$ from (\ref{QCD-eps}). For
cubic vertices, shifted momenta contribute a factor of $z$, which
may render the amplitude non-vanishing when $z\to\infty$. Feynman
diagrams of leading $z$ behavior are those that along the line
connecting $i,j$, only cubic vertices are attached. Here we have $m$
cubic vertices and $m-1$ propagators, thus the overall factor is
${z^m/ z^{m-1}}\sim z$. To get on-shell amplitudes, we need to
multiply $\epsilon_{i}, \epsilon_j$, thus we have ${z/ z^2}\sim
z^{-1}$, i.e.,  vanishing behavior when $z\to \infty$.\footnote{See
also Peskin's lecture \cite{Peskin:2011in}.}

Having gained some intuition in this analysis, we move to insightful arguments
given in \cite{ArkaniHamed:2008yf}.
Note that momenta $p_i(z), p_j(z)\to \infty$ when $z\to \infty$.
If we take one particle as ingoing and another as outgoing, the
process is as if a hard light-like particle is
shooting through a soft background,  created by all
un-deformed particles.  The process can be analyzed by
working with quadratic fluctuations of the soft background.

For illustration, we analyze the amplitude of two scalars and $n$-photons in scalar QED,
by deforming momenta of two scalar particles.
The Lagrangian is $L_{s}= D_\mu \phi^* D^\mu \phi$, so the
$z$-dependent vertex is $\partial_\mu \phi^* A^\mu \phi\sim (q\cdot A) \phi^* \phi$.
If we choose the $q$-light cone gauge
\bea  q\cdot A=0~~~\label{q-light-cone}\eea
there is no vertex which contributes $z$ factors, since quartic vertices are $z$-independent.
However, each scalar
propagator contributes a $ z^{-1}$. One thus arrives at
\bea M_{n=2}(z)\to z^0,~~~~M_{n>2}(z)\to z^{-1} \eea
where the subscript $n$ means $n$-photons.  This result depends
crucially on the fact that there are only two-derivatives in the
Lagrangian. If terms like $D_\mu \phi^* D_\nu \phi F^{\mu\a} F^{\nu
\b} \eta_{\a\b}$ are introduced, the $z$-dependence will be totally
different.

There is a very important subtle point in the gauge choice (\ref{q-light-cone}).
This choice cannot be fulfilled in certain special cases.
From the equation for the gauge choice $\Lambda(p)$
\bea q_\mu A^\mu(p)+ i q_\mu p^\mu \Lambda(p)=0 \eea
one sees that the $\Lambda(p)$ can be solved if and only if $q\cdot p\neq0$.
In most Feynman diagrams, the momentum $p$ is the
sum of some external momenta and for generic momentum configurations
we do have $q\cdot p\neq 0$.
However, in diagrams where all other external particles interact with these two deformed particles
through a single cubic vertex, one actually has $q\cdot p=0$.
In scalar QED, we do not have such particular diagrams since there is no photon self-interaction.
In scalar-YM theory, we do have them, due to non-Abelian self-interactions.
For this type of diagrams in scalar-YM theory, we find its $z$-behavior
directly by inspection of Feynman rules
\bea M(z)\to \left\{ \begin{array}{ll} z~~~, & (i,j)~{\rm nearby}\\ z^0~~,
&  (i,j)~{\rm not~nearby} \end{array}\right.\eea

In gauge theories, the enhanced spin ``Lorentz" symmetry plays an important role in
controlling the large $z$-behavior (plus Ward-identities).
To see it, we decompose the gauge field ${\cal A}= A+a$  in terms of a background $A$ and the fluctuation $a$.
The Lagrangian becomes then
\bea L_{\rm YM}= -{1\over 4} {\rm tr} D_{[\mu} a_{\nu]}D^{[\mu}
a^{\mu]}+{i\over 2}{\rm tr} [a_\mu, a_\nu] F^{\mu\nu}+(D_\mu
a^\mu)^2= -{1\over 4} {\rm tr} D_{\mu} a_{\nu}D^{\mu}
a^{\mu}+{i\over 2}{\rm tr} [a_\mu, a_\nu]
F^{\mu\nu}~~~\label{YM-Lag-exp}\eea
where $(D_\mu a^\mu)^2$ has been added to fix the gauge of $a$.
In (\ref{YM-Lag-exp}) only the first term can potentially
have ${\cal O}(z)$-vertices and dominate the large $z$-behavior.
But the first term has also the enhanced spin
symmetry, i.e., {\sl a Lorentz transformation acting only on the
$\nu$-indices of $a_\nu$ alone while $D_\mu$ is
untouched}. To make it clear, we rewrite the Lagrangian as
\bea L_{\rm YM}=-{1\over 4} {\rm tr} \eta^{ab} D_{\mu} a_{a}D^{\mu}
a_{b}+{i\over 2}{\rm tr} [a_a, a_b]
F^{ab}~~~\label{YM-Lag-exp-1}\eea
where the second term break the enhanced spin symmetry explicitly.
With this symmetry argument, the amplitude will be of the general form
\bea M^{ab}= (c_1 z+c_0+c_{-1}{1\over z}+...) \eta^{ab}+
A^{ab}+{1\over z} B^{ab}+...\eea
where $A^{ab}$ comes from the second term of (\ref{YM-Lag-exp-1}) and is antisymmetric.
To get on-shell amplitudes, one needs to contract $M^{ab}$ with polarization vectors $\eps_i, \eps_j$.
Using the Ward-identity
\bea (p_i+zq)_a M^{ab} \eps_{jb}=0,~~~~~\Longrightarrow q_a
M^{ab}={-1\over z} p_{ia} M^{ab} \eps_{jb} \eea
For example, with $h_i=-, h_j=+$, we have $\eps_i^-(z)=\eps_j^+(z)=q$, thus
\bea M^{-+} & = &\eps_{ia}^- M^{ab} \eps_{jb}^+  = q_a M^{ab} q_b=
{-1\over z} p_{ia} \left[(c_1 z+c_0+c_{-1}{1\over z}+...) \eta^{ab}+
A^{ab}+{1\over z} B^{ab}+...\right] q_b \nn
& = & {-1\over z} p_{ia} A^{ab} q_b\to {1\over z}\eea
where $p_i\cdot q=0$ has been used. With $h_i=+, h_j=-$, we
have $\eps_{i}^+(z)=q^*-zp_j$, $\eps_j^-(z)=q^*+zp_i$, so
\bea M^{+-} & = &\eps_{ia}^+ M^{ab} \eps_{jb}^-  = (q^*-zp_j)
\left[(c_1 z+c_0+c_{-1}{1\over z}+...) \eta^{ab}+ A^{ab}+{1\over z}
B^{ab}+...\right]( q^*+zp_i) \nn
& \to &z^3\eea
We can do better by using the $q$-light-cone gauge to eliminate
${\cal O}(z)$-vertices up to a unique set of diagrams, which can exist only when $i,j$ are nearby.
If $i$ and $j$ are separated by at least two insertions of $F^{ab}$,
we will have $M^{-+}, M^{--}, M^{++}\to z^{-2}$ and $M^{+-}\to z^2$.

For pure gravity theory\footnote{The large $z$ behavior of gravity
is not so easy to discuss using Feynman diagrams directly, see
several discussions  in \cite{Bedford:2005yy, Cachazo:2005ca,
Benincasa:2007qj}.}, we again expand around the background and add a
de-Donder gauge fixing term. We then have
\bea L= \sqrt{-g}\left[{1\over 4}g^{\mu\nu} \nabla_\mu
h^\b_\a\nabla_\nu h^\a_\b-{1\over 8}g^{\mu\nu} \nabla_\mu
h^\a_\a\nabla_\nu h^\b_\b-h_{\a\b} h_{\mu\nu} {1\over 2}
R^{\b\mu\a\nu}+{1\over 2}g^{\mu\nu} \partial_\mu \phi\partial_\nu
\phi\right]\eea
where $\phi$ is the dilaton field and $R_{\mu\nu}=0$ by background field equation.
With field redefinition $h_{\mu\nu}\to
h_{\mu\nu}+g_{\mu\nu} \phi\sqrt{2\over D-2}$, $\phi\to {1\over
2}g^{\mu\nu} h_{\mu\nu}+\phi\sqrt{D-2\over 2}$, the Lagrangian is
simplified to
\bea L= \sqrt{-g}\left[{1\over 4}g^{\mu\nu}
g^{\a\rho}g^{\b\sigma}\nabla_\mu h_{\a\b} \nabla_\nu
h_{\rho\sigma}-{1\over 2} h_{\a\b}  h_{\mu\nu}
R^{\b\mu\a\nu}+{1\over 2}g^{\mu\nu}\partial_\mu \phi\partial_\nu
\phi\right] \eea
We will drop the dilaton field in later discussions. To make two
copies of enhanced spin symmetry,\footnote {Actually  $e=\W e$ and
$\omega=\W \omega$, but two copies make things transparent.} we
introduce the left vielbein $e$ and the right vielbein $\W e$, so
that
\bea h_{\mu\nu}= e^a_\mu \W e^{\W a}_\nu h_{a\W a},~~~~\nabla_\a
h_{\mu\nu}= e^a_\mu \W e^{\W a}_\nu D_\a h_{a\W a}=e^a_\mu \W e^{\W
a}_\nu\left( \partial_\a h_{a\W a}+ \omega^b_{\a a} h_{b \W a}+ \W
\omega^{\W b}_{\a\W a}h_{a\W b}\right)\eea
The final Lagrangian is given by
\bea L=\sqrt{-g}\left[ {1\over 4} g^{\mu\nu} \eta^{ab}\W\eta^{\W a\W
b} D_\mu h_{a\W a} D_\nu h_{b\W b}-{1\over 2} h_{a\W a} h_{b\W b}
R^{ab \W a\W b} \right]\eea
Taking the light-cone gauge $\omega_{ab}^+=\W\omega_{\W a\W
b}^+= g^{++}=g^{+i}=0$ and $g^{+1}=1$, the general pattern of
two-leg off-shell amplitude is
\bea M^{a\W a b\W b}=c z^2 \eta^{ab} \W\eta^{\W a\W b}+z(\eta^{ab}\W
A^{[\W a\W b]}+A^{[ab]}\W \eta^{\W a\W b})+A^{[ab][\W a\W
b]}+(\eta^{ab}\W B^{\W a\W b}+B^{ab} \W\eta^{\W a\W b})+{1\over z}
C^{ab\W a\W b} ~~~\label{Gra-M}\eea
where square brackets denote anti-symmetrization of indices. The form
(\ref{Gra-M}) manifestly has the form of ``squaring" a Yang-Mills theory,
similar to the KLT relation $M_{\rm gravity}\sim M_{\rm gauge}\times M_{\rm gauge}$.
Using Ward-identity to replace $q_a M^{a\W a, b\W b} \eps_{j, b\W
b}=- z^{-1} p_{ia}M^{a\W a, b\W b} \eps_{j, b\W b}$, so that
\bea M^{--,--}(z) & = & \eps_{i,a\W a}^{--} M^{a\W a, b\W b}
\eps_{j, b\W b}^{--}= {1\over z^2} p_{ia} p_{i \W a} M^{a\W a, b\W
b} (q^*_b+z p_{ib})( q^*_{\W b}+z p_{i\W b})\nn
& = &  {1\over z} C^{ab\W a\W b} p_{ia} p_{i \W a} p_{ib} p_{i\W
b}\to {1\over z}\eea
In fact, with a little extra work one can show that $C^{a b\W a\W
b}$ is the sum of terms antisymmetric in $(ab)$ and in $(\W a\W b)$.
The leading scaling behavior is actually $ z^{-2}$.

In short, the large $z$-behavior
of amplitudes under the BCFW-deformation is a nontrivial
intrinsic property of a theory. The understanding of this property
provides a new way to calculate  on-shell amplitudes,
as to be discussed now.

\subsection{On-shell recursion relations of gluons}

We now derive on-shell recursion relations
\cite{Britto:2004ap,Britto:2005fq} for partial
amplitudes.\footnote{There are also efforts to understand on-shell
recursion relations by using Feynman diagrams
\cite{Draggiotis:2005wq,Vaman:2005dt}.}
The starting point is the meromorphic function $A(z)$ of a single complex variable $z$,
obtained by picking up a pair of particles $(i,j)$ and
doing the BCFW-deformation (\ref{BCFW-deform-mom}), with proper
choice of $q$ such that $A(z)\to 0$ when $z\to \infty$.
From (\ref{Gluon-Bound}), one sees that no matter which
helicity configuration $(i,j)$ is, there is at least one choice of
$q$ satisfying the vanishing requirement of $A(z)$ at infinity of $z$.

The function $A(z)$ has a simple single-pole structure for general
momenta, due to propagators of the form
\bea {1\over (p+p_i(z))^2}= {1\over (p+p_i)^2+ z (2q\cdot
(p+p_i))}~~~~\label{BCFW-pole-form}\eea
where $p_j$ is not inside the momenta sum $p$. For any function
$A(z)$ has only single poles at finite $z$, we can consider  the
following contour integration
\bea I & = & \oint {dz \over z} A(z)~~~\label{BCFW-contour}\eea
where the contour is a big enough circle including all finite poles.
The integration can be evaluated in two different ways. The first is
to deform the contour to the infinity and we denote the result as
the boundary contribution $B$. If $A(z)\to 0$ as $z\to\infty$, the
boundary contribution $B=0$. The second is to deform the contour to
encircle all finite poles, so we have
\bea I & = & A(z=0) +\sum_{z_\a}{\rm Res} \left( {A(z)\over z}
\right)_{z_\a}\eea
where nonzero finite poles come from propagators of the form (\ref{BCFW-pole-form})
and $A(z=0)$ is the  tree-level partial amplitude we intend to calculate.
Identifying both evaluations, one has
\bea A(z=0)= B- \sum_{z_\a}{\rm Res} \left( {A(z)\over z}
\right)_{z_\a}~.~~~\label{BCFW-boundary}\eea
Now the key is to calculate  residues,  which can be obtained via
the  {\sl factorization property} (\ref{Factorization}). The residue
depends on products of two on-shell sub-amplitudes,
\bea \left( {A(z)\over z} \right)_{z_\a}= -\sum_{h=\pm} A_L(
p_i(z_\a), p^h(z_\a)) {1\over P_\a^2} A_R(-p^{-h}(z_\a), p_j(z_\a))
\eea
where one sums over two helicities of the inner on-shell propagator.
Putting all together we obtain a recursion relation for the gluon amplitude
\bea A_n & = &  \sum_{z_\a,h=\pm} A_L( p_i(z_\a), p^h(z_\a)) {1\over
p_\a^2} A_R(-p^{-h}(z_\a), p_j(z_\a))+B~~~~\label{BCFW-ij}\eea
where $B$ will be zero if the proper deformation makes $A(z)\to 0$
when $z\to \infty$.

For color-ordered partial amplitudes, allowed propagators always have momenta of
the form $p_{kt}=p_k+p_{k+1}+p_{k+2}+...+p_t$.
Thus,  the number of terms in the recursion formula depends on the choice of pair $(i,j)$.
When $i,j$ are adjacent, one has minimum number of terms.
For example, taking $(n-1,n)$ as the
BCFW-deformation pair with $q=\la_{n-1}\W\la_n$, we have
\bea & & A_n(1,2,... , (n-1)^-,n^+)\nn & =  &
\sum_{i=1}^{n-3}\sum_{h=+,-} \left ( A_{i+2}({\hat n},1,2,...,
i,-{\hat p}^h_{n,i} ) {1\over p^2_{n,i}} A_{n-i}(+{\hat
p}^{-h}_{n,i}, i+1,... , n-2, {\hat{n-1} } )
\right),~~~\label{BCFW-paper}\eea
where
\bea p_{n,i} & = & p_n+p_1+\ldots + p_i, \nn
\WH p_{n,i} & = & p_{n,i} +{p_{n,i}^2\over \gb{n-1|p_{n,i}|n}}
\lambda_{n-1} \tilde\lambda_{n}, \nn
\WH p_{n-1} & = & p_{n-1} -{p_{n,i}^2\over \gb{n-1|p_{n,i}|n}}
\lambda_{n-1} \tilde\lambda_{n} , \nn
 \WH p_{n} & = & p_{n}
+{p_{n,i}^2\over \gb{n-1|p_{n,i}|n}} \lambda_{n-1}
\tilde\lambda_{n}.~~~\label{BCFW-para}\eea
Shown in Figure \ref{fig:BCFW} is a pictorial representation of
(\ref{BCFW-paper}). Although it is  obvious that $\WH p_{n,i}$ is
null, it is not so easy to read out its spinor and anti-spinor
components. After some algebra, it can be shown that
\bea \ket{\WH p_{n,i}}= \a \bket{p_{n,i}|n},~~~~\bket{\WH
p_{n,i}}=\b \ket{p_{n,i}|n-1},~~~\a\b={1\over
\Spab{n-1|p_{n,i}|n}}~~~\label{P-comp} \eea
which are very useful in practical calculations.

\EPSFIGURE[ht]{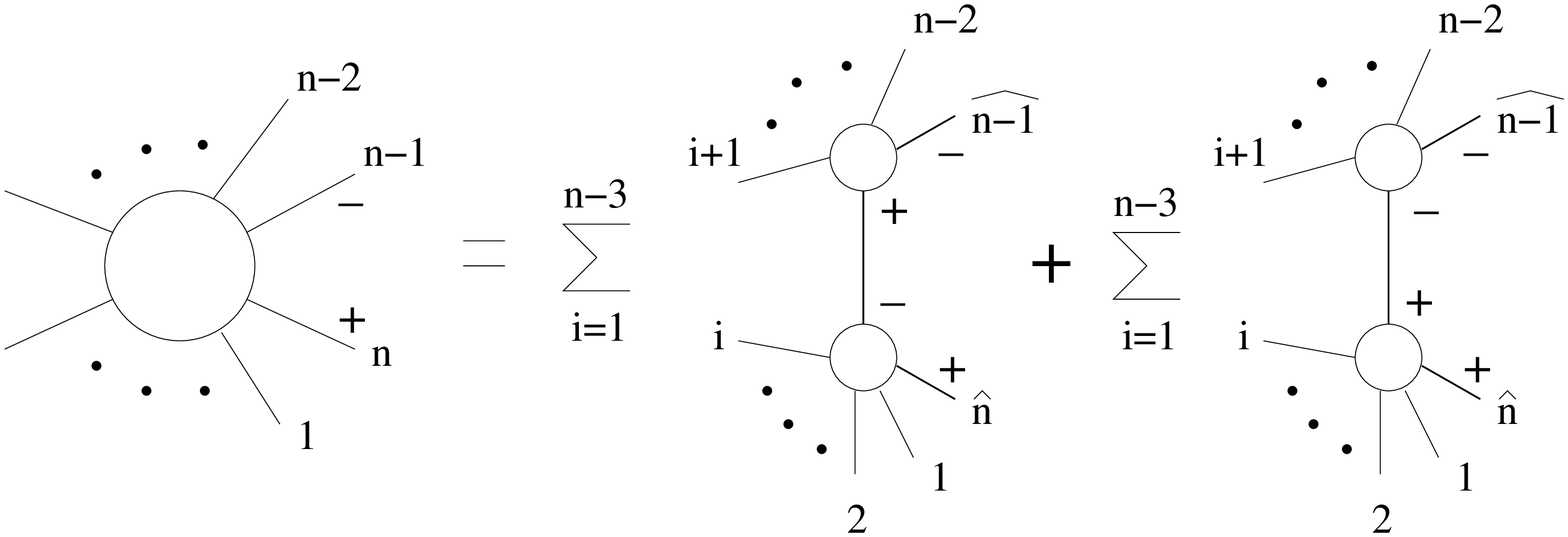,width=14.5cm} {Pictorial representation
of the recursion relation (\ref{BCFW-paper}). Note that the
difference between terms in the two sums is the helicity
assignment of the internal line.
    \label{fig:BCFW} }

We now demonstrate the usefulness of on-shell recursion relations in the example of six gluons.
It is the simplest case where non-MHV amplitudes show up.
We will calculate three helicity configurations: $(---+++)$, $(++-+--)$, $(+-+-+-)$.

We start with $A(1^-,2^-,3^-,4^+,5^+, 6^+)$, which is actually the simplest.
The reference gluons are chosen to be $\hat 3$ and $\hat 4$. There
are three possible configurations of external gluons, as shown in Figure \ref{fig:six}.
The middle graph vanishes. In the other two graphs, only one
helicity configuration of the internal gluon gives a nonzero answer.
We are left with only two graphs to evaluate. Moreover, the two graphs are
related by a flip of indices and a complex conjugation. Therefore,
only one computation is needed.

\EPSFIGURE[ht]{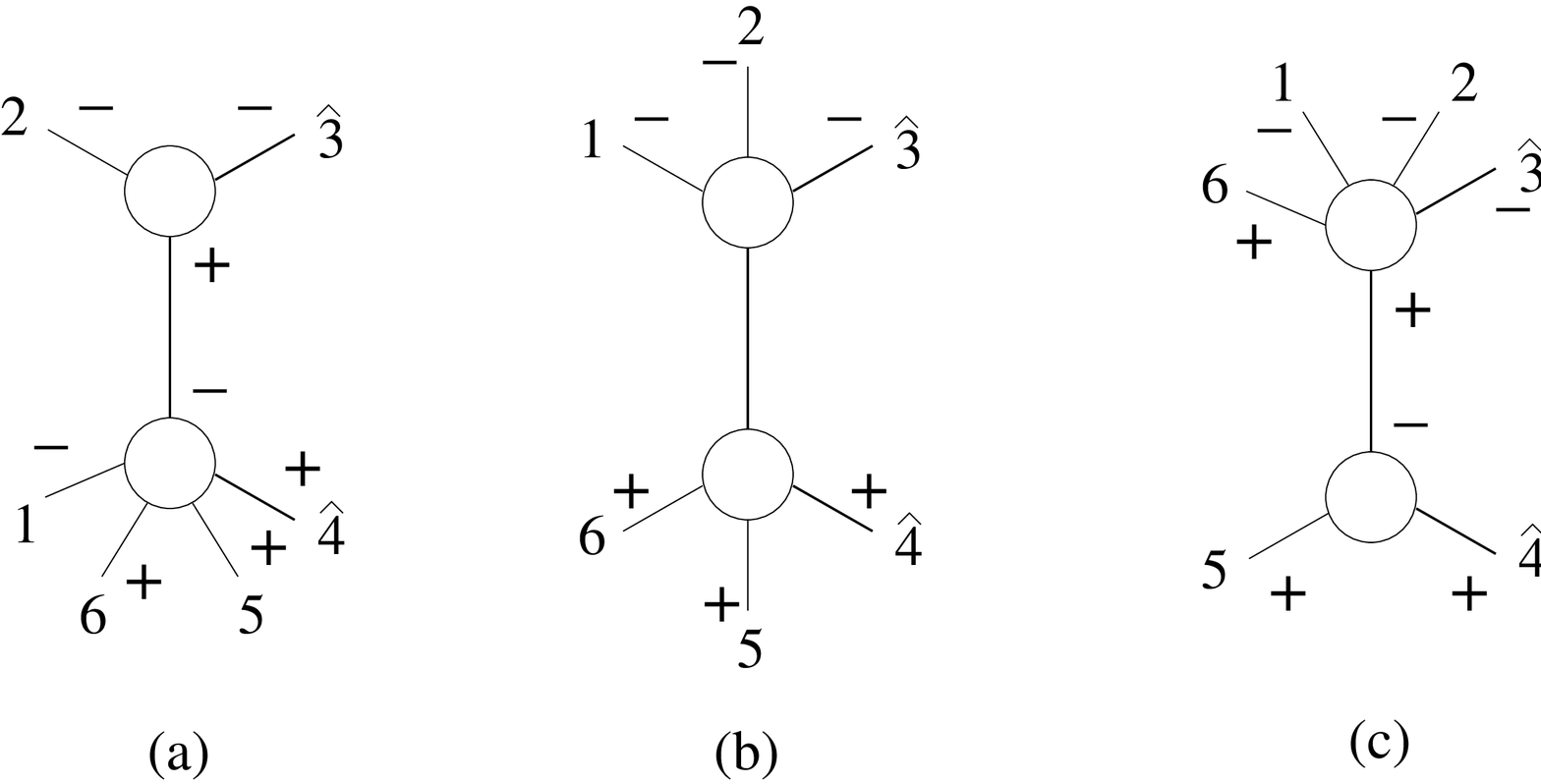,width=14.5cm} {Configurations contributing to
the six-gluon amplitude $A_6(1^-,2^-,3^-,4^+,5^+,6^+)$. Note that
$(a)$ and $(c)$ are related by a flip of indices and a complex
conjugation, $(b)$ vanishes for either helicity of the internal
line.
    \label{fig:six} }

Let us  work out the details of Figure \ref{fig:six}(a). It is the product of two MHV amplitudes and a
propagator,
\bea \left( \vev{2~\hat 3}^3\over \vev{\hat 3 ~\hat p}\vev{\hat p ~
2} \right) {1\over t_2^{[2]}}\left( {\vev{1~\hat p}^3 \over
\vev{\hat p ~ \hat 4}\vev{\hat 4~ 5}\vev{5~6}\vev{6~1} }
\right)~~~\label{6-split-1}\eea
where $t_i^{[k]}=\sum_{j=i}^{i+k-1} p_j$. Note that
\bea \lambda_{\hat 3} & = & \lambda_3,~~~ \lambda_{\hat 4}  =
\lambda_4 - {t_2^{[2]}\over \vev{3~2}[2~4]}\lambda_3,
~~~\vev{\bullet ~ \hat p}  = -{ \gb{ \bullet | 2+3 | 4} \over [\hat
P ~ 4]}.\eea
(\ref{6-split-1}) can straightforwardly be simplified to
\bea {\Spab{ 1 | 2+3 | 4}^3 \over [2~3][3~4]\Spaa{5~6}\Spaa{6~1}
t_2^{[3]}\Spab{ 5 | 3+4 | 2} } .~~~\label{6-split-2}\eea

Finally, performing the flip: $i\to i+3$ and  the complex
conjugation: $\Spaa{~~}  \leftrightarrow [~~]$ in (\ref{6-split-2}),
we obtain the expression for  Figure \ref{fig:six}(c). Adding them
up and factoring out a common term, we get
\bea A(1^-,2^-,3^-,4^+,5^+,6^+) = {1\over \gb{ 5 | 3+4 | 2}}\left(
{\gb{ 1 | 2+3 | 4}^3 \over [2~3][3~4]\vev{5~6}\vev{6~1} t_2^{[3]}} +
{\gb{ 3 | 4+5 | 6}^3 \over [6~1][1~2]\vev{3~4}\vev{4~5} t_3^{[3]}
}\right).\eea
Actually, this expression was first found \cite{Roiban:2004ix}
by taking a collinear limit of a seven-gluon amplitude in \cite{Bern:2004ky}.

For the other two helicity configurations, there are three terms.
Specifically,
\bea A(1^+,2^+,3^-,4^+,5^-,6^-) & = &{[2~4]^4\vev{5~6}^3\over
[2~3][3~4] \vev{6~1} t_{2}^{[3]} \gb{1|2+3|4} \gb{5|3+4|2}}\nn & & +
{ \gb{3|1+2|4}^4\over \vev{1~2}\vev{2~3}[4~5][5~6]  t_{1}^{[3]}
\gb{1|2+3|4} \gb{3|1+2|6}} \nn &  & + { [1~2]^3 \vev{3~5}^4\over
[6~1]\vev{3~4}\vev{4~5} t_{3}^{[3]} \gb{5|3+4|2} \gb{3|4+5|6}}~,
~~~\label{6-NMHV-2}\eea
from configurations $(2,\hat{3}|\hat{4},5,6,1)$,
$(1,2,\hat{3}|\hat{4},5,6)$, and $(6,1,2,\hat{3}|\hat{4},5)$, respectively; and
\bea A(1^+,2^-,3^+,4^-,5^+,6^-) & = & { [1~3]^4 \vev{4~6}^4 \over
[1~2] [2~3] \vev{4~5} \vev{5~6} t_{1}^{[3]} \gb{6|1+2|3}
\gb{4|2+3|1}} \nn & & + { \vev{2~6}^4 [3~5]^4\over
\vev{6~1}\vev{1~2} [3~4] [4~5] t_{3}^{[3]} \gb{6|4+5|3}
\gb{2|3+4|5}} \nn &  & + { [1~5]^4 \vev{2~4}^4 \over
\vev{2~3}\vev{3~4} [5~6] [6~1] t_{2}^{[3]} \gb{4|2+3|1}
\gb{2|3+4|5}}~,~~~\label{6-NMHV-3} \eea
from configurations
$(1,\hat{2}|\hat{3},4,5,6)$, $(6,1,\hat{2}|\hat{3},4,5)$, and
$(5,6,1,\hat{2}|\hat{3},4)$, respectively.
In $A(1^+,2^-,3^+,4^-,5^+,6^-)$, the second and the third terms can
be obtained from the first by shifting all indices: $i\to i+2$ and
$i\to i+4$.

The recursion relation (\ref{BCFW-ij}) expresses any amplitude in
terms of amplitudes of fewer gluons but of generic different
helicity configurations. It is hard to find closed-form solutions of
(\ref{BCFW-ij}) for general $n$.\footnote{In \cite{Dixon:2010ik},
compact analytical formulae for all tree-level color-ordered gauge
theory amplitudes involving any number of external gluons and up to
three massless quark-anti-quark pairs have been given by projecting
the known expressions \cite{Drummond:2008cr} for ${\cal N}=4$ SYM
theory. The result for ${\cal N}=4$ SYM theory   will be presented
shortly. }

However, there is a set of amplitudes that closes under the recursion procedure.
In other words, a given amplitude in the set is determined by amplitudes in the set only.
They are amplitudes of the form
\bea A_{p,q} = A(1^-,2^-,\ldots , p^-, (p+1)^+ , \ldots ,
(p+q)^+)~~~\label{Split-pq}\eea
for any integers $p\geq 1$ and $q\geq 1$, which will be referred to
as split helicity amplitudes \cite{Britto:2005dg}. Let us apply the
recursion formula (\ref{BCFW-ij}) to (\ref{Split-pq}) by taking
$p^-$ and $(p+1)^+$ to be reference gluons. Only two terms appear
and they are
\bea & &((p-1), \hat{p}~|~ \widehat{p+1},\ldots ,(p+q),1,\ldots ,
(p-2)), \nn & & ((p+3),\ldots , (p+q), 1,\ldots
,\hat{p}~|~\widehat{p+1},(p+2)).\eea
The first term  depends on $A_{p-1,q}\times A_{2,1}$,  while the second on $A_{p,q-1}\times A_{1,2}$.
Thus the set of amplitudes (\ref{Split-pq}) closes under (\ref{BCFW-ij}).

Denote the number of terms in $A_{p,q}$ as $N_{p,q}$. It satisfies
also a recursion relation: $ N_{p,q}= N_{p-1,q}+N_{p,q-1} $ with
boundary conditions $N_{2,q}=1,~\forall q\geq 1$ and
$N_{p,2}=1,~\forall p\geq 1$. This relation can be solved by a
binomial coefficient: $N_{p,q} = C_{p+q-4}^{p-2}$. A general
solution for split helicity amplitudes will be provided in section
5.

\section{On-shell recursion relations: further developments}


The derivation of gluon on-shell recursion relations depends on the
following observations: (1) for tree-level amplitudes, there are
only single poles from propagators under BCFW-deformation; (2) the
residues of single poles  are determined by factorization
properties; (3) with proper choice of deformation, the boundary
contribution is zero. Among these observations, the first two are
universal for all local quantum field theories. One naturally
generalizes on-shell recursion relations to other quantum field
theories, by carefully taking care of boundary
contributions.\footnote {For more general discussions, see for
example,  \cite{ArkaniHamed:2008yf, Cheung:2008dn}.} These include,
gravity theory \cite{Bedford:2005yy,
Cachazo:2005ca,Benincasa:2007qj}, gluons coupled with fermions\footnote
{In \cite{Luo:2005my}, applications to fermions in fundamental representations were emphasized,
as they are of particular importance in LHC physics.}
\cite{Luo:2005rx, Luo:2005my, Quigley:2005cu, Ozeren:2006ft}, gluons
coupled with massive scalars \cite{Badger:2005zh}, gluons coupled
with massive vector bosons \cite{Badger:2005jv}, quiver gauge
theories \cite{Park:2006va}, gluons coupled with Higgs particles
\cite{Berger:2006sh}, one-loop integral coefficients
\cite{Bern:2005hh, Brandhuber:2007up}, one-loop rational part
\cite{Bern:2005hs,Bern:2005ji, Bern:2005cq},  supersymmetric
theories \cite{Brandhuber:2008pf,ArkaniHamed:2008gz}, other
dimensions \cite{Cheung:2009dc, Gang:2010gy}, off-shell currents
\cite{Feng:2011tw}, etc.
These generalizations have forms  similar to (\ref{BCFW-ij}), of
expressions in term of sub-amplitudes $A_L, A_R$ and helicity sums
of middle propagators. But there are new features in different
situations and some of them will be presented in later subsections.
Before going to explicit generalizations, we  discuss two issues.

When we deal with massive particles instead of massless ones, we
need a null momentum $q$ satisfying conditions (\ref{BCFW-con-mom})
\cite{Badger:2005zh} to define deformation (\ref{BCFW-deform-mom}).
In the massless case, one simply has $q=\la_i\W\la_j$ or
$q=\la_j\W\la_i$. For massive momenta, things are more complicated.
For the case where $p_i^2=0$ and $p_j^2\neq 0$, one finds
\bea q_{\a \dot \a}= \ket{i} ( \ket{p_j|i})= \la_{i\a} p_{j\dot \a
\b}\la^\b_i,~~~~{\rm or}~~~q_{\a \dot \a}= \bket{i} ( \bket{p_j|i})=
\W\la_{i\dot \a} p_{j\a \dot
\b}\W\la^{\dot\b}_i~.~~~\label{BCFW-q-1mass}\eea
For the case $p_i^2\neq 0$, $p_j^2\neq 0$, we first construct two
null momenta by linear combinations $\eta_\pm=(p_i+x_\pm p_j)$ with
$x_\pm= \left(-2p_i\cdot p_i\pm \sqrt{ (2p_i\cdot p_j)^2-4 p_i^2
p_j^2}\right)/ 2p_j^2$.  The solution can then be written as
\bea q= \la_{\eta_+}\W\la_{\eta_-},~~~~~{\rm or}~~~~~~q=
\la_{\eta_-}\W\la_{\eta_+}~.~~~\label{BCFW-q-2mass}\eea

When fermion propagators are involved, extra care should be taken
for signs. In sub-amplitudes $A_L, A_R$,  we usually take all
momenta to be out-coming (or in-going),. Depending on the choice of
the momentum $p$ of the middle propagator, we will have either
$A_L(p,..) A_R(-p,...)$ or $A_L(-p,...) A_R(p,..)$. The definition
of spinors of negative momentum $-p$ is not unique. One option is to
take $\la_{p}=\la_{-p}$, $\W\la_{p}=-\W\la_{-p}$.

Notice that the propagator
used in the on-shell recursion relation is always the scalar-like ${i/ p^2}$.
This form is natural for scalars or vectors,  but not for fermions.
The familiar fermion propagator is ${i\cancel{p}/ p^2}$.
In the sum $\sum_\pm A_L(p) A_R(-p)$, one term will yield $\bket{p}\bra{-p}=\bket{p}\bra{p}$,
while the other $\ket{p}\bbra{-p}=-\ket{p}\bbra{p}$.
However,  $\ket{p}\bbra{p}+\bket{p}\bra{p}=\cancel{p}$, according to identity (\ref{Project}).
Thus to reproduce the desired $\cancel{p}$, there must be a relative sign between the two
terms in the sum.  More details about signs
can be found in the Appendix of \cite{Georgiou:2004wu}.


Now we are going to present several generalizations. Each has its
unique points. The generalization to supersymmetric theories groups
fermions and bosons together as well as different helicity states.
The generalization to off-shell current and amplitudes with
un-vanishing boundary contributions extend the scope of
applications. Bonus relation provides more relations other than
on-shell recursion relations. The rational part of one-loop
amplitudes has double poles, which do not exist in tree-level
amplitudes. The generalization to three dimensional space-time will
have the feature that  propagators are quadratic function of $z$.
Finally, the proof of CSW rule uses different analytic deformation
comparing to BCFW-deformation.

\subsection{Generalization to supersymmetric theories}

In a supersymmetric theory, bosonic and fermionic fields can be
grouped together in super-multiplets with the help of Grassmannian
coordinates. External lines of supersymmetric scattering amplitudes
can be represented in super-fields. To get scattering amplitudes of
component fields, we just need  to expand Grassmannian coordinates.

To make the discussion more concrete, we focus on the ${\cal N}=4$
theory. First, we can write down a super-wave-function of Grassmann
variables $\eta^{A}$ ($A=1,2,3,4$ is for R-symmetry $SU(4)$ of
${\cal N}=4$ theory)
\bea
  \Phi(p,\eta) &=& G^{+}(p) + \eta^A \Gamma_A(p) + \frac{1}{2}\eta^A \eta^B S_{AB}(p)
  + \frac{1}{3!}\eta^A\eta^B\eta^C \epsilon_{ABCD} \bar\Gamma^{D}(p) \nonumber \\
  &&\  + \frac{1}{4!}\eta^A\eta^B\eta^C \eta^D \epsilon_{ABCD}
  G^{-}(p)\,,~~~~\label{super-wave}
\eea
which incorporates  all on-shell states of the theory.
The $\mathcal{N}=4$ version of MHV tree amplitudes is \cite{Nair:1988bq}
\bea
  {\cal A}^{\rm MHV}_{n}(\lambda,\tilde\lambda,\eta) = \frac{ \delta^{(4)}(\sum_i \la_i \W\la_i)\,
    \delta^{(8)} (\sum_{i=1}^n\
\lambda_{i}^{\alpha}\, \eta^A_i)}{\vev{1\, 2}\vev{2\,
3}\ldots\vev{n\, 1}}~. ~~\label{intro-MHV-n}\eea
 The combination $\sum_{i=1}^n\
\lambda_{i}^{\alpha}\, \eta^A_i$ is called the ``super-momentum" and
the appearance of $ \delta^{(8)}$ is dictated by ${\mathcal N}=4$
supersymmetry to impose super-momentum conservation, just as
$\delta^{(4)}$ ensures ordinary momentum conservation. To get
amplitudes for component fields, we expand (\ref{intro-MHV-n}) and
read out corresponding Grassmannian components. For example,
\bea \delta^{(8)} (\sum_{i=1}^n\ \lambda_{i}^{\alpha}\,
\eta^A_i)=\delta^4 (\sum_{i<j} \Spaa{i|j} \eta_i^A\eta_j^A)\to
\Spaa{i|j}^4 \prod_{A=1}^4 \eta_i^A \eta_j^A \eea
$\prod_{A=1}^4 \eta_i^A\eta_j^A$ corresponds to the amplitude of
all particles are gluons of positive helicity, except $i,j$ of negative helicity.
If  all particles are gluons of positive helicity
except that $i$ is a gluon of negative helicity, $j$ a fermion of
positive spin and $k$ a fermion of negative spin, we need the
term $\prod_{A=1}^4\eta_i^A \eta_j^1 \prod_{B=2}^4 \eta_k^B$.
It is given by
\bea \delta^4 (\sum_{i<j} \Spaa{i|j} \eta_i^A\eta_j^A)\to
\Spaa{i|k}^3 \Spaa{i|j} \eta_i^1 \eta_j^1\prod_{A=2}^4 \eta_i^A
\eta_j^A\eea
where we need to be careful about signs when exchanging
Grassmanian variables.

As one sees, the helicity information is now hidden in $\eta$'s.
One thus needs to generalize the BCFW-deformation to the following
\bea \la_i(z)=\la_i+z\la_j,~~~\W\la_j(z)=\W\la_j-z\W\la_j,
~~~\eta_j(z)=\eta_j-z\eta_i,~~~~\label{BCFW-def-susy}\eea
so both momentum and super-momentum conservations are kept. The
large $z$ behavior of super-amplitudes is universal.
It is $z^{-1}$ when $i,j$ are nearby and $ z^{-2}$ when $i,j$ are not
nearby.  The on-shell recursion relation in ${\cal N}=4$ theory is \cite{Brandhuber:2008pf,ArkaniHamed:2008gz}
  \bea
 \mathcal{A} = \sum_{\rm split~\a} \int d^{4}\eta_{P_{i}}
\mathcal{A}_{L}(p_i(z_\a), p_{\a}(z_\a)) \frac{1}{p_{\a}^2}
\mathcal{A}_{R}(p_j(z_\a), -p_{\a}(z_\a))\,.
~~~\label{BCF-super}\eea
where the integration is over Grassmanian variables and the sum is
over all possible inner propagators.

There is one nice thing in (\ref{BCF-super}). All helicity
configurations are packed together as one-object, the recursion
relation is closed. That is, all sub-amplitudes are of the  same
type, just like the case of split helicity amplitudes. Thus it is
possible to solve the recursion relation explicitly, as we will see
in the next section.

\subsection{Recursion relations for off-shell currents}

Before the discovery of on-shell recursion relations, an off-shell
 recursion relation (also known as Berends-Giele recursion relation)
for gluon current was proposed \cite{Berends:1987me} based directly
from Feynman diagrams, where all external lines except one are
on-shell. The off-shell recursion relation of current $J^\mu$
is\footnote {The factor ${-i}/{p_{1,k}^{2}}$ indicates that the
gluon propagator is in the Feynman gauge.}
 \bea
J^{\mu}\left(1,2,...,k\right) & = & \frac{-i}{p_{1,k}^{2}}
\left[\sum_{i=1}^{k-1}V_{3}^{\mu\nu\rho}\left(p_{1,i},p_{i+1,k}\right)J_{\nu}
\left(1,...,i\right)J_{\rho}\left(i+1,...,k\right)\right.\nonumber\\
 &  & \left.+\sum_{j=i+1}^{k-1}\sum_{i=1}^{k-2}V_{4}^{\mu\nu\rho\sigma}
J_{\nu}\left(1,...,i\right)J_{\rho}\left(i+1,...,j\right)J_{\sigma}
\left(j+1,...,k\right)\right]\label{2.4}\eea
where $p_{i,j}=p_{i}+p_{i+1}+\cdots+p_{j}$.
$p_{1,k}$ is the momentum of the off-shell line and vertices are
 \bea
V_{3}^{\mu\nu\rho}\left(p,q\right) & = & \frac{i}{\sqrt{2}}
\left(\eta^{\nu\rho}\left(p-q\right)^{\mu}+2\eta^{\rho\mu}q^{\nu}-2\eta^{\mu\nu}p^{\rho}\right)\nonumber\\
V_{4}^{\mu\nu\rho\sigma} & = &
\frac{i}{2}\left(2\eta^{\mu\rho}\eta^{\nu\rho}-\eta^{\mu\nu}\eta^{\rho\sigma}-
\eta^{\mu\sigma}\eta^{\nu\rho}\right)~.~~~~\label{2.3}\eea
Recursion (\ref{2.4}) starts with
$J^{\mu}(1)=\epsilon^{\pm\mu}(p_1)$, which is the current with only
one on-shell gluon. Shown in Fig \ref{Fig:division} is a graphic
presentation of (\ref{2.4}). As we have mentioned, the off-shell
recursion relation (\ref{2.4}) has been used to prove the validity of Parke-Taylor formula of
MHV-amplitudes.

   \begin{figure}[hbt]
  \centering
  \scalebox{1.13}[1.13]{\includegraphics{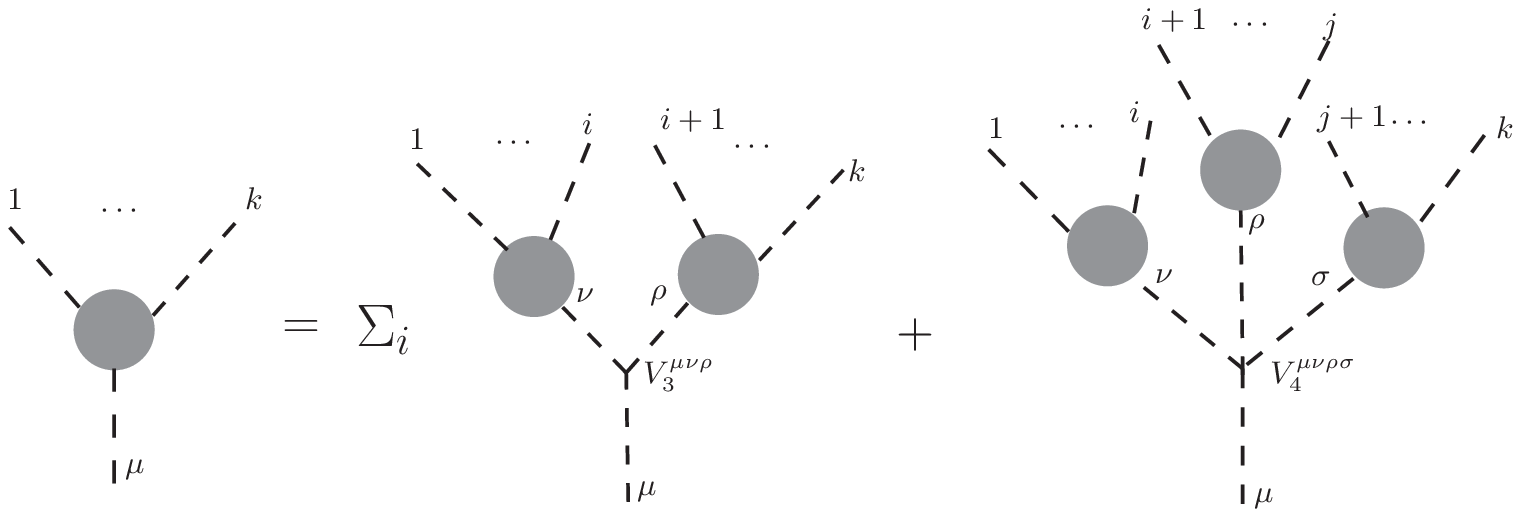}}\quad
  \caption{A graphic description for the off-shell recursion relation of a gluon current. \label{Fig:division}}
\end{figure}

Different from gauge invariant on-shell amplitudes, $J^{\mu}(1,2,...,k)$ is gauge dependent,
as there is a line which is not on-shell and not contracted with a physical polarization vector.
The gauge freedom arises in several places.
The first is in the choice of null reference momentum to define polarization vectors of external on-shell gluons
\bea \epsilon_{i\mu}^{+}= \frac{\Spab{ r_i|\gamma_{\mu}|p_i}}{\sqrt2
\Spaa{r_i|p_i}},~~~~~\epsilon_{i\mu}^{-}= \frac{\Spba{
r_i|\gamma_{\mu}|p_i}}{\sqrt2 \Spbb{r_i|p_i}}~~~~~\label{2.1}\eea
where $p_i$ is the momentum of the $i$-th gluon and $r_i$ is its
null reference momentum. The second is in the choice of gluon
propagator and we always use the Feynman gauge. To deal with the
gauge dependence, we need to define two more polarization
vectors\footnote{The $\eps^L, \eps^T$ have different mass-dimensions
from those of $\eps^\pm$. This can be fixed by proper factors such as
$\sqrt{p_i^2}$ when $p_i^2\neq 0$. For our purpose, this factor does
not matter and we will use \eqref{2.1} for simplicity.}
\bea \epsilon_\mu^L =
p_i,~~~~\epsilon_\mu^T={\Spab{r_i|\gamma_\mu|r_i}\over 2 p_i\cdot
r_i}~~~\label{2.1.1}\eea
Using {\bf Fierz rearrangements} (\ref{Fierz}), one sees that
\bea 0 & = & \epsilon^+\cdot \epsilon^+= \epsilon^+\cdot
\epsilon^L=\epsilon^+\cdot \epsilon^T=\epsilon^-\cdot
\epsilon^-=\epsilon^-\cdot \epsilon^L=\epsilon^-\cdot \epsilon^T
=\epsilon^T\cdot \epsilon^T=\epsilon^L\cdot \epsilon^L\nonumber \\
1 &= & \epsilon^+\cdot \epsilon^- =\epsilon^L\cdot
\epsilon^T~~~\label{2.1.2}\eea
These four $\epsilon$ vectors form a basis of the four-dimension space-time and we have
\bea g_{\mu\nu}= \epsilon_\mu^+ \epsilon_\nu^- +\epsilon_\mu^-
\epsilon_\nu^+ +\epsilon_\mu^L \epsilon_\nu^T+\epsilon_\mu^T
\epsilon_\nu^L ~~~\label{2.1.3}\eea
which is the numerator of the gluon propagator in Feynman gauge.

Now we are ready to write down the on-shell recursion relation for
$J^\mu$ \cite{Feng:2011tw}. Since $J^\mu(1,2,...,k)$ has $k$
on-shell gluons,  we can take a pair of on-shell gluons to do the
BCFW-deformation and write down the corresponding on-shell recursion
relation. The boundary behavior under the $\Spba{i|j}$-deformation
will be same as those of on-shell amplitudes.
That is,  $z^{-1}$ for
helicity configurations $(-,+),(+,+),(-,-)$ and $z^3$ for the
helicity configuration $(+,-)$.\footnote{The boundary behavior is in
fact more subtle. For example, if $(i,j)$ are not nearby, we will
have ${1/ z^2}$ behavior for $(-,+),(+,+),(-,-)$. For our purpose,
naive counting is enough.}$^{,}$\footnote{The boundary behavior of
off-shell particles has also been analyzed in \cite{Boels:2011mn}
 directly by using Feynman diagrams. } The off-shell line causes no extra problem.

Taking $(i,j)=(1,k)$, one has
 \bea
J^{\mu}\left(1,2,...,k\right) & = & \sum_{i=2}^{k-1}\sum_{h,\W h}
\left[A\left(\WH{1},...,i,\WH{p}^{h}\right)\cdot\frac{1}{p_{1,i}^{2}}
\cdot J^{\mu}\left(-\WH{p}^{\W h},i+1,...,\WH{k}\right)\right.\nonumber\\
+J^{\mu}\left(\WH{1},...,i,\WH{p}^{h}\right)& &\left.
\cdot\frac{1}{p_{i+1,k}^{2}}\cdot A\left(-\WH{p}^{\W
h},i+1,...,\WH{k}\right)\right],~~
 (h,\W h)=(+,-),(-,+),(L,T),(T,L)~~\label{2.16}\eea
There are several things in (\ref{2.16}) which need to be mentioned.
First, since $J^{\mu}\left(1,2,...,k\right)$ is gauge dependent, all
reference momenta in sub-currents on the right-handed side of
(\ref{2.16}) must  be the same as those on the left-handed side of (\ref{2.16}).
Secondly, for the on-shell momentum $\WH p$ on the right-handed side
of (\ref{2.16}), we must sum over all four polarization vectors in
(\ref{2.1})  and (\ref{2.1.1}) (not just vectors in (\ref{2.1})). As
we see in (\ref{2.1.3}), they decompose the $g_{\mu\nu}$ factor of
gluon propagator in Feynman gauge. We can neglect vectors in
(\ref{2.1.1}) for on-shell amplitudes, because $\epsilon^L=\WH p$
and when all other particles are on-shell and with physical
polarizations, $\WH p\cdot A=0$ by virtue of Ward identities. For
configurations $(h,\W h)=(L,T),(T,L)$ in (\ref{2.16}), we have
either $\WH p\cdot A_L=0$ or $\WH p\cdot A_R=0$, so we are left with
only two familiar  helicity configurations in recursion relations
for on-shell amplitudes. For current $J^\mu$, we do not have
$\epsilon^{L,T}\cdot J\neq 0$, thus we can not neglect the sum over
$(h,\W h)=(L,T),(T,L)$. However, we will show that usually these two
terms vanish by a special choice of gauge. Also we can use Ward
identities to simplify calculations. For example,  for the
$(h,\tilde{h})=(T,L)$ configuration, the second term in (\ref{2.16})
vanishes due to Ward identity
\bea
A\left(-\WH{p}^{L},i+1,...,\WH{k}\right)=-\WH{p}^L_{\mu}\cdot
M^{\mu}\left(i+1,...,\WH{k}\right)=0 \eea

The starting point is of course the two-point off-shell currents of various helicity configurations
\bea J^{\mu}\left(1^{-},2^{+}\right) & = & \frac{1}{\sqrt{2}s_{12}}
\left(\frac{[r_{1}2]\langle1r_{2}\rangle}{[r_{1}1]\langle
r_{2}2\rangle}
\left(1-2\right)^{\mu}+\frac{[2r_{1}]\langle21\rangle}{[r_{1}1]\langle
r_{2}2 \rangle}\langle r_{2}|\gamma^{\mu}|2]+\frac{[12]\langle
r_{2}1\rangle}{[r_{1}1]
\langle r_{2}2\rangle}[r_{1}|\gamma^{\mu}|1\rangle\right)\label{3.17}\\
J^{\mu}\left(1^{+},2^{+}\right) & = &
\frac{1}{\sqrt{2}s_{12}}\left(\frac{[12]\langle r_{2}r_{1}
\rangle}{\langle r_{1}1\rangle\langle
r_{2}2\rangle}\left(1-2\right)^{\mu}+\frac{[21] \langle
r_{1}2\rangle}{\langle r_{1}1\rangle\langle r_{2}2\rangle}\langle
r_{2}|\gamma^{\mu}|2] +\frac{[21]\langle r_{2}1\rangle}{\langle
r_{1}1\rangle\langle r_{2}2\rangle}\langle r_{1}|
\gamma^{\mu}|1]\right)~~\label{2.18}\\
J^{\mu}\left(1^{-},2^{T}\right) & =
&\frac{1}{\sqrt2s_{12}}\left(-\frac{\Spaa{12}\Spbb{2r_1
}}{\Spbb{r_11}}\left(1+2\right)^{\mu}+\frac{\Spaa{12}\Spbb{21}}{\Spbb{r_11}}\Spab{1|\gamma^{\mu}|r_1}\right)
~~\label{2.18-1}\\
J^{\mu}\left(1^{+},2^{T}\right) & =
&\frac{1}{\sqrt2s_{12}}\left(\frac{\Spaa
{r_12}\Spbb{21}}{\Spaa{r_11}}\cdot\left(1+2\right)^{\mu}
-\frac{\Spaa{12}\Spbb{21}}{\Spaa{r1}}\Spab{r|\gamma^{\mu}|1}\right)~~\label{2.18-2}
\eea
where the gauge dependence of each on-shell gluon is kept.
Now we use these building blocks to calculate one example of three on-shell gluons.

\subsection*{The example of $J^{\mu}\left(1^{-},2^{+},3^{+}\right)$}
Under the $\Spba{i|j}$-deformation, the recursion relation is
\bea J^{\mu}\left(1^{-},2^{+},3^{+}\right)&  = &
J^{\mu}\left(\WH{1}^{-},\WH{p}^{+}\right) \cdot\frac{1}{s_{23}}\cdot
A\left(-\WH{p}^{-},\WH{2}^{+},3^{+}\right)
 +J^{\mu}\left(\WH{1}^{-},\WH{p}^{-}\right)\cdot\frac{1}{s_{23}}\cdot A\left(-\WH{p}^{+},
 \WH{2}^{+},3^{+}\right)\nonumber\\
& &
+J^{\mu}\left(\WH{1}^{-},\WH{p}^{L}\right)\cdot\frac{1}{s_{23}}\cdot
A\left(-\WH{p}^{T},\WH{2}^{+},3^{+}\right)
 +J^{\mu}\left(\WH{1}^{-},\WH{p}^{T}\right)\cdot\frac{1}{s_{23}}\cdot A\left(-\WH{p}^{L},\WH{2}^{+},3^{+}\right)
\label{3.23}\eea
Here there is only one cut $s_{23}$ due to color-ordering.
The second term in (\ref{3.23}) vanishes if all helicities are positive while
the fourth term vanishes due to Ward identity. Using a general reference
null-momentum $q$ for the internal gluon $\WH{p}$, the first and third terms
are given as
\bea
 &&J^{\mu}\left(\WH{1}^{-},\WH{p}^{+}\right)\cdot\frac{1}{s_{23}}\cdot
 A\left(-\WH{p}^{-},\WH{2}^{+},3^{+}\right)\nonumber\\
&&=\frac{1}{\sqrt2s_{\WH{1}\WH{p}}}\left[-\frac{\Spaa{\WH{1}q}\Spbb{\WH{p}\WH{2}}}
{\Spbb{\WH{2}\WH{1}}\Spaa{q\WH{p}}}\left(\WH{1}-\WH{p}\right)^{\mu}-
\frac{\Spaa{\WH{p}\WH{1}}\Spbb{\WH{2}\WH{p}}}
{\Spaa{q\WH{p}}\Spbb{\WH{2}\WH{1}}}\Spab{q|\gamma^{\mu}
|\WH{p}}+\frac{\Spaa{\WH{1}q}\Spbb{\WH{p}\WH{1}}}
{\Spbb{\WH{2}\WH{1}}\Spaa{q\WH{p}}}\Spab{\WH{1}|\gamma^{\mu}|
\WH{2}}\right]\cdot\frac{1}{s_{23}}\cdot\frac{\Spbb{\WH{2}3}^3}
{\Spbb{\WH{2}\WH{p}}\Spbb{\WH{p}3}}\nonumber\\
&&J^{\mu}\left(\WH{1}^{-},\WH{p}^{L}\right)\cdot\frac{1}{s_{23}}\cdot A\left(-\WH{p}^{T},
\WH{2}^{+},3^{+}\right)\nonumber\\
&&=\frac{1}{\sqrt2s_{\WH{1}\WH{p}}}\left[-\frac{\Spaa{\WH{1}\WH{p}}\Spbb{\WH{p}\WH{2}
}}{\Spbb{\WH{2}\WH{1}}}\left(\WH{1}+\WH{p}\right)^{\mu}+\frac{\Spaa{\WH{1}
\WH{p}}\Spbb{\WH{p}\WH{1}}}{\Spbb{\WH{2}\WH{1}}}\Spab{\WH{1}|\gamma^{\mu}|\WH{2}}\right]
\cdot\frac{1}{s_{23}}\cdot\frac{\Spaa{\WH{p}\WH{1}}\Spaa{q\WH{1}}\Spbb{\WH{2}3}}
{\Spaa{\WH{p}q}\Spaa{\WH{1}\WH{2}}\Spaa{\WH{1}3}}
~~~\label{1.33-1.33}\eea
On can numerically check that the result is $q$-gauge invariant.
For this helicity configuration, a good gauge choice is $r_1=p_2$, $r_2=r_3=p_1$.
When we choose $q=p_1$ in (\ref{1.33-1.33}), it is easy to see that many terms vanish.
Plugging all these in, we get immediately
\bea
J^{\mu}\left(1^{-},2^{+},3^{+}\right)=\frac{\Spbb{32}\Spaa{1|\gamma^{\mu}{k}_{123}|1}}
{\sqrt{2}s_{12}s_{123}\Spaa{23}} \eea

In formula \eqref{2.16} we have taken a pair of on-shell particles
to do the BCFW-deformation. However,
one can also take an on-shell particle and the off-shell leg to do
the BCFW-deformation and  write down corresponding recursion
relation. More details can be found in \cite{Feng:2011tw}.

\subsection{Recursion relations with nonzero boundary
contributions} 

The original version of on-shell recursion relation was constructed
and proved under the assumption that the boundary contribution $B$
in (\ref{BCFW-boundary}) vanishes. This is indeed the case if $A(z)$
vanishes as $z\to \infty$.\footnote{More accurately, if $A(z)$ has
the expansion $A(z)=\sum_{i} {c_i/ (z-z_i)}+ B_0+B_1 z+...+B_k z^k$,
the boundary contribution is zero as long as $B_0=0$ even if
$B_i\neq 0$ for some $i\in [1,k]$.} It is also the most used
version. In cases of  $B\neq 0$, one asks whether it can be
calculated recursively starting from lower-point on-shell
amplitudes. So far we do not have complete answer for this question,
but interesting results have been uncovered
\cite{Feng:2009ei,Feng:2010ku, Feng:2011tw, Benincasa:2011kn,
Feng:2011jxa}.

For some theories, we can recursively
calculate boundary contributions by analyzing structures of
Feynman diagrams.
One may also trade boundary contributions with roots of amplitudes
\cite{Benincasa:2011kn}. In the following, we will present these results separately.


\subsubsection{Dealing with boundary contributions by analyzing Feynman diagrams}

We now calculate boundary contributions recursively by analyzing
Feynman diagrams. The first example is the $\la \phi^4$
theory.\footnote{There are other ways to deal with this theory by
using auxiliary fields \cite{Benincasa:2007xk,Boels:2010mj}.} We
will use the pair $(1,2)$ to do the BCFW-deformation. As shown in
Figure \ref{Fig:division}, there are two possible types of diagrams:
(a) particles $1,2$ are attached to different vertices and (b)
particles $1,2$ are attached to the same vertex. For diagrams in
category (a), there is at least one propagator on the line
connecting $1,2$.
Its momentum depends on $z$ linearly, so we have a
factor ${1/(p^2- z\Spab{1|p|2})}$ in the expression. Under the limit
$z\to \infty$, contributions in category (a) go to zero and they do
not give boundary contributions.

   \begin{figure}[hbt]
  \centering
  \includegraphics[viewport=195 583 556 713,clip]{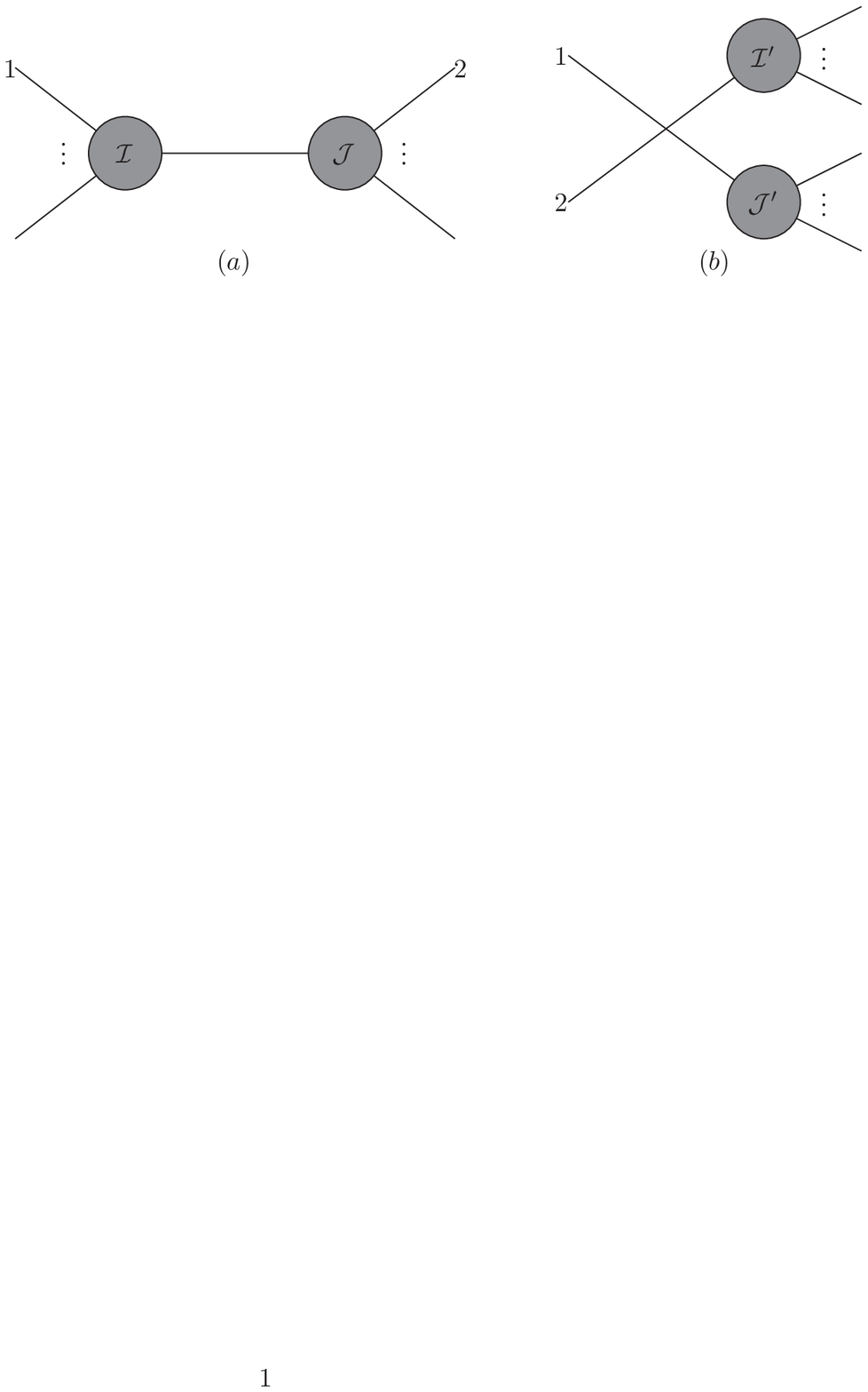}
  \caption{(a) Diagrams not giving boundary contributions
      (b)  Diagrams giving boundary contributions  \label{Fig:division}}
\end{figure}

In category (b), the whole contribution does not depend on $z$ at
all, since $1,2$ are attached to the same vertex. One has nonzero
boundary contributions from this category. By this analysis, one
sees also that boundary contributions can be  calculated  by
attaching lower-point amplitudes to this vertex. Immediately, we can
write down the on-shell recursion relation  with $(i|j)$-deformation
for this theory \cite{Feng:2009ei}
\bea A= A_{\rm b}+ A_{\rm pole}~~~~\label{Phi4-BCFW-1}\eea
Here contributions from poles are presented by the standard recursion formula
\bea A_{\rm pole} & = & \sum_{i\in{\cal I},j\cancel\in {\cal I}}
A_{\cal I}\left( \{K_{\cal I^{\prime }}\}, p_{i}(z_{\cal I}),
-p_{\cal I}(z_{\cal I}) \right)\frac{1}{p_{\cal I}^2}A_{\cal
J}\left( \{K_{\cal J^{\prime }}\}, p_{j}(z_{\cal I}), p_{\cal
I}(z_{\cal I}) \right)~~~~\label{Phi4-BCFW-2}\eea
and the boundary contribution is
\bea A_{\rm b} & = & (-i\lambda)\sum_{ {\cal I^{\prime }\bigcup
J^{\prime }}=\{n\}\backslash \{i,j\}}A_{\cal I^{\prime
}}\left(\{K_{\cal I^{\prime }})\right)\frac{1}{p_{\cal I^{\prime
}}^2}\frac{1}{p_{\cal J^{\prime }}^2}A_{\cal J^{\prime
}}\left(\{K_{\cal J^{\prime }}\}\right)~~~~\label{Phi4-BCFW-3}\eea
(\ref{Phi4-BCFW-3}) states simply the fact that sets
${\cal I^{\prime }}$, ${\cal J^{\prime }}$ and particles $i,j$
are attached to the same vertex of coupling constant $-i\la$.
These two contributions are pictorially represented in Figure \ref{Fig:division}
$(a)$ and $(b)$, where we have set $i,j=1,2$.

Now we give an example of color-ordered six-point amplitudes with the $\Spab{1|2}$-deformation.
%
%
There is only one contribution from the pole part
\bea A_{6,\rm pole}^{\Spab{1|2}}(1,\dots,6)&=&
A_4(5,6,\widehat{1},-\widehat{p})\frac{1}{p_{561}^2}A_4(\widehat{p},\widehat{2},3,4)
   =(-i\lambda)^2\left(\frac{1}{p_{156}^2}\right)
\eea
while there are two contributions from the boundary\footnote
{For simplicity we have defined
$A_2(a,b)=\delta^4(p_a-p_b) p_a^2$ and $p_{ijk}=p_i+p_j+p_k$. }
\bea
 A_{6,\rm b}^{\Spab{1|2}} (1,\dots,6) & = &
 A_4(1,2,-p_1,-p_2)\left(\frac{1}{p_3^2}A_2(p_1,3)\right)
 \left(\frac{1}{p_{123}^2}A_4(p_2,4,5,6)\right)
  \nonumber\\
 && +A_4(1,2,-p_1,-p_2)\left(\frac{1}{p_{3,4,5}^2}A_4(p_1,3,4,5)\right)
   \left(\frac{1}{p_6^2}A_2(p_2,6)\right)
    \nonumber\\
   & = & (-i\lambda)^2 \left(\frac{1}{p_{123}^2}+\frac{1}{p_{126}^2}\right)
      \eea
Putting all together, we have
\bea
 A_6^{\rm FD}(1,\dots,6)=(-i\lambda)^2\left(\frac{1}{p_{123}^2}
    +\frac{1}{p_{126}^2}+\frac{1}{p_{156}^2}\right)
\eea
  which agrees with the result directly from evaluating Feynman diagrams.

\begin{figure}[hbt]
  \centering
  \includegraphics[viewport=160 607 465 705,clip]{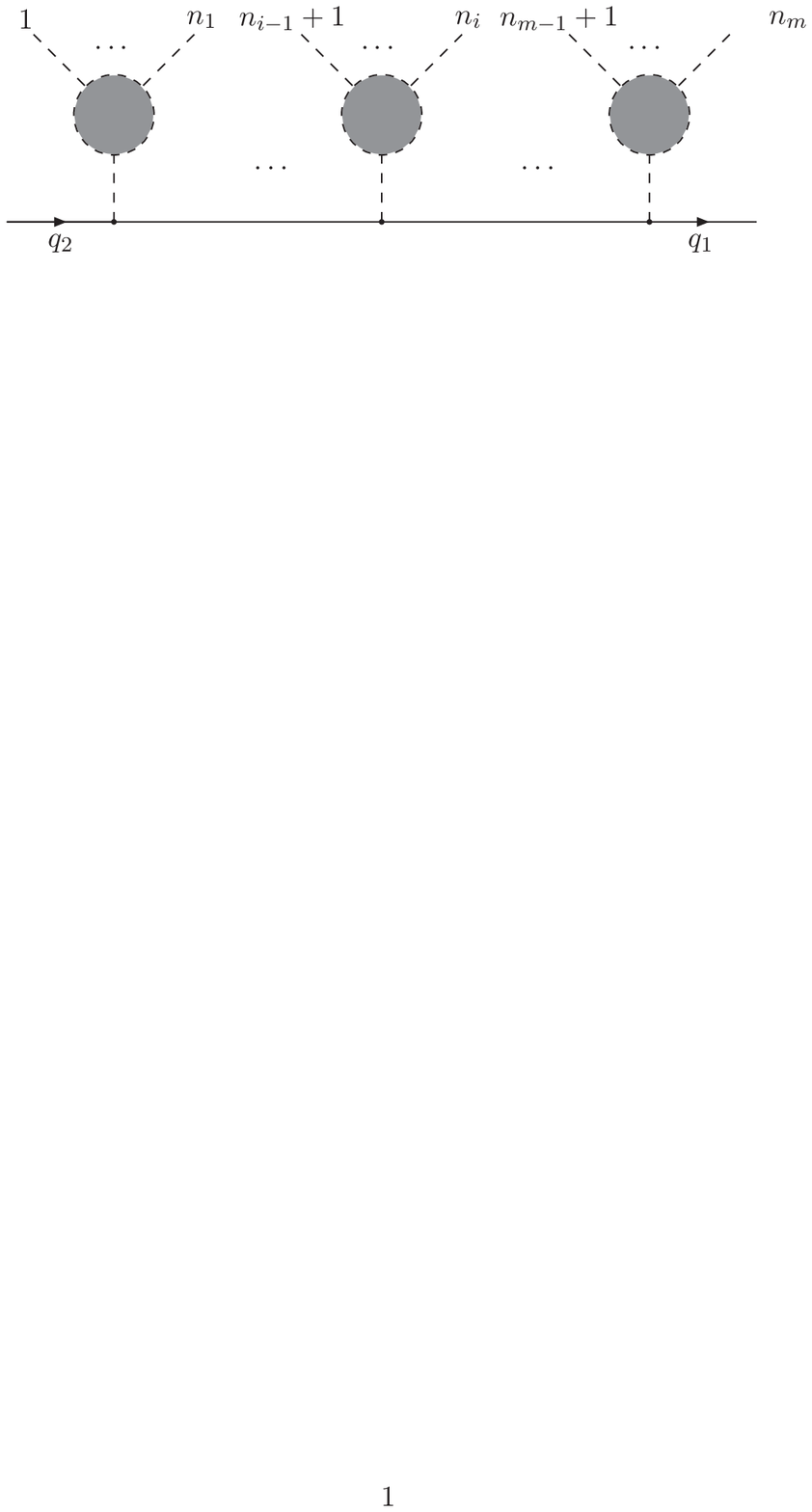}
  \caption{General Feynman diagrams for the Yukawa theory. \label{Fig:general}}
\end{figure}

Our second example is the Yukawa theory, where fermions are coupled to scalars.
For the interaction between two fermions of momenta $q_1,q_2$ and $n$ scalars of momenta $p_1,...,p_n$,
the ordered amplitude is $A(q_1,p_1,...,p_n,q_2)$.
As shown in Figure \ref{Fig:general}, there is one common feature in general Feynman diagrams:
{\sl  a single  fermion line connecting two fermions
while scalars are attached through Yukawa coupling at the same side}.
Using the fermion propagator ${i\cancel{p}/p^2}$, the amplitude can be written as
\bea  A= \sum_{diagrams}{\cal S}_i {\cal Q}_i  ~~~~\label{A-SQ}\eea
where $S_i$ is the contribution from scalar part and $Q_i$ is of the form
\bea {\cal Q}(q_1^-, q_2^+; R_1,...,R_m)\sim
i^m{\Spab{1|R_1|R_2|...|R_m|2}\over R_1^2 R_2^2 .... R_m^2} \eea
by assuming helicities of $q_1,q_2$ to be $(-,+)$ and $m$ fermion propagators along the line.
When $h_{q_1}=h_{q_2}$,
we must have even number of fermion propagators (i.e., $m$ is even)
while when $h_{q_1}=-h_{q_2}$,
we must have odd number of fermion propagators (i.e., $m$ is odd) to get nonzero amplitudes.

Taking the two fermions as reference lines and using the $\Spab{1|2}$-deformation,
%
%
one sees  that ${\cal S}_i$-factors in (\ref{A-SQ}) do
not depend on $z$ and all $z$-dependences are inside ${\cal Q}_i$.
The details  depend on helicity configurations and
we take the configuration $(h_{q_1}, h_{q_2})=(+,+)$ for illustration.
Now the number of fermion propagators should be even and we have either $m=0$ or $m\geq 2$.
For $m\geq 2$, we have
\bea & & {\cal Q}(q_1^+, q_2^+; R_1,...,R_m)\sim
i^m{\Spbb{1|(q_1+R_1+z \la_2\W\la_1)(\prod_{j=1}^{m-2} (q_1+R_j+z
\la_2\W\la_1))(q_1+R_m+z \la_2\W\la_1)|\W\la_2-z \W\la_1}\over
(q_1+R_1+z \la_2\W\la_1))^2(\prod_{j=1}^{m-2} (q_1+R_j+z
\la_2\W\la_1)^2)(q_1+R_m+z \la_2\W\la_1)^2 }\nn
 & = & i^m {\Spbb{1|(q_1+R_1)(\prod_{j=1}^{m-2} (q_1+R_j+z \la_2\W\la_1))(q_1+R_m+z
\la_2\W\la_1)|\W\la_2-z \W\la_1}\over (q_1+R_1+z
\la_2\W\la_1))^2(\prod_{j=1}^{m-2} (q_1+R_j+z
\la_2\W\la_1)^2)(q_1+R_m+z \la_2\W\la_1)^2 }\nn
 & = & i^m {\Spbb{1|(q_1+R_1)(\prod_{j=1}^{m-2} (q_1+R_j+z \la_2\W\la_1))
 (q_1+R_m)|\W\la_2-z \W\la_1}\over (q_1+R_1+z
\la_2\W\la_1))^2(\prod_{j=1}^{m-2} (q_1+R_j+z
\la_2\W\la_1)^2)(q_1+R_m+z \la_2\W\la_1)^2 }\nn
 & &+ i^m {\Spbb{1|(q_1+R_1)(\prod_{j=1}^{m-2} (q_1+R_j+z \la_2\W\la_1))(z
\la_2\W\la_1)|\W\la_2}\over (q_1+R_1+z
\la_2\W\la_1))^2(\prod_{j=1}^{m-2} (q_1+R_j+z
\la_2\W\la_1)^2)(q_1+R_m+z \la_2\W\la_1)^2 } \eea
Under the limit $z\to \infty$, it vanishes,
since each term has $m-1$ of $z$ in the numerator and $m$ of $z$ in the denominator.

For $m=0$, we have
\bea {\cal Q}(q_1^+, q_2^+)=\Spbb{1|2-z 1}=\Spbb{1|2}\eea
which is independent of $z$.
Thus we found a source of nonzero boundary contributions. The
on-shell recursion relation is
\bea & & A_{n+2}(q_1^+;p_1,...,p_n;q_2^+) \nn
& = & \sum_{i=1,h=\pm}^{n-1}
A_{i+2}(q_1^+(z_i);p_1,...,p_i;q_i^h(z_i)){1\over (q_1+\sum_{j=1}^i
p_j)^2}A_{n-i+2}(-q_i^{-h}(z_i);p_{i+1},...,p_n;q_2^+(z_i))\nn
& & + {(-i g)\Spbb{1|2}\over (\sum_{i=1}^n p_i)^2}
A_{n+1}(p_1,...,p_n,p_{\phi}) ~~~~\label{12++BCFW} \eea
where $A_{n+1}$ is  the amplitude of $n+1$ scalars.


%

In comparison with the case of all scalars, the analysis of boundary
contributions becomes more complicated for fermions and vectors.
 It is hard to figure out boundary
contributions directly from Feynman diagrams in general.
Partly this is due to the wave functions of fermions and vectors.
The wave function of scalars is simply $1$. To do the analytic
continuation of  amplitude from on-shell to off-shell, we just move
momentum $p_\mu$ to off-shell. For fermions and vector bosons, their
wave functions are nontrivial.

\subsubsection{Expressing boundary contributions in terms of roots of amplitudes}

We now present an analysis of boundary contributions from a
different perspective \cite{Benincasa:2011kn, Feng:2011jxa}. In
general, on-shell recursion relations with boundary contributions
can be written as
\bea M_n(z)=\sum_{k\in \mathcal{P}^{(i,j)}}{M_L(z_k)M_R(z_k)\over
p_k^2(z)}+C_0+\sum_{l=1}^{v}C_l z^l~,~~~\label{M-BCFW-z1}\eea
where we have explicitly kept the deformation parameter $z$ in the expression and assumed that $i\in
k$ so $p_k^2(z)= (-2p_k\cdot q) (z-z_k)$ with $z_k={p^2_k / 2p_k\cdot q}$.
Pulling all denominators in $M_n(z)$ together, one has
\bea M_n(z)= c{\prod_{s}(z-w_s)^{m_s}\over
\prod_{k=1}^{N_p}p_k^2(z)}~,~~~\sum_s
m_s=N_z=N_p+v~.~~~\label{bcfw-factorization}\eea
where $w_s$ are roots of the shifted amplitude. Unlike results
without boundary contributions, (\ref{M-BCFW-z1}) has not only
single poles but also a pole at $z=\infty$ of degree $v+1$. To
determine $M_n(z)$ completely, we need all coefficients related to
the pole at $z=\infty$, in addition to residues of single poles at
finite $z$.

With $N_z\geq N_p$  in (\ref{bcfw-factorization}) we can split
all roots into two groups ${\cal I},{\cal J}$ with $n_{\cal I}$ and
$n_{\cal J}$ (so $N_z=n_{\cal I}+n_{\cal J}$) roots,
respectively. For $n_{\cal I}< N_p$, we have
\bea c{\prod_{s=1}^{n_{\cal I}}(z-w_s)\over
\prod_{k=1}^{N_p}p_k^2(z)}=\sum_{k\in \mathcal{P}^{(i,j)}} {c_k\over
p_k^2(z)}~,~~~\label{Np-zero-pole}\eea
where $c_k$ are unknown $z$-independent coefficients and so
\bea M_n(z)= \sum_{k\in \mathcal{P}^{(i,j)}} {c_k\over
p_k^2(z)}\prod_{t=1}^{n_{\cal J}}(z-w_t)~~~~\label{M-zero-z1}\eea
To find $c_k$, we perform a contour integration around the pole $z_k$ over (\ref{M-BCFW-z1}),
(\ref{M-zero-z1}) and obtain
\bea  {M_L(z_k)M_R(z_k)\over (-2p_k\cdot q)}= {c_k\over (-2p_k\cdot
q)}\prod_{t=1}^{n_{\cal J}}(z_k-w_t), ~~\Longrightarrow
c_k={M_L(z_k)M_R(z_k)\over \prod_{t=1}^{n_{\cal J}
}(z_k-w_t)},~~~\label{ck-sol}\eea
Putting this back we have
\bea M_n(z)= \sum_{k\in \mathcal{P}^{(i,j)}} {M_L(z_k)M_R(z_k)\over
p_k^2(z)}\prod_{t=1}^{n_{\cal J}}{(z-w_t)\over
z_k-w_t}~~~~\label{M-zero-z2}\eea
So far we have not used the information of pole at infinity, which
will give relations among $N_z$ roots. A simple use of the
information is to set $n_{\cal I}=N_p-1$ and we arrive
\bea M_n(z)= \sum_{k\in \mathcal{P}^{(i,j)}} {M_L(z_k)M_R(z_k)\over
p_k^2(z)}\prod_{t=1}^{v+1}{(z-w_t)\over
z_k-w_t}~~~~\label{Mz-new}\eea
which is the one written down in \cite{Benincasa:2011kn} and obtained by a new method.
Setting $z=0$, we get the on-shell recursion relation with nonzero boundary contributions
\bea M_n= \sum_{k\in \mathcal{P}^{(i,j)}} {M_L(z_k)M_R(z_k)\over
p_k^2}\prod_{t=1}^{v+1}{w_t\over w_t-z_k}~~~~\label{Mz=0-new}\eea

We now make a few remarks on (\ref{Mz-new}) and (\ref{Mz=0-new}).
Firstly,  the divergent degree $v$ is  a function of $n$ in general (effective) quantum field theories.
But for gauge theory, gravity theory or other well-defined renormalizable theories, $v$ is independent of $n$.
Secondly, both poles and roots are important to determine amplitudes.
However, they differ in one crucial point: {\sl Poles are local property and easier to determine
while roots are (quasi)global property and harder to analyze}.
The (quasi)global feature of roots can be easily seen from the MHV-amplitude
$A(1^-,2^+,..,(j-1)^+, j^-, (j+1)^+,...,n^+)$.
With the $\Spba{i|1}$-deformation where $i$ is another particle of positive helicity,
we have a root $w=-{\Spaa{1|j}/ \Spaa{i|j}}$  of multiplicity four,
which changes locations  with the choice of $i$.

Thirdly, due to their (quasi)global nature, roots are difficult to
get recursively from lower-point amplitudes. The best we can get is
a set of consistent conditions, under various collinear or multiple
particle limits. Under these limits, higher-point amplitudes
factorize into products of two lower-point ones. With the help of
consistent conditions thus obtained, we can get roots under various
limits. These limits may help to determine the roots, but  they
cannot guarantee explicit solutions in general. The difficulty in
practical applications can be seen from following examples.

Now we see how to use consistent condition to find information of roots.
Note that (\ref{Mz-new}) is true for all $z$ since $M_n(z)$ is an on-shell amplitude for all $z$.
Thus  we have the following factorization relation
\bea \lim_{p_\a^2(z)\to 0} p_\a(z)^2 \sum_{k\in \mathcal{P}^{(i,j)}}
{M_L(z_k)M_R(z_k)\over p_k^2(z)}\prod_{s=1}^{v+1}{(z-w_s)\over
(z_k-w_s)} = M_L(z) M_R(z)~~~~\label{Gen-cons}\eea
We can compare rational functions of $z$ on both sides to
find the number and values of roots. For this to work, we must
ensure that $p_\a^2(z)\to 0$ can be realized for all $z$.
$p_\a(z)$ can be divided to two types. The first type is that $p_\a$
does not depend on $z$ at all, thus this condition holds.
The second type is that $p_\a$ has only $i$, thus $p_\a^2(z)=p_\a^2-
2z p_\a\cdot q$.
For given external momenta in general, if $p_\a^2(z)=0$ is true for a given value of $z$,
it cannot be true for another value of $z$.
However, collinear limits $p^2_{ik}(z)\to 0$ and
$p^2_{jk}(z)\to 0$ are exceptions.
The reason is that for massless particles we
have $p_{ik}^2(z)=\Spaa{i|k}( \Spbb{i|k}-z\Spbb{j|k})$, thus we can
take either $\Spaa{i|k}\to 0$ or $( \Spbb{i|k}-z\Spbb{j|k})\to 0$.
Although $( \Spbb{i|k}-z\Spbb{j|k})\to 0$ can be
true only for a given value of $z$, $\Spaa{i|k}\to 0$ is true for all $z$.

We can write down rational functions of $z$ and obtain information
on roots from lower point amplitudes, by using following
factorization channels: (1) $p_\a^2\to 0$ limit where $i,j\cancel\in
\a$. One particular channel of this type is that $p_{ij}^2\to 0$,
which corresponds to two possible limits $\Spaa{i|j}\to 0$ and
$\Spbb{i|j}\to 0$. (2) $\Spaa{i|k}\to 0$ or $\Spbb{j|k}\to 0$
limits. General discussions on these limits can be found in
\cite{Benincasa:2011kn, Feng:2011jxa} and here we present two
examples to demonstrate the idea.

{\bf Example One: MHV amplitude}

Taking the $\Spba{2|1}$-deformation, 
one gets from $A_n(1^-,2^+,3^-,..,n^+)$
\bea A_n(z) & = & {-1\over \Spaa{1|2}\Spaa{2|3}...\Spaa{n|1}} \left(
{\Spaa{n|3}\Spaa{1|2}\over \Spaa{n|2}} \right)^4 {\Spaa{n|1}\over
\Spaa{n|1}+z\Spaa{n|2}}\prod_{j=1}^k { w_j-z\over w_j-
z_\a}~~~\label{MHV-wrong-z}\eea
with $z_\a=-{\Spaa{1|n}/ \Spaa{2|n}}$.
Now we consider the collinear limit\footnote
{There is only one nontrivial choice in the limit, $\Spaa{a|a+1}\to 0$ and there
is no multiple particle channel.}
of $ a^+ (a+1)^+$ with $4\leq a\leq n-1$.
The right-handed side of (\ref{Gen-cons}) has the factor
\bea  {\Spaa{1(z)|3}^4\over
\Spaa{1|2}\Spaa{2|3}...\Spaa{a-1|p_{a,a+1}}\Spaa{p_{a,a+1}|a+2}...\Spaa{n|1(z)}}\eea
so we find  a root  $w= -{\Spaa{1|3}/ \Spaa{2|3}}$ of multiplicity four.
In the original amplitude, we should have
\bea w_j= -{\Spaa{1|3}\over \Spaa{2|3}}\left( 1+f_j\right)\eea
$f_j$ should satisfy these requirements:
(1) it has a factor $\Spaa{a|a+1}$, so vanishes under the collinear limit;
(2) it is helicity neutral for all particles, so there is either the extra factor $\Spbb{a|a+1}$
or the combination ${\Spaa{a|a+1}\Spaa{t|s}/ \Spaa{a|s}\Spaa{a+1|t}}$ with spinors $\la_t,\la_s$;
(3) it is dimensionless;
(4) it is consistent with all collinear limits $\Spaa{a|a+1}\to 0$;
(5)  it does not produce un-physical poles in physical amplitudes.
Under these requirements, we should take $f_j=0$ for all $j=1,2,3,4$.

Another interesting limit is $\Spaa{2|3}\to 0$. Under this limit,
the would-be root $w=-{\Spaa{1|3}/ \Spaa{2|3}}\to \infty$, thus the
combination ${(w-z)/(w-z_{n1})}\to 1$. The degree of $z$ is then
reduced at the left-handed side of (\ref{Gen-cons}).

{\bf Example Two: The six-gluon amplitude
$M_6(1^-,2^-,3^-,4^+,5^+,6^+)$}

The known six-gluon amplitude $M_6(1^-,2^-,3^-,4^+,5^+,6^+)$ is
\bea M_6(1^-,2^-,3^-,4^+,5^+,6^+)={1\over
\Spab{5|3+4|2}}\bigg({\Spab{1|2+3|4}^3\over
[23][34]\Spaa{5|6}\Spaa{6|1}(p_2+p_3+p_4)^2}\nn
+{\Spab{3|4+5|6}^3\over
[61][12]\Spaa{3|4}\Spaa{4|5}(p_3+p_4+p_5)^2}\bigg),~~~\label{A6-split-hel}
\eea
For our purpose we will use the $\Spba{5|3}$-deformation.\footnote{For deformation $\Spba{4|3}$,
there is no pole and the recursion relation should be modified accordingly.}
The boundary on-shell recursion relation gives
following $z$-dependent amplitudes
 \bea
& &
M_6(1^-,2^-,\wh{3}(z)^-,4^+,\wh{5}(z)^+,6^+)\nn&=&{\Spba{6|5+3|4}^3
\Spaa{3|5}^3 \over
\Spba{2|3+4|5}\Spaa{4|5}^4\Spbb{6|1}\Spbb{1|2}p_{345}^2(\Spaa{3|4}+z\Spaa{5|4})
}\prod_l{w_l-z\over w_l-z_{34}}\nn
&+&{\Spba{4|2+3|5}^3\Spba{3|5+6|1}^3\over
\Spba{2|3+4|5}\Spba{3|2+4|5}^3\Spbb{2|3}\Spbb{4|3}\Spaa{5|6}\Spaa{6|1}
(p_{234}^2+z\Spba{3|2+4|5})}\prod_l{w_l-z\over
w_l-z_{234}}.\nn\label{6nmhv-z}\eea
where $z_{34}=-{\Spaa{4|3}/\Spaa{4|5}}$ and
$z_{234}=-{p_{234}^2/\Spba{3|2+4|5}}$. The pole structure of
six-gluon amplitude is the following. There are three
three-particles poles, $s_{123}=s_{456}$, $s_{234}=s_{561}$,
$s_{345}=s_{612}$. Among them $s_{123}=s_{456}$ is trivial.
For two particle poles we have $\Spbb{1|2}$, $\Spbb{2|3}$,
$\Spaa{3|4}$, $\Spbb{3|4}$, $\Spaa{4|5}$, $\Spaa{5|6}$, $\Spaa{6|1}$
and $\Spbb{6|1}$, after considering the holomorphic and
anti-holomorphic part .

Now we can read out information of roots under various limits. For
example, when $p_{216}^2\to 0$, factorization limit leads to
\bean & & M_4(6^+,1^-,
2^-,-p_{612}^+)M_4(p_{612}^-,\wh{3}^-,4^+,\wh{5}^+)={\Spaa{1|2}^3\Spbb{4|5-z3}^3\over
\Spbb{3|4}\Spaa{6|1}\Spab{2|p_{612}|3}\Spba{5-z3|p_{612}|6}}\eean
which leads to triple roots $ w_l^{(3)}= {\Spbb{4|5}/\Spbb{4|3}}$.
This shows the power of $z$-dependent factorization limits because
we do not need to work out detailed comparisons.  By similar method
we can read out values of roots under various factorization limits
\bea   &&\Spbb{1|2}\to
0,~~w_l^{(3)}=-{\Spba{\mu|1+2|3}\over\Spba{\mu|1+2|5}}=-{\Spba{6|4+5|3}\over
\Spba{6|3+4|5}},\nn
&&\Spaa{1|6}\to0,~~ w_l^{(3)}={\Spbb{4|5}\over
\Spbb{4|3}}={\Spba{4|2+3|1}\over \Spbb{4|3} \Spaa{1|5}},\nn
&&\Spbb{1|6}\to0,~~ w_l^{(3)}=-{\Spaa{2|3}\over
\Spaa{2|5}}=-{\Spba{6|4+5|3}\over \Spba{6|3+4|5}},\nn
&&p_{216}^2\to 0,~~w_l^{(3)}={\Spbb{4|5}\over
\Spbb{4|3}}=-{\Spba{6|4+5|3}\over \Spba{6|3+4|5}},\nn
&&\Spbb{2|3}\to
0,~~w_l^{(3)}=-{\Spba{\mu|2+3|1}\over\Spbb{\mu|3}\Spaa{5|1}}={\Spba{4|2+3|1}\over
\Spbb{4|3} \Spaa{1|5}},\nn
&&\Spbb{3|4}\to 0,~~w_l^{(3)}\to \infty,\nn
&&\Spaa{5|6}\to 0,~~w_l^{(3)}=-{\Spba{4|5+6|\mu}\over
\Spbb{4|3}\Spaa{\mu|5}}={\Spba{4|2+3|1}\over \Spbb{4|3}
\Spaa{1|5}},\nn
&&\Spaa{5|4}\to 0,~~w_l^{(3)}=-{\Spba{6|4+5|\mu}\over
\Spbb{6|3}\Spaa{\mu|5}}=-{\Spba{6|4+5|3}\over
\Spba{6|3+4|5}}.~~\label{6-limit-zero}\eea

However,  \eqref{6-limit-zero} can not help to find true
expressions of roots. To see this, notice that numerator from
expression (\ref{A6-split-hel}) is given by
\bea N & = & T_1+ T_2 \nn
T_1 & = & -\Spaa{4|5}\Spbb{2|1}\Spbb{6|1}
s_{345}\Spaa{4|5}\Spaa{1|5}^3\Spbb{4|3}^3\left(z+{\Spaa{3|4}\over\Spaa{5|4}}\right)
\left(-{\Spba{4|p_{23}|1}\over \Spaa{1|5}\Spbb{4|3}}+z\right)^3\nn
T_2 & = &
-\Spaa{1|6}\Spaa{5|6}\Spbb{3|2}\Spbb{4|3}\Spab{5|p_{234}|3}\Spab{5|p_{345}|6}^3\left(
{ s_{234}\over \Spab{5|p_{234}|3}}+z \right)
\left({\Spab{3|p_{345}|6}\over \Spab{5|p_{345}|6}}+z
\right)^3.\label{6-num}\eea
For general momentum configurations where $T_1$ and $T_2$ are
not zero, we have a polynomial of degree four. The
analytic expressions of its roots are very complicated and not {\sl
rational functions of spinors} in general.\footnote{We have checked this via
numerical method by setting all spinor components to be integer numbers.}
Due to the irrationality, it is very hard to find explicit expressions, even with (\ref{6-limit-zero}) to help.

\subsection{Bonus relations}

On-shell recursion relations depend crucially on the behavior of
$A(z)$ when $z\to \infty$, which can be divided into three
categories. The vanishing of boundary contributions requires only
$A(z)\to  z^{-1}$, which we will call the standard type. Opposite to
the standard type, we have other two types: one with $A(z)\not\to 0$
and another  with $A(z)\to  z^{-a},~~a\geq 2$. The former provides
nonzero boundary contributions, as discussed in the previous
subsection. The latter leads to ``bonus relations", to be discussed
here.

There are several places where bonus relations can be established.
The first place is among tree-level amplitudes of gravitons, where the
large $z$ behavior of ${\cal M}(z)$ is $z^{-2}$, as explained in \cite{ArkaniHamed:2008yf}. In
\cite{ArkaniHamed:2008gz} this fact was emphasized and in
\cite{Spradlin:2008bu} some
applications of bonus relations are given.\footnote{Bonus  relations were discussed in
\cite{Benincasa:2007qj, ArkaniHamed:2008yf,
Spradlin:2008bu,Badger:2008rn, Feng:2010my,Badger:2010eq,
He:2010ab} where their usefulness was demonstrated from various
aspects. Some applications  can be found in later section where BCJ
and KLT relations are proved.}
Another place for bonus relations is among
tree-level amplitudes in QED \cite{Badger:2010eq}.

Bonus relations can be derived  from the observation
\bea 0= \oint {dz \over z}  z^b
A(z),~~~b=1,1,...,a-1,~~~if,~~A(z)\to {1\over
z^a}~.~~~\label{bonus-key}\eea
Because the $z^b$ factor, there is no pole at $z=0$. Taking
contributions from other poles, we have bonus relations
\bea 0 =\sum_\a \sum_h  A_L(p^h(z_\a)) {z_\a^b\over p^2} A_R(
-p^{-h}(z_\a))~~~~\label{bonus-rel-gen}\eea
for $b=1,...,a-1$.

Having established \eqref{bonus-rel-gen}, we present a simple
application \cite{Spradlin:2008bu}. All known formulas of $n$-graviton MHV amplitudes
in literature \cite{Berends:1988zp, Mason:2008jy,
Bern:2007xj, Bedford:2005yy, Nair:2005iv, Elvang:2007sg} can be
divided into two categories: those having manifest permutation
symmetry for $(n-2)$ elements and those having manifest permutation
symmetry for $(n-3)$ elements. One example of manifest
$(n-2)!$-permutation symmetric expression\footnote{In fact, results
in \eqref{N-gra-BCFW} and \eqref{N-gra-BGK} are for
 ${\cal N}=8$ super-gravity, where factor $\delta^{(4)}(\sum_i \la_i
\W\la_i)\, \delta^{(8)} (\sum_{i=1}^n\ \lambda_{i}^{\alpha}\,
\eta^A_i)$ has been neglected.}
 is  \cite{Elvang:2007sg}
\bea {\cal M}_n =\sum_{\sigma\in P(3,...,n)}
F(1,2,\sigma(3,...,n))~~~\label{N-gra-BCFW}\eea
with
\bea F(1,2,3,...,n) & = &
\Spaa{1|n}\Spbb{n|1}\left(\prod_{s=4}^{n-1}\b_s \right)
A(1,2,3,...,n)^2,~~~~\b_s=-{\Spaa{s|s+1}\over
\Spaa{2|s+1}}\Spab{2|p_{345...(s-1)}|s}\eea
One example of manifest $(n-3)!$-permutation symmetric expression
 is \cite{Berends:1988zp, Bern:2007xj}
\bea {\cal M}_n = \sum_{\sigma\in P(4,...,n)}
{\Spaa{1|2}\Spaa{3|4}\over \Spaa{1|3}\Spaa{2|4}}
F(1,2,3,(4,...,n))~. ~~~\label{N-gra-BGK}\eea

To prepare for later proof,  we use the bonus relation $b=1$ of
\eqref{bonus-rel-gen} under $\Spba{2|1}$-deformation. For MHV
amplitudes, only cut $(p_1+p_k)^2$ gives nonzero contributions.
Defining sub-amplitudes in on-shell recursion relations
\bea M_k=\int d^8 \eta {\cal M}_L(\WH 1, k, -\WH p(z_k)){1\over
(p_1+p_k)^2} M_R(\WH p(z_k), \WH
2,3,...,k-1,k+1,...,n),~~~z_k=-{\Spaa{1|k}\over
\Spaa{2|k}}~~\label{Bonus-Mk}\eea
on-shell recursion relations give
\bea {\cal M}_n=M_3+M_4+...+M_n~~~\label{Mn-Mk}\eea
and the bonus relation gives
\bea 0= z_3 M_3+z_4 M_4+...+z_n M_n~. ~~~\label{zk-Mk}\eea
Solving, for example, $M_3$ from \eqref{zk-Mk} and plugging back to
\eqref{Mn-Mk}, we find
\bea {\cal M}_n = \sum_{k=4}^n {\Spaa{1|2}\Spaa{3|k}\over
\Spaa{1|3}\Spaa{2|k}} M_k~.~~~\label{Mn-new}\eea

Now we prove the equivalence of \eqref{N-gra-BCFW} and
\eqref{N-gra-BGK} by induction.
As shown in \cite{Elvang:2007sg}, $n=4,5$ are true.
Assuming our claim is true for amplitudes with
$k$-gravitons ($k\leq n$), using \eqref{Mn-new} for
$(n+1)$-gravitons we have
\bea {\cal M}_{n+1} &= & \sum_{k=4}^{n+1} {\Spaa{1|2}\Spaa{3|k}\over
\Spaa{1|3}\Spaa{2|k}} M_k= {1\over (n-2)!} \sum_{\sigma\in
P(4,...,n+1)}\sum_{k=\sigma(4)}^{\sigma(n+1)}
{\Spaa{1|2}\Spaa{3|\sigma(k)}\over \Spaa{1|3}\Spaa{2|\sigma(k)}}
M_{\sigma(k)}\nn
& = & {1\over (n-3)!} \sum_{\sigma\in
P(4,...,n+1)}{\Spaa{1|2}\Spaa{3|\sigma(n+1)}\over
\Spaa{1|3}\Spaa{2|\sigma(n+1)}} M_{\sigma(n+1)}~.~\label{Mn-mm} \eea
Recall \eqref{Bonus-Mk} for the definition of
$M_{\sigma(n+1)}$. Replacing $M_R$ by \eqref{N-gra-BGK} for
$n$-gravitons and using
\bea \int d^8 \eta {\cal M}_L(\WH 1, n+1, -\WH p(z_{n+1}){1\over
(p_1+p_{n+1})^2} F( \WH 2,3,..,n,\WH p(z_{n+1} )=
F(1,2,3,...,n+1)~,\eea
which comes from the on-shell recursion relation \eqref{N-gra-BCFW}, we obtain
\bea M_{n+1}=\sum_{\sigma\in P(4,...,n)}{\Spaa{\WH p(z_{n+1}
)|2}\Spaa{3|4}\over \Spaa{\WH p(z_{n+1} )|3}\Spaa{2|4}}
F(1,2,...,n+1) ~~~\label{Mn+1-F}\eea
Putting \eqref{Mn+1-F} back to \eqref{Mn-mm} and after some algebraic
simplification, we obtain
\bea {\cal M}_{n+1} = \sum_{\sigma\in P(4,...,n+1)}
{\Spaa{1|2}\Spaa{3|4}\over \Spaa{1|3}\Spaa{2|4}}
F(1,2,3,(4,...,n+1))~. \eea
Thus completes the inductive proof of equivalence.

\subsection{Recursion relations for the rational part of one-loop amplitudes}

On-shell recursion relation is  powerful for tree-level amplitudes
due to their simple analytic properties. For one-loop amplitudes,
there are branch cuts, in addition to poles.
It is highly nontrivial to have recursion relations. However, there are well defined objects,
the so-called ``rational parts" of one-loop amplitudes, for which it is possible to write down
on-shell recursion relations.

``Rational parts of one-loop amplitudes" come from the following.
Using Pasarino-Veltman reduction \cite{Passarino:1978jh}, one-loop
amplitudes can always be written as linear combinations of scalar
basis with rational coefficients. If we keep results to all order of
$\eps$ in dimensional regularization, the scalar basis are
pentagons, boxes, triangles, bubbles and tadpoles in
$(4-2\eps)$-dimension. If we keep results only to the order of
${\cal O}(\eps)$, the scalar basis are boxes, triangles, bubbles and
tadpoles in $4$-dimension, plus ``rational terms".

\subsubsection{Color structure}

Similar to the color decomposition \eqref{color-decom} for
tree-level amplitudes, $n$-point one-loop amplitude for $U(N)$ gauge
theory can be written as \cite{Bern:1994zx}
\bea {\cal A}_n^{\rm full}\left( \{k_i,\lambda_i,a_i\}\right) =
  \sum_{J} n_J\,\sum_{m=0}^{\lfloor{n/2}\rfloor}
      \sum_{\sigma \in S_n/S_{n;m}}
     \mbox{Gr}_{n-m,m}\left( \sigma \right)\,A_{n-m,m}^{[J]}(\sigma_1,\sigma_2,
     \ldots,\sigma_{n-m};\sigma_{n-m+1},\ldots,\sigma_n)~,
~~~\label{one-loop-color-decomposition} \eea
where ${\lfloor{x}\rfloor}$ is the largest integer $\leq x$ and
$n_J$ is the number of particles of spin $J$. Color factors for
primitive amplitudes are\footnote{For convenience we abbreviate
$\mbox{Tr}\left( T^{a_1}\cdots T^{a_n}\right)$ as $\mbox{Tr}\left(
a_1,\cdots,a_n\right)$.}
\bea \mbox{Gr}_{n,0} = N_c\mbox{Tr}\left( a_1,\cdots
,a_n\right)~,~~~\nonumber\eea
and for other partial amplitudes
\bea \mbox{Gr}_{n-m,m} = \mbox{Tr}\left(a_1,\cdots,n-m\right)\,
\mbox{Tr}\left( n-m+1,\cdots ,n\right)~.~~~\nonumber\eea
$S_n$ is the set of all permutations of $n$ objects, and $S_{n;m}$
is the subset leaving $\mbox{Gr}_{n-m,m}$ invariant.

Partial amplitudes $A_{n-m,m}$ of double trace structure
$\mbox{Gr}_{n-m,m}(m\neq 0)$ are algebraically related to primitive
amplitudes $A_{n,0}$ of single trace. $A_{n-m,m}$ can be expressed
as linear combinations of $A_{n,0}$, thus computing of primitive
amplitudes is enough to construct full one-loop amplitudes. The
relation is \cite{Bern:1994zx}
\bea
 A_{n-m,m}(\alpha_1,\alpha_2,\ldots,\alpha_{n-m};\beta_{1},\ldots,\beta_m)\ =\
 (-1)^{m} \sum_{\sigma\in COP\{\alpha\}\bigcup \{\beta^T\}} A_{n,0}(\sigma)
~,~~~\label{sublanswer} \eea
where $\beta^T$ is the set of $\beta$ with reversed ordering, and
$COP\{\alpha\}\bigcup \{\beta^T\}$ is the set of all permutations of
$\{\alpha,\beta^T\}$ preserving the cyclic ordering inside the set
$\a$ and $\b^T$, but allowing all possible relative orderings
between $\alpha$ and $\b^T$.\footnote{This  equation has been
understood from various point of views
\cite{Bern:1990ux,DelDuca:1999rs,Luo:2004ss,
Luo:2004nw,Feng:2011fja}.}

\subsubsection{Two special cases}

In gauge theories, tree-level amplitudes with at most one negative
helicity vanish, but one-loop amplitudes of these helicity
configurations do not. The latter are actually rational
functions.\footnote{Other one-loop amplitudes of rational functions
were discussed in \cite{Bern:2005ji} by using similar on-shell
recursion relations.} For example, the one-loop amplitude of all
positive helicities is \cite{Bern:1993sx,Bern:1993qk}
\bea A_{n;1}(1^+,2^+,...,n^+) & = & { \sum_{1\leq i_1<
i_2<i_3<i_4\leq n} \Spaa{i_1|i_2}\Spbb{i_2|i_3}\Spaa{i_3|i_4}
\Spbb{i_4|i_1}\over
\Spaa{1|2}\Spaa{2|3}...\Spaa{n|1}}~.~~~\label{1loop-All-Plus}\eea
Taking the $\Spba{j|n}$-deformation, the on-shell recursion relation
for (\ref{1loop-All-Plus}) can be written as \cite{Bern:2005hs}
\bea A_{n;1}(+++...+) & = & A_{n-1;1}(1^+,...,\WH j^+,...,(n-2)^+,
\WH{K}_{n-1,n}^+) {1\over K_{n-1,n}^2} A_{3}^{\rm tree}
(-\WH{K}_{n-1,n}^-, (n-1)^+, \WH n^+)\nn
& & +A_{n-1;1}(2^+,...,\WH j^+,...,(n-1)^+, \WH{K}_{n,1}^+) {1\over
K_{n,1}^2} A_{3}^{\rm tree} (-\WH{K}_{n,1}^-,  \WH n^+,1^+)\nn
& & + { -\Spab{j|K_{jn}|n}^2\over
\Spaa{1|2}\Spaa{2|3}...\Spaa{n|1}}~~~~\label{1loop-All-Plus-Rec}\eea
Now some explanations are in order for \eqref{1loop-All-Plus-Rec}.  Different from tree amplitudes, the
factorization property of one-loop amplitudes for residues is:
\bea A_n^{\rm 1-loop}\to A_{L}^{\rm 1-loop} A_R^{\rm tree}+ A_L^{\rm
tree} A_R^{\rm 1-loop}+ A_L^{\rm tree} {\cal S} A_R^{\rm
tree}~.~~~\label{1loop-factor}\eea
Among the three terms, the first two are  clear.
The third term is not understood well. Fortunately, for the configuration
of all positive helicities, the third term in (\ref{1loop-factor}) is zero.
$A^{\rm tree}$ is nonzero only for the three-point
amplitude, thus the first two lines in
(\ref{1loop-All-Plus-Rec}) are contributions from poles. The
third line in (\ref{1loop-All-Plus-Rec}) is the boundary
contribution since under our choice of deformation, the amplitude
does not vanish when $z\to \infty$. The boundary contribution is not
easy to derive if we do not know the explicit expression
(\ref{1loop-All-Plus}).

Things become more complicated for one-loop amplitudes of only one
negative helicity and general results can be found in
\cite{Mahlon:1993si}. In the  previous example and general
tree-level amplitudes, there are only single poles under the
BCFW-deformation. Now we have double poles such as ${\Spaa{a|b}/
\Spbb{a|b}^2}$, which can be seen in the example of five gluons
\bea A_{5;1}(1^-,2^+, 3^+, 4^+, 5^+) & = & {1\over \Spaa{3|4}^2}
\left[ -{\Spbb{2|5}^3\over
\Spbb{1|2}\Spbb{5|1}}+{\Spaa{1|4}^3\Spbb{4|5}\Spaa{3|5}\over
\Spaa{1|2}\Spaa{2|3}\Spaa{4|5}^2}-{\Spaa{1|3}^3\Spbb{3|2}\Spaa{4|2}
\over \Spaa{1|5}\Spaa{5|4}\Spaa{3|2}^2}\right]\eea
Under the  $\Spba{1|2}$-deformation, one gets a double pole from
$\Spaa{2|3}^2$ in the denominator. The double pole structure raises
the following questions \cite{Bern:2005hs}: (1) how to get the
double pole structure from factorization properties; (2) how to find
single pole contributions hidden inside the double pole.

The answer to the first question comes from the structure of
one-loop three-point amplitude of same helicity. When two momenta
are collinear, the result is divergent as follows \cite{Bern:2005hs}
\bea  A_{3;1}(1^+, 2^+, 3^+) & = &
{\Spbb{1|2}\Spbb{2|3}\Spbb{3|1}\over K_{12}^2}\eea
The double pole structure can then be obtained as
\bea A_L^{\rm tree} {1\over K_{a,a+1}^2} A_{3;1}(-\WH K_{a,a+1}^+, \W
a^+, (a+1)^+)\to A_L^{\rm tree} {1\over K_{a,a+1}^2}{1\over K_{a,a+1}^2}
\Spbb{\WH K_{a,a+1}|\WH a}\Spbb{\WH a|a+1} \Spbb{a+1|\WH
K_{a,a+1}}\eea
The second question is solved by multiplying  a dimensionless
function $K_{cd}^2 {\cal S}^{(0)} (a, s^+, b){\cal S}^{(0)} (c, s^-,
d)$  \cite{Bern:2005hs},
 where the {\sl soft factor} is given in \eqref{soft-factor} and we
 recall them here
\bea {\cal S}^{(0)} (a, s^+, b)= {\Spaa{a|b}\over
\Spaa{a|s}\Spaa{s|b}},~~~~{\cal S}^{(0)} (c, s^-,
d)=-{\Spbb{c|d}\over \Spbb{c|s}\Spbb{s|d}}~.~~~\label{Soft-tree}\eea

Thus under the $\Spba{1|2}$-deformation, the on-shell recursion
relation for the configuration of one negative helicity is
\cite{Bern:2005hs}
\bea & & A_{n;1}(1^-,2^+,...,n^+)= A_{n-1;1}(4^+,...,n^+, \WH 1^-,
\WH K_{23}^+) {1\over K_{23}^2} A_{3}^{\rm tree} (\WH 2^+, 3^+, -\WH
K_{23}^-)\nn
& & +\sum_{j=4}^{n-1} A_{n-j+2}^{\rm tree}((j+1)^+,...,n^+, \WH 1^-, \WH
K_{2...j}^-) {1\over K_{2...j}^2} A_{j;1}(\WH 2^+, 3^+,...,j^+, \WH
K_{2...j}^+) \nn
& & + A_{n-1}^{\rm tree}(4^+,...,n^+, \WH 1^-, \WH K_{23}^-) {1\over
K_{23}^2} A_{3;1} (\WH 2^+, 3^+, -\WH K_{23}^+)\left( 1+ K_{23}^2
{\cal S}^{(0)} (\WH 1, \WH K_{23}^+, 4){\cal S}^{(0)} (3,-\WH
K_{23}^-, \WH 2)\right)\eea
where the last line represents contributions from the hidden single
pole in the double pole.

\subsubsection{Recursion relations of general one-loop amplitudes}

We now discuss on-shell recursion relations for rational parts of
general one-loop amplitudes, as proposed in \cite{Bern:2005cq}. The
general form of one-loop amplitudes is $\sum R_i {\rm
Li}_2(R_j)+\sum R_k \ln(R_k)+R_n$ where $R$ are rational functions.
Poles (from $R_i, R_k, R_n$) and branch cuts\footnote {Branch cuts
can be chosen as lines from zero or the pole of $R_j, R_k$ to
infinity.} will show up in the analytic structure of one-loop
amplitudes, thus we will have
\bea B= \oint{dz \over z} A_n^{\rm 1-loop}(z)=A_n^{\rm
1-loop}(z)+\sum_{{\rm poles}~~z_\a} {\rm Res} \left( {A_n^{\rm
1-loop}(z)\over z}\right)+ \sum_{\rm branch~cuts}\int_{b_i}^{\infty}
{dz \over z}{\rm Disc} A_n^{\rm
1-loop}(z)~~~\label{1-loop-contour}\eea
where $B$ is the boundary contribution (which can be set to zero by
proper choice of deformation), $b_i$'s are starting points of branch
cuts and ${\rm Disc}$ is the discontinuity of $A_n^{\rm 1-loop}$
crossing the branch cut.  For simplicity, we choose paths of
 branch cuts carefully so they will not intersect with each other.
The pole and the staring point of an branch cut could overlap. In this case, we can move the pole
away from the staring point by a small amount $\delta$ and take
$\delta\to 0$ at the end of all calculations.

To calculate the pure rational part of one-loop amplitude, we need
to separate contributions of rational part and cut part in
(\ref{1-loop-contour}).
The cut part is assumed to be
known by other methods, for example, the unitarity cut method
\cite{Bern:1994cg,Bern:1994zx,Britto:2004nc,Britto:2005ha}. That is,
under the deformation we have
\bea A_n(z)= C_n(z)+ R_n(z) ~~~\label{1loop-An-1}\eea
with known $C_n(z)$. In this separation, the cut part may contain
spurious singularities. For example, it may contain ${\ln(r)/
(1-r)^2}$ with $r=s_1/s_2$ and $r\to 1$ is not a physical pole. To
remedy this, one may add some rational terms to get rid of these
spurious singularities. For example, ${\ln(r)/r (1-r)^2}\to
{[\ln(r)+(1-r)]/  (1-r)^2}$. In the end, we get a cut part $\WH
C(z)$ free of spurious singularities and
\bea A_n(z)= \WH C_n(z)+ \WH R_n(z) ~~~\label{1loop-An-2}\eea
where $\WH C_n(z)=C_n(z)+\WH {CR}(z)$ and $\WH{CR}(z)$ is the
rational part. Observe that the discontinuity part of $A_n$ comes
from $C_n$ only, so we have
\bea \WH C_n(0)= -\sum_{\rm poles} {\rm Res}\left( {\WH C_n(z)\over
z}\right)+ \sum_{\rm branch~cuts}\int_{b_i}^{\infty} {dz \over
z}{\rm Disc} \WH C_n(z) \eea
and (\ref{1-loop-contour}) becomes
\bea A_n(0)= \WH C_n(0)-\sum_{\rm poles} {\rm Res}\left( {\WH
R_n(z)\over z}\right)~~~\label{1loop-A-R}\eea
where boundary contributions are assumed to vanish. With $\WH C_n$,
there are only physical poles in (\ref{1loop-A-R}).

Now we consider residues in (\ref{1loop-A-R}), by using the
factorization property (\ref{1loop-factor}). But
(\ref{1loop-factor}) is for the whole amplitude. We need to see how
the factorization works for the cut and the rational part, respectively. As
explained in \cite{Bern:2005cq}, in general, the pure cut part and
rational part are factorized separately. (\ref{1loop-factor}) is
true for the pure cut part as well as the rational part, except for
two particle channels of a particular helicity configuration, which
can be avoid if we choose the deformation pair carefully. With this
understanding, we can write down the on-shell recursion relation for
rational part as
\bea R_n^D & \equiv & -\sum_{\rm poles} {\rm Res}\left( {
R_n(z)\over
z}\right)~~~~\label{R-D-rec}\\
& = & \sum_p\sum_{h=\pm } R(..\WH p_i, .., -p^h) {1\over p^2}
A^{\rm tree}(p^{-h},..,\WH p_j,...)+ A^{\rm tree}(..\WH p_i, .., -p^h)
{1\over p^2} R(p^{-h},..,\WH p_j,...)\nonumber \eea
Having $R_n^D$ and the explicit result for $\WH C_n$ we can find
\bea -\sum_{\rm poles} {\rm Res}\left( {\WH R_n(z)\over
z}\right)=R_n^D+\sum_{\rm pole~\b}  {\rm Res}\left( {\WH
{CR}_n(z)\over z}\right)~~~\label{Res-WH-R}\eea
where poles at the right-handed side of (\ref{Res-WH-R}) should
include spurious poles as well.

Using above frame of recursion relation,  rational parts of various
six gluons amplitudes have been calculated \cite{Bern:2005cq}.
Same results for rational parts are also obtained using improved
Feynman diagram methods \cite{Xiao:2006vr,Su:2006vs,Xiao:2006vt}.

\subsection{On-shell recursion relations in 3D}

As we have mentioned, solution (\ref{BCFW-con-mom}) exists for the
BCFW deformation when and only when the dimension of space-time is
four or above. In 3D, there is a non-linear  generalization of the
BCFW-deformation \cite{Gang:2010gy}.

In 3D, we can use $\vec\sigma$-matrices to write momentum in matrix form
\bea p^{\a\b}= x^\mu (\sigma_\mu)^{\a\b},~~\label{3D-p} \eea
with
\bea \sigma^0=\left( \begin{array}{cc} -1 & 0 \\ 0 & -1 \end{array}
\right),~~~\sigma^1=\left( \begin{array}{cc} -1 & 0 \\ 0 & 1
\end{array} \right)~~,
\sigma^2=\left( \begin{array}{cc}  0 & 1 \\ 1 & 0 \end{array}
\right),~~~\label{3D-sigma}\eea
For null momentum, we have
\bea  p^{\a \b}=\la^\a \la^\b,~~~~2p_i\cdot
p_j=-\Spaa{i|j}^2,~~~\label{3D-p-spinor}\eea
where, unlike the case in 4D, there is only one spinor $\la$. A key
requirement for on-shell deformation is to preserve momentum
conservation. Noticing the quadratic form of momentum in
\eqref{3D-p-spinor}, it is suggested that
 the on-shell BCFW-deformation should be considered as matrix  transformation over two
spinors \cite{Gang:2010gy}
\bea  \left( \begin{array}{c} \la_i(z) \\ \la_j(z) \end{array}
\right)= R(z) \left( \begin{array}{c} \la_i \\
\la_j
\end{array} \right),~~\label{3D-BCFW-def}\eea
where the $R(z)$ is a two-by-two matrix depending on $z$. Under this
transformation, the on-shell condition for momentum $p_i, p_j$ is
kept automatically, while  conservation of momentum then reduces to
\bea  \left( \begin{array}{cc} \la_i(z) & \la_j(z) \end{array}
\right)\left( \begin{array}{c} \la_i(z) \\ \la_j(z) \end{array}
\right)= \left( \begin{array}{cc} \la_i & \la_j \end{array}
\right)\left( \begin{array}{c} \la_i \\ \la_j \end{array}
\right),~~~~{\rm or}~~~ R^T(z) R(z)=I,~~~\label{3D-R-con} \eea
or, $R(z)\in SO(2,C)$. Generally, $SO(2,C)$ can be parameterized as
\bea R(z) =\left( \begin{array}{cc} {z+z^{-1}\over 2}
 & -{z-z^{-1}\over 2i}\\  {z-z^{-1}\over 2i} & {z+z^{-1}\over 2}\end{array}\right).~~
 \label{3D-R-form}\eea
If we write $z= e^{i\theta}$, $R(z)$ is a familiar rotation in 2D.
With this deformation, for example, of the pair $(1,\ell)$, we can
calculate $\WH p_f(z)=p_I+...+ p_l(z)+...+p_J$ with $1<I \leq \ell
\leq J$
\bea \WH p_f^2(z)= a_f z^{-2} +b_f+ c_f z^2  \eea
where
\bea a_f &= & -2 \W q\cdot (p_f-p_\ell),~~~b_f= (p_f+p_1)\cdot
(p_f-p_\ell),~~~a_f=-2 q\cdot (p_f-p_\ell)\nn
q^{\a\b}& = & {1\over 4} (\la_1+i\la_\ell)^\a (\la_1+i\la_\ell)^\b,
~~~~\W q^{\a\b} =  {1\over 4} (\la_1-i\la_\ell)^\a
(\la_1-i\la_\ell)^\b~.\eea
Comparing to propagators in 4D which have linear dependence on
$z$, $\WH p_f^2(z)$ is much more complicated and on-shell condition
gives four solutions $\{ \pm z_{1,f}^\star, \pm
z_{2,f}^\star\}$
\bea \{  z_{1,f}^2,  z_{2,f}^2\} =\left\{ {(p_f+p_1)\cdot
(p_f-p_\ell)\pm \sqrt{ (p_f+p_1)^2 (p_f-p_\ell)^2}\over 4 q\cdot
(p_f-p_\ell)} \right\}~~\label{3D-four-sol}\eea
after using the Schouten identity
\bea \Spaa{r|p|s}^2=\Spaa{r|p|r}\Spaa{s|p|s}+p^2\Spaa{r|s}^2~.  \eea

Thus prepared, we can write down the on-shell recursion relation in
3D. Starting from contour integration
\bea A(z=1) & = & \oint_{z=1} {dz \over z-1} A(z)  \eea
where the contour is a small circle around $z=1$. Deforming the
contour to region outside of small circle, we will evaluate residues
at $\WH p_f(z)=0$ and at the infinity. After some short
calculations, we have
\bea A(z=1)= B+ \sum_f H(z_{1,f}^\star, z_{2,f}^\star) A_L
(z_{1,f}^\star){1\over p_f^2} A_R(z_{1,f}^\star) +\{ z_{1,f}^\star
\leftrightarrow z_{2,f}^\star\}~,~~~\label{3D-BCFW}\eea
where $H(a,b)$ is defined by
\bea H(a,b)= \left\{ \begin{array}{ll} {a^2(b^2-1)\over a^2-b^2},~~~
& l={\rm odd} \\  {a(b^2-1)\over a^2-b^2},~~~ & l= {\rm even}
\end{array}\right.\eea
Now a few remarks are in order for (\ref{3D-BCFW}). First, different
from the 4D BCFW recursion relation, there are four poles
(\ref{3D-four-sol}) for a given propagator and we need to sum up all
contributions from them. The summation of four solutions is counted
by the factor $H(a,b)$ plus $\{ z_{1,f}^\star \leftrightarrow
z_{2,f}^\star\}$. Second, the boundary behavior is  important when
we try to write down the recursion relation. In the case considered
in \cite{Gang:2010gy}, the boundary contribution is zero. In fact,
since when $z\to \infty$ each propagator contributes $ z^{-2}$ under
our deformation, the boundary behavior is better than in 4D
for many theories.

The application of (\ref{3D-BCFW}) to 3D ABJM theory is presented in
\cite{Gang:2010gy} where one may find more details.

\subsection{Cachazo-Svrcek-Witten (CSW) Rules}

The last generalization of on-shell recursion relations to be
discussed is, in fact, a little off the main line of this article.
All previous generalizations are  based on the same BCFW-deformation
in \eqref{BCFW-deform-mom}. Now we make another generalization by
defining a different deformation. Using the newly defined
deformation, we will prove CSW rules.\footnote{There are many works
on CSW rules, for which one can found in the review
\cite{Brandhuber:2011ke}. In this review, we just give the proof of
CSW rule by on-shell recursion relation. CSW rules  have also been
understood from Lagrangian by field redefinitions
\cite{hep-th/0511264,hep-th/0510111}.}

As we have mentioned in the introduction,  Witten's twistor program
\cite{Witten:2003nn} provides the geometric picture  that the
MHV-amplitude (\ref{MHV}) locates on a straight line in twistor
space. This point of view has trigged many important works of
studying scattering amplitudes. One is the conjectured CSW rules
\cite{Cachazo:2004kj} for calculations of scattering amplitudes. The
proposal was originally for tree-level amplitudes, but quickly
generalized to one-loop and recently to all planar loop-amplitudes.

CSW rules use MHV amplitudes as vertices and scalar
propagator ${i/ p^2}$ to construct all allowed Feynman-like diagrams.
There is a technical problem to use MHV amplitudes as vertices.
The momentum of propagator is not null and there is no natural definition for its spinors.
To deal with this, it was proposed to define the spinor
$\ket{p}\equiv \bket{p|\eta}$ with the help of an auxiliary anti-spinor $\bket{\eta}$.
This introduces an $\bket{\eta}$-dependence in each diagram.
However, when we sum all diagrams up, the final result  is $\bket{\eta}$-independent.

One way to view the definition of $\ket{p}$ is as follow.
Any momentum $p$ can always be decomposed as
\bea p= {\cal P}+z\eta\eea
where $\eta$ is an auxiliary null momentum and we require ${\cal P}^2=0$.
The null condition on ${\cal P}$ fixes $z={p^2/ 2p\cdot \eta}$.
Since ${\cal P}$ is null, it has natural spinor $\ket{{\cal P}}$.
On the other hand,
\bea \bket{p|\eta}= \bket{{\cal P}|\eta}+z\bket{\eta|\eta}=
\ket{{\cal P}}\Spbb{\cal P|\eta}\Longrightarrow \ket{{\cal
P}}={\bket{p|\eta}\over \Spbb{\cal P|\eta}}~~~\label{CSW-spinor}\eea
There are same numbers of $\ket{\cal P}$ in the numerator and the denominator,
so the degree of $\ket{\cal P}$ is zero.
Thus we can simply use  $\ket{{\cal P}}\sim \bket{p|\eta}$.

Having defined spinors of off-shell momenta, we are ready to prove the CSW rules
\cite{Risager:2005vk} by using analytic properties of scattering
amplitudes, like those used for on-shell recursion relations. We
start with the case of NMHV amplitudes and the following holomorphic
deformations
\bea \bket{i(z)}= \bket{i}+z\Spaa{j|k}\bket{\eta},~~~\bket{j(z)}=
\bket{j}+z\Spaa{k|i}\bket{\eta},~~~\bket{k(z)}=
\bket{k}+z\Spaa{i|j}\bket{\eta},~~~\label{NMHV-def}\eea
where $i,j,k$ are three particles of negative helicity. In the
BCFW-deformation, one takes a pair of particles. One particle is
deformed in the spinor component and the other deformed in the
anti-spinor component. In (\ref{NMHV-def}), only anti-spinor
components are deformed. The null condition is automatically kept
while momentum conservation is preserved by using the  Schouten
identity (\ref{Schouten}). Directly inspecting Feynman diagrams, one
sees the shifted amplitude has the boundary behavior $z^{-2}$. There
is no residue at the infinity point.

Just like the derivation of on-shell recursion relations, considering
the contour integral $\oint (dz/ z) A(z)$, we immediately write down
\bea A & = & \sum_{\a, i\in A_L} A_L( z_\a) {1\over p_\a^2} A_R(
z_\a) \eea
where at least one of $j,k$ is in $A_R$.
Notice that under deformation (\ref{NMHV-def}), there is no
nonzero three-point $\O {\rm MHV}$-amplitude for general momentum configuration.
Thus if both $j,k\in A_R$, $p_\a$ in $A_L$ should be of negative helicity to have nonzero contribution.
If only one of $j,k$ is in $A_R$, $p_\a$ in $A_L$ should be of positive helicity to have nonzero contribution.
This decomposes the NMHV-amplitude into two MHV-diagrams, as suggested by the CSW rule.
We still need to work out spinor components, which turns out to be $\ket{p_\a}\sim \bket{p_\a|\eta}$.
This completes CSW rules for NMHV amplitudes.

For general ${\rm N}^{n-1}{\rm MHV}$-amplitudes, we make the deformation
\bea \bket{m_i(z)}=m_i+z r_i \bket{\eta},
~~~i=1,...,n+1,~~~\label{CSW-gen-shift}\eea
for $n+1$ particles of negative helicity.
Here $\sum_i r_i \ket{m_i}=0$ to ensure momentum conservation.
To avoid degeneracy, we require that the sum of any subset of $m_i$'s is not  zero.
The shifted amplitude has the boundary behavior $ z^{-n}$.
On-shell recursion relations tell that $A=\sum_\a A_L (1/ p_\a^2) A_R$ where $A_L, A_R$
are ${\rm N}^r {\rm MHV}$ amplitudes with $r<n-1$ .
Via induction, the amplitude can be calculated by CSW rules.

There is a subtle issue in calculating $A_L, A_R$ by using CSW rules.
Every ${\rm N}^{n-1}{\rm MHV}$ CSW diagram appears $n-1$ times and at each time,
the propagator is shifted differently by deformation.
For the result from CSW rules to match up the one from recursion, we need to prove
\bea \sum_{i=1}^{n-1} {1\over p_i^2} \prod_{j=1,j\neq i}^{n-1}
{1\over \WH p_{ji}^2}=  \prod_{i=1}^n {1\over
p_i^2},~~~\label{CSW-key}\eea
where
\bea \WH p_{ji}^2=(p_j+z_i
\ket{\la_j}\bket{\eta})^2=p_j^2-p_i^2{\Spab{\la_j|p_j|\eta}\over
\Spab{\la_i|p_i|\eta}}\eea
is the shifted momentum of $j$-th propagator when $i$-th propagator
goes on-shell, which is the location of pole under deformation (\ref{CSW-gen-shift}).
 (\ref{CSW-key}) can be proved simply by
defining an analytic function $I(z)=\prod_{i=1}^n {1/ (p_i+
z\ket{\la_i}\bket{\eta})^2}$ and calculating the contour integral $\oint(dz/ z) I(z)$.
With this settled, CSW rules can be explained by using the analytic
property of amplitudes under the generalized deformation
(\ref{CSW-gen-shift}).

\subsubsection*{Another example of holomorphic deformation}

As we have seen in previous subsection that for one-loop amplitude
with all positive helicities, the result is a pure rational function
given in (\ref{1loop-All-Plus}). Due to the structure of the
denominator, $A_{n;1}$ can only have poles  in two-particle
channels.

To make the amplitude vanish faster at $z=\infty$, we want to
increase the power of $z$  in the denominator. Thus we make the
following pure holomorphic deformations\footnote{This example was
studied by Britto, Cachazo and Feng around 2005-2006, but has not
been published.}
\bea \la_1(z)&=&\la_1 -z [2~3]\eta,~~~ \la_2(z)  =  \la_2-z [3~1]
\eta,~~  \la_3(z)  =  \la_3-z[1~2] \eta.~~~\label{plus-1loop-def}
\eea
Four two-particle channels $p_{n1}, p_{12}, p_{23},
p_{34}$ contain $z$-dependence. The possible $z$-dependence of
numerators in (\ref{1loop-All-Plus}) is given by
\bean && \braket{1~2}[2~3]\braket{3~i}[i~j],~~~~~\braket{1~2}[2~i]
\braket{i~j}[j~k],~~~~~~\braket{1~3}[3~i]\braket{i~j}[j~k],\\ &&
\braket{2~3}[3~i]\braket{i~j}[j~k],~~~~~\braket{2~i}[i~j]\braket{j~k}[k~l],
~~~~~\braket{3~i}[i~j]\braket{j~k}[k~l]\eean
with  all $i,j,k\geq 4$. Only the first term has possible
$z^2$-dependence. So the amplitude vanishes as  ${z^2/z^4}=  z^{-2}$
when $z\to\infty$.

With the $z^{-2}$ behavior, it is natural to seek bonus relations.
We can reduce the complexity of computations by choosing
$\ket{\eta}=\bket{p_{12}|3}$. With this choice, one reduces  one
power of $z$ in both $s_{12}$ of the denominator and
$\braket{1~2}[2~3]\braket{3~i}[i~j]$ of the numerator. Now consider
the following contour integration\footnote{Now $z=0$ is a pole, so
we can obtain the original $A_n$.}
\bea 0 & = &\int {dz\over z} A_n(z)
\Spaa{\la_2(z)|\la_3(z)}~~\label{z23-1loop}\eea
where the factor $\Spaa{\la_2(z)|\la_3(z)}$ makes contributions from
the pole $s_{23}$ vanishing. We are left with only two poles
$s_{n1}$ and $s_{34}$ in the bonus relation:
\bea  & & A(1^+,2^+,..., n^+) ~~~~~~\label{Using-Z23}\\
& = &  {\gb{n|p_{12}|3}\over \gb{1|p_{12}|3}\braket{n~1}} A_{n-1}(
p_{n1}^+, \WH 2_a^+, \WH 3^+, 4^+,..., (n-1)^+) \left(
-{\gb{n|p_{123} p_{23} p_{123} |3} \over p_{23}^2 \gb{n|p_{123}|3}}
\right)\nonumber \\
& & +{\gb{4|p_{12}|3}\over \gb{3|p_{12}|3}\braket{3~4}} A_{n-1}(\WH
1^+, \WH 2_b^+, p_{34}^+,5^+,...,n^+)
 \left( {\braket{3|p_{12}|3} \gb{4|p_{23}|1}\over \gb{4|p_{12}|3} [1~2]\braket{2~3}} \right)
\nonumber \eea
where hatted variables are calculated from the corresponding poles.

Now we  apply (\ref{Using-Z23})  to the case $n=5$, starting with
the four gluon amplitude
$$ A_4(1^+,2^+,3^+,4^+)= {[2~3] [4~1]\over
\braket{2~3}\braket{4~1}}={[1~2][3~4]\over
\braket{1~2}\braket{3~4}}$$ The first term is given by
\bean { [4|5+1|1+2|3]\over
\braket{1~2}\braket{2~3}\braket{4~5}\braket{5~1}} \eean
while  the second is given by
\bean - {[5~1][1~2]\over \braket{2~3}\braket{3~4}\braket{4~5}}\eean
In total, we have
 \bea A_5 & = & { [4~1] [2~3] \over \braket{2~3}\braket{4~5}\braket{5~1}}
+{-[3~4][4~5]\over \braket{1~2}\braket{2~3}\braket{5~1}}+
{-[5~1][1~2]\over \braket{2~3}\braket{3~4}\braket{4~5}}\eea
which is simpler than the one given by
(\ref{1loop-All-Plus}).

\section{Applications of on-shell recursion relations}

Having discussed several generalizations, we now present some applications of on-shell recursion relations. We will
divide these applications into two categories: those calculating
desired amplitudes and those proving intrinsic properties of amplitudes.

\subsection{Split  helicity amplitudes}

Our first explicit results calculated from on-shell recursion
relation are  split helicity amplitudes mentioned in
\eqref{Split-pq}. One good property of this kind of  amplitudes is
that they are closed under on-shell recursion relations and can be
solved explicitly. Split NMHV amplitude
$A(1^-,2^-,3^-,4^+,\ldots,n^+)$ was obtained via on-shell recursion
relations in \cite{Luo:2005rx}. For general split helicity
amplitudes, the full solution was worked out in \cite{Britto:2005dg}
and given by
\bea A(1^-,\ldots,q^-,(q{+}1)^+,\ldots,n^+) =
\sum_{k=0}^{\min(q-3,n-q-2)} \sum_{A_k,B_{k+1}} {N_1 N_2 N_3 \over
D_1 D_2 D_3}~.~~\label{fullzig} \eea
Here $A_k$ and $B_{k+1}$ range over all subsets of indices
$\{2,\ldots,q-2\}$ and $\{q+1,\ldots,n-1\}$ of cardinality $k$ and $k+1$, respectively.
In increasing numerical order, the elements are labeled by
$a_1,a_2,\ldots,a_k$ and $b_{k+1},\ldots,b_1$. There are a total of
$C_{n - 4}^{q - 2}$ terms in the sum as we have counted before.
$N_i$ and $D_i$ are defined by (where we have used
$p_{x,y}=p_x+p_{x+1}+...+p_y$)
\bea N_1 &= & {\langle 1 | p_{2,b_1} p_{b_1 + 1, a_1} p_{a_1 + 1,
b_2} \cdots p_{b_{k+1} + 1,q - 1}|q\rangle^3}~,\nn N_2 &= & \langle
b_1{+}1~b_1 \rangle \langle b_2{+}1~b_2 \rangle \cdots \langle
b_{k+1}{+}1~b_{k+1} \rangle~,\nn N_3 &= & [a_1~a_1{+}1] \cdots
[a_k~a_k{+}1]~,\nn D_1 &= & p^2_{2,b_1} p^2_{b_1 + 1, a_1} p^2_{a_1
+ 1, b_2} \cdots p^2_{b_{k+1} + 1,q - 1}~,\nn D_2 &= & F_{q,1}
\overline{F}{}_{2,q-1}~,\nn D_3 &= & [2|p_{2,b_1}|b_1{+}1 \rangle
\langle b_1|p_{b_1 + 1,a_1}|a_1]
[a_1{+}1|p_{a_1+1,b_2}|b_2{+}1\rangle \cdots \langle
b_{k+1}|p_{b_{k+1}+1,q-1}|q{-}1]~, ~~~\label{pieces}\eea
 where  $F_{x,y}$ is given by
\bea F_{x,y} = \langle x~x{+}1\rangle \langle x{+}1~x{+}2\rangle
\cdots \langle y{-}1~y\rangle~,~~\label{aaa}\eea
 and $\overline{F}_{x,y}$ is given by the same expression but of the inner product $[\cdot~\cdot]$.

The best way to illustrate terms in (\ref{fullzig}) is to use the so-called zigzag diagrams,
as shown in Figure \ref{fig:zigzag} \cite{Britto:2005dg}.
First draw a big circle. Then
arrange gluon indices in clockwise order around the circle,
with negative helicities $\{1,\ldots,q\}$ on the top side and
positive helicities $\{q+1,\ldots,n\}$ on the bottom side. A
zigzag is a connected collection of non-self-intersecting line
segments which begins at $(1,2)$ and ends at $(q-1,q)$, alternating
at each step between the top and bottom sides. It is clear that
there is a one-to-one correspondence between zigzag diagrams
and choices of subsets $A_k$ and $B_{k+1}$. The line segments in
a zigzag diagram are in one-to-one correspondence with the momenta
$P_{x,y}$ appearing in (\ref{pieces}).
The rule for transforming any given zigzag diagram into a formula is clear
from (\ref{pieces}).

\EPSFIGURE[h]{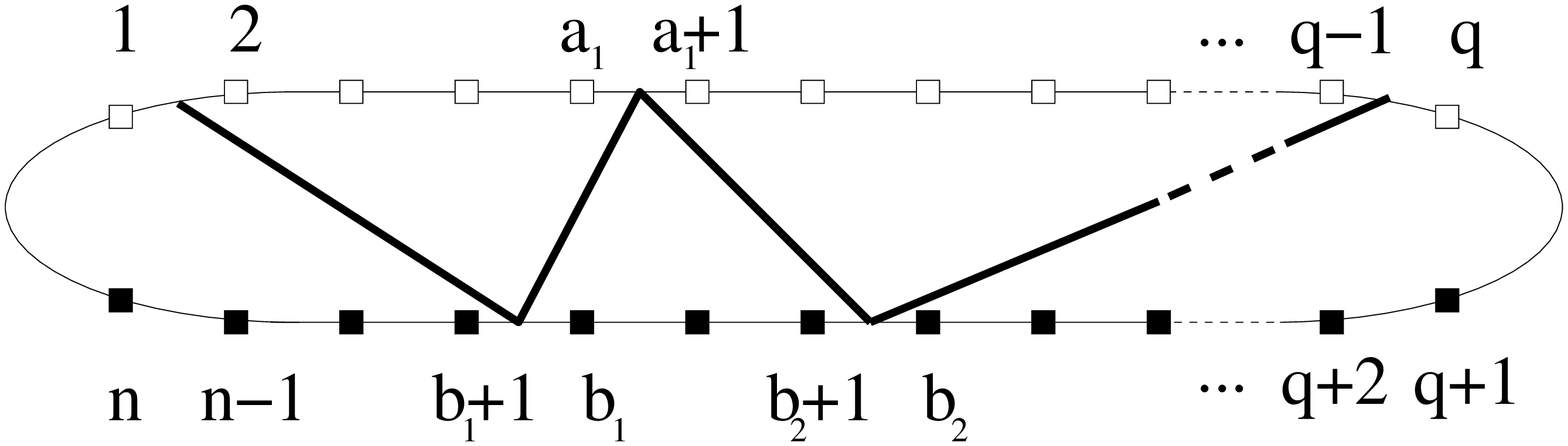,width=14.5cm} {Zigzag diagrams for
split helicity.
    \label{fig:zigzag}
    }

\begin{figure}[h]
\begin{center}$
\begin{array}{ccc}
\includegraphics[width=1.5in]{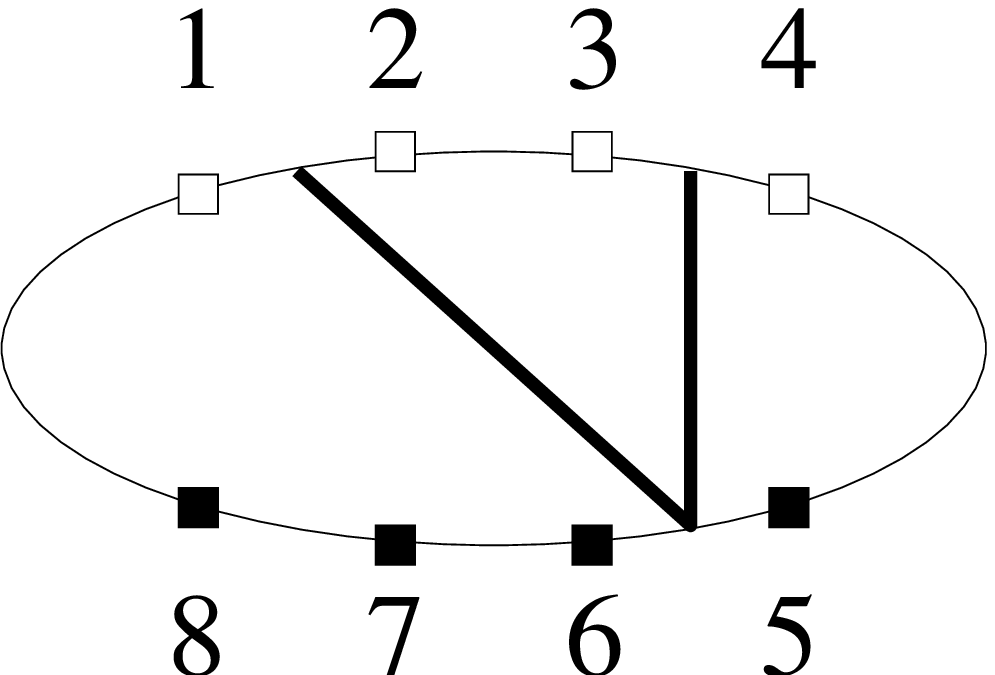} &
\includegraphics[width=1.5in]{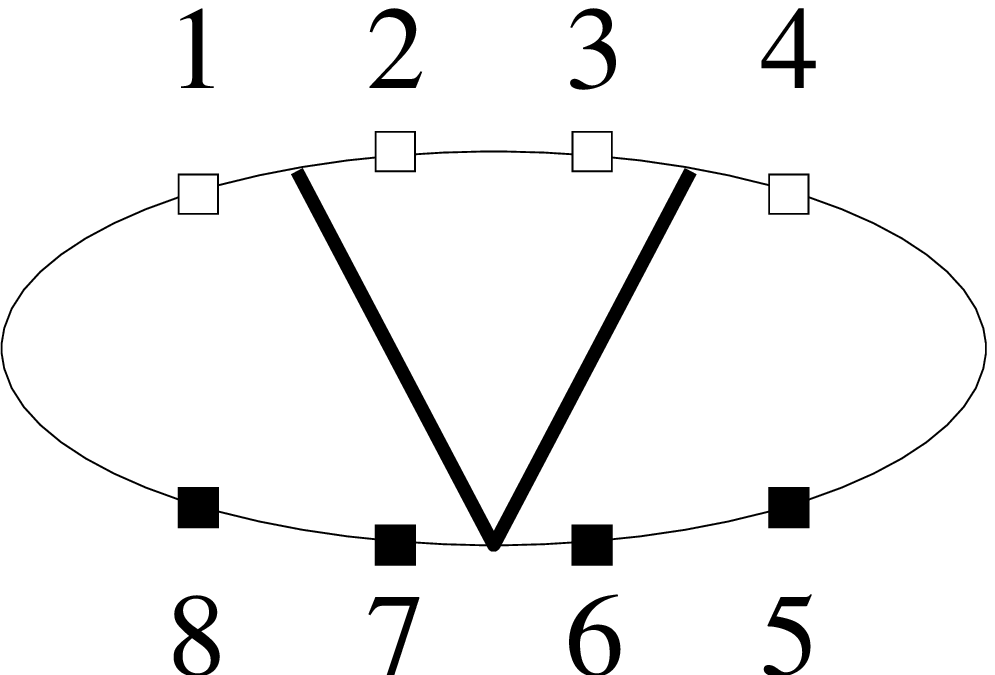} &
\includegraphics[width=1.5in]{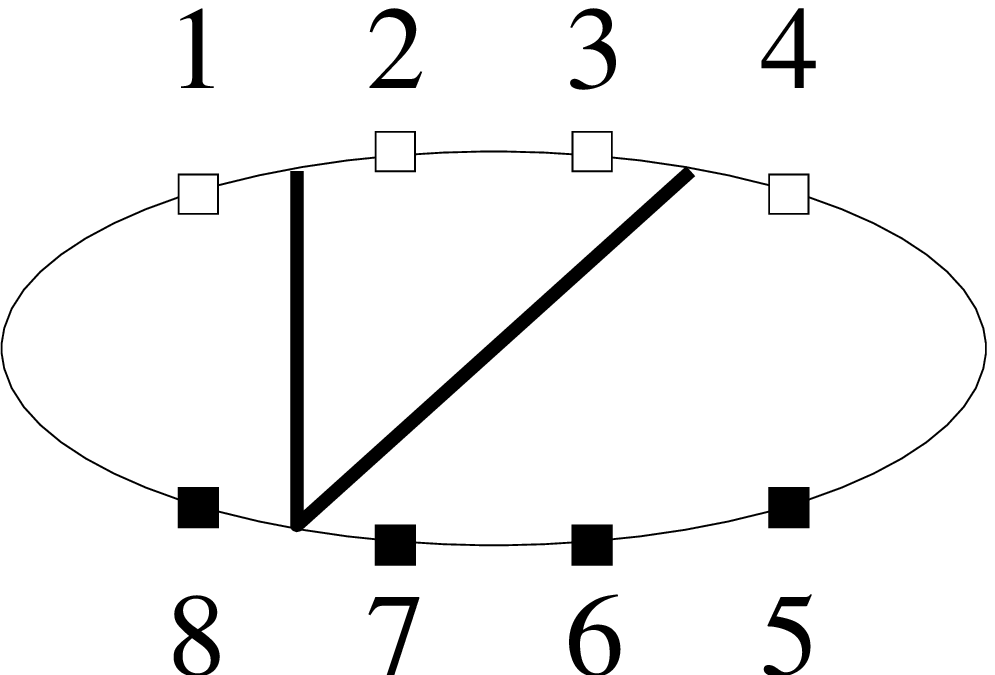}\\
(a) & (b) & (c)\\
\includegraphics[width=1.5in]{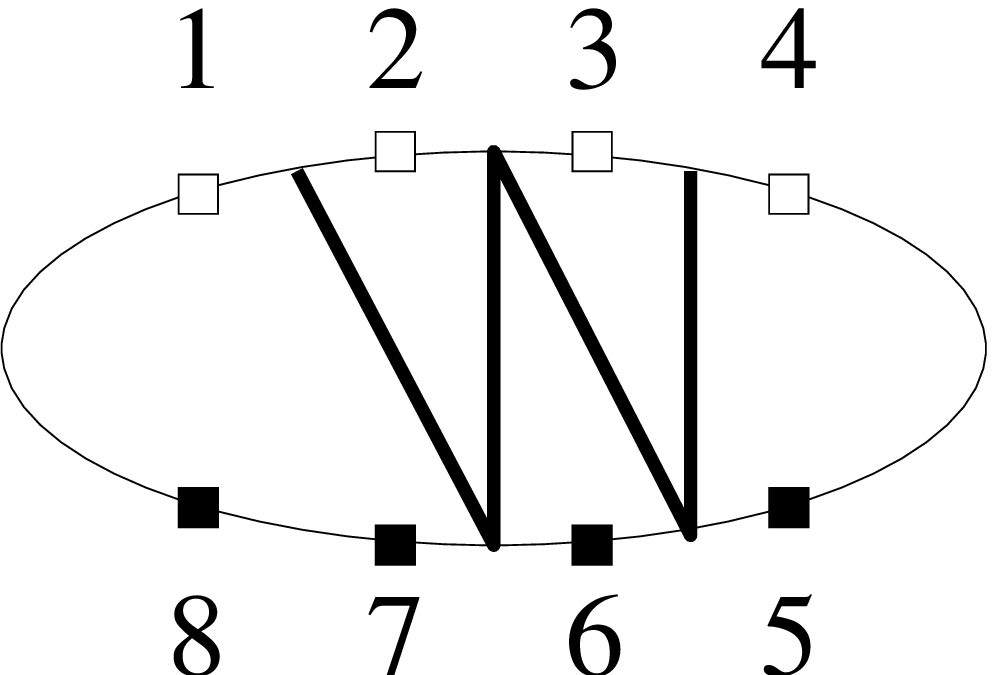} &
\includegraphics[width=1.5in]{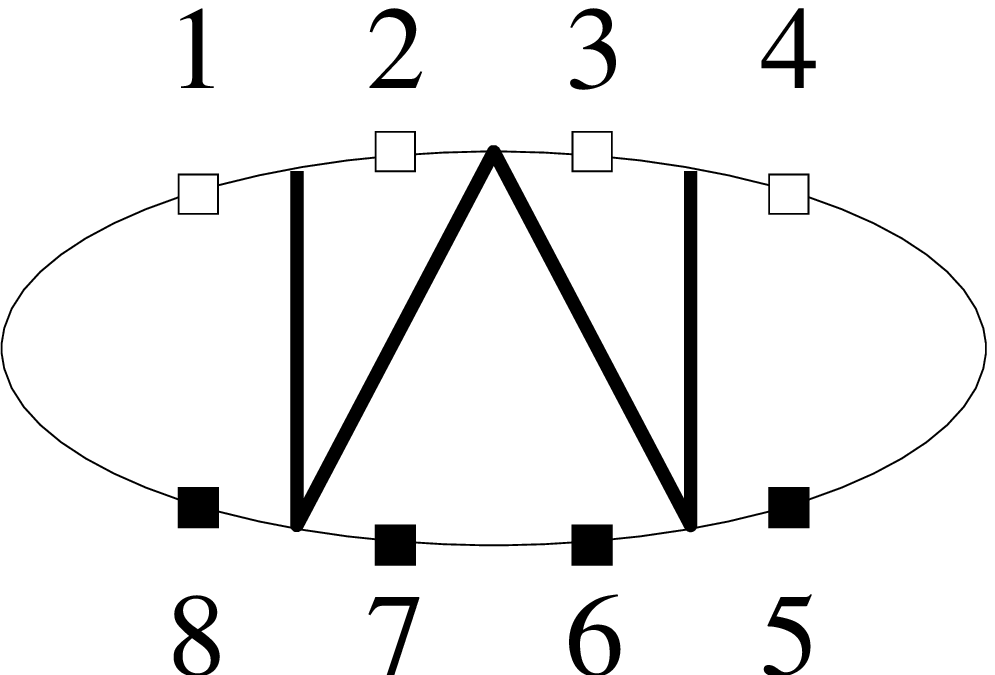} &
\includegraphics[width=1.5in]{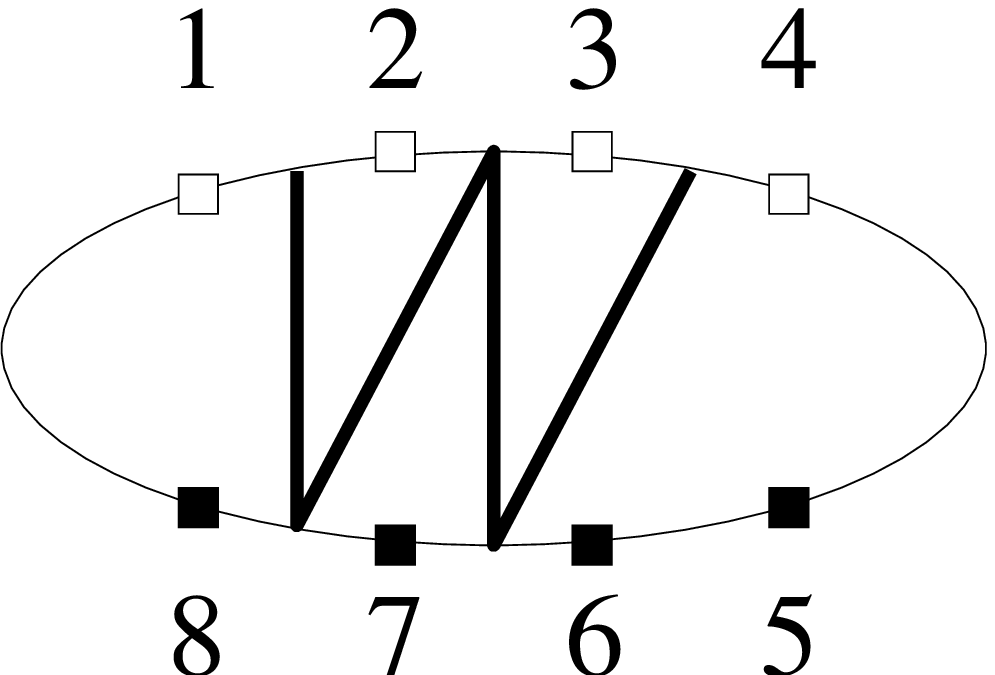}\\
(d) & (e) & (f)
\end{array}$
\end{center}
\caption{Six zigzag diagrams for $A_8(1^-, 2^-, 3^-, 4^-, 5^+, 6^+,
7^+, 8^+)$.} \label{fig:8split}
\end{figure}

To demonstrate the use of zigzag diagrams, we give the result for
the split helicity amplitude of eight gluons $A_8(1^-, 2^-, 3^-,
4^-, 5^+,6^+, 7^+, 8^+)$. As shown in Figure \ref{fig:8split}, there
are six zigzag diagrams and their corresponding expressions are
\bea (a) & = & {1 \over F_{4,1} \overline{F}{}_{2,3}} {\langle 1 |
P_{2,5} P_{6,3}|4\rangle^3 \over P_{2,5}^2 P_{6,3}^2 } { \langle 6~5
\rangle \over [2|P_{2,5}|6\rangle \langle 5|P_{6,3}|3]}~,\nn
(b) & = & {1 \over F_{4,1} \overline{F}{}_{2,3}} {\langle 1 |
P_{2,6} P_{7,3}|4\rangle^3 \over P_{2,6}^2 P_{7,3}^2 } { \langle 7~6
\rangle \over [2|P_{2,6}|7\rangle \langle 6|P_{7,3}|3]}~,\nn
(c) & = & {1 \over F_{4,1} \overline{F}{}_{2,3}} {\langle 1 |
P_{2,7} P_{8,3}|4\rangle^3 \over P_{2,7}^2 P_{8,3}^2 } { \langle 8~7
\rangle \over [2|P_{2,7}|8\rangle \langle 7|P_{8,3}|3]}~,\nn
(d) & = & {1 \over F_{4,1} \overline{F}{}_{2,3}} { \langle 1 |
P_{2,6} P_{7,2} P_{3,5} P_{6,3} |4\rangle^3 \over P^2_{2,6}
P^2_{7,2} P^2_{3,5} P^2_{6,3}} {\langle 7~6 \rangle \qquad [2~3]
\qquad \langle 6~5 \rangle \over [2|P_{2,6}|7\rangle \langle
6|P_{7,2}|2] [3|P_{3,5}|6\rangle \langle 5|P_{6,3}|3]}~,\nn
(e) & = &{1 \over F_{4,1} \overline{F}{}_{2,3}} { \langle 1 |
P_{2,7} P_{8,2} P_{3,5} P_{6,3} |4\rangle^3 \over P^2_{2,7}
P^2_{8,2} P^2_{3,5} P^2_{6,3}} {\langle 8~7 \rangle \qquad [2~3]
\qquad \langle 6~5 \rangle \over [2|P_{2,7}|8\rangle \langle
7|P_{8,2}|2] [3|P_{3,5}|6\rangle \langle 5|P_{6,3}|3]}~,\nn
(f)& =& {1 \over F_{4,1} \overline{F}{}_{2,3}} { \langle 1 | P_{2,7}
P_{8,2} P_{3,6} P_{7,3} |4\rangle^3 \over P^2_{2,7} P^2_{8,2}
P^2_{3,6} P^2_{7,3}} {\langle 8~7 \rangle \qquad [2~3] \qquad
\langle 7~6 \rangle \over [2|P_{2,7}|8\rangle \langle 7|P_{8,2}|2]
[3|P_{3,6}|7\rangle \langle 6|P_{7,3}|3]}~. ~~\label{8split}\eea

The proof of (\ref{fullzig}) was given in \cite{Britto:2005dg} where one may find more details.

\subsection{The solution of ${\cal N}=4$ SYM theory at tree level}

As we have remarked in the previous section, amplitudes in ${\cal
N}=4$ SYM theory are similar to split helicity amplitudes. They are
closed under on-shell recursion relations \eqref{BCF-super} and can
be solved explicitly \cite{Drummond:2008cr}.\footnote {The solution
for tree-level $\mathcal{N}=8$ supergravity was given
in~\cite{arXiv:0901.2363}}.  The solution is very important. It
inspired many important works, such as the dual superconformal
symmetry \cite{Drummond:2008vq, Berkovits:2008ic, Beisert:2008iq,
Brandhuber:2008pf} and the Yangian symmetry \cite{Drummond:2009fd}.
It has been understood from new point of views, such as from the
leading singularity of primitive higher loop amplitudes
\cite{Bullimore:2009cb, Kaplan:2009mh}.

The starting point of on-shell recursion relation is the three-point
MHV-amplitude in (\ref{intro-MHV-n}) and three-point anti-MHV-amplitude
\bea
  {\cal A}^{\rm \widebar{MHV}}_{3}(\lambda,\tilde\lambda,\eta) = \delta^{(4)}(p)
  \frac{\delta^{(4)}(\eta_1[23]+\eta_2[31]+\eta_3[12])}{[12][23][31]}\,.
~~\label{bar-MHV3}\eea
Using the super version of on-shell recursion relation (\ref{BCF-super}),
one gets the NMHV-amplitude as
\bea
 {\cal A}^{\rm {NMHV}}_{n} = \mathcal{A}_n^{\rm MHV} \mathcal{P}_n^{\rm NMHV}
 =\frac{ \delta^{(4)}(p) \, \delta^{(8)}(q)}{\vev{1\, 2}\vev{2\, 3}
 \ldots\vev{n\, 1}} \, \sum_{2\leq s<t  \leq n-1} \!\!\!\!\!\! R_{n;st} \,,~~~\label{ansatzNMHV}
\eea
where $p=\sum_i p_i$, $q=\sum_i\la_i \eta_i$ and $R_{r;st}$ is a
dual superconformal invariant
 \bea
 R_{r;st} = \frac{\l<s \,\, s-1\r> \l<t \,\, t-1\r>
}{x_{st}^2\l<r|x_{rs}x_{st}|t\r>\l<r|x_{rs}x_{st}
|t-1\r>\l<r|x_{rt}x_{ts}|s\r>\l<r|x_{rt}x_{ts}|s-1\r>}\delta^{(4)}(\Xi_{r;st})\,.~~~\label{def-Rrst}
\eea
with\footnote{In the definitions of $p_i,\eta_i$, $x_i$ and
$\theta_i$ are solved up to arbitrary $x_1,\theta_1$ and the
momentum and supermomentum conservation are satisfied after we
impose periodic conditions $x_1=x_{n+1}$ and
$\theta_1=\theta_{n+1}$.}
\bea p_i=
x_i-x_{i+1},~~~\la_i\eta_i=\theta_i-\theta_{i+1},~~~x_{ab}=x_a-x_b
\eea
 and the Grassmannian odd quantity $\Xi_{r;st}$
\bea \Xi_{r;st} = \l<r|x_{rs}x_{st}|\theta_{tr}\r> +
\l<r|x_{rt}x_{ts}|\theta_{sr}\r>\,.~~~\label{defxi} \eea

To express general ${\rm N}^k{\rm MHV}$-amplitudes, we introduce
several structures. The first is a function of $(r+1)$-pair of
subscripts
\bea R_{n;b_1a_1;b_2a_2;\ldots;b_ra_r;ab} = \frac{\l< a\,\,a-1\r>
\l<
  b\,\,b-1\r> \delta^{(4)}(\l< \xi | x_{a_r a}x_{ab} | \theta_{ba_r} \r> +
  \l< \xi |x_{a_r b} x_{ba} | \theta_{aa_r} \r>)}{x_{ab}^2 \l< \xi |
  x_{a_r a} x_{ab} | b \r> \l< \xi | x_{a_r a} x_{ab} |b-1\r> \l< \xi
  | x_{a_r b} x_{ba} |a\r> \l< \xi | x_{a_r b} x_{ba} |a-1\r>}\,,
  \label{generalR}
\eea
where the chiral spinor $\langle \xi |$ is given by
\bea \l< \xi | = \l<n | x_{nb_1}x_{b_1 a_1} x_{a_1 b_2} x_{b_2 a_2}
\ldots x_{b_r a_r} \,. \eea
 The sum of the pair $a_i, b_i$ is over $L_i\leq a_i< b_i< U_i$.
At the boundary where $a_i=L_i$ or $n_i=U_i$,
we need to introduce a second function of two groups of superscripts to handle the speciality
\bea \sum_{L\leq a<b \leq U} \!\!\!\!\!\!
R_{n;b_1a_1;\ldots;b_ra_r;ab}^{l_1\ldots l_p;u_1\ldots u_q} \, .
\label{superscripts} \eea
Modifications indicated by these superscripts are as follows.
For terms in the sum where $a=L$ we replace the explicit dependence on $\langle L-1|$ in
(\ref{generalR}) by,
\bea \langle L-1 | \longrightarrow \langle n|x_{nl_1} x_{l_1l_2}
x_{l_2l_3} \ldots x_{l_{p-1} l_p} \, . \label{Lrep} \eea
 Similarly, for terms in the sum where $b=U$ we replace the explicit
dependence on $\langle U|$ in (\ref{generalR}) by,
\bea \langle U | \longrightarrow \langle n | x_{nu_1} x_{u_1u_2}
x_{u_2u_3} \ldots x_{u_{q-1}u_q} \, . \label{Urep} \eea
In the term where $a=L$ and $b=U$, both replacements occur.
When no replacement is made on one boundary, we will use the superscript $0$.

With these notations, we write
\bea \mathcal{A}_n^{\rm N^k MHV} = \mathcal{A}_n^{\rm MHV}
\mathcal{P}_n^{\rm N^k MHV}.~~ \label{ANNMHV} \eea
For $k=1$, we have (\ref{ansatzNMHV}). For $k=2$, we have
\begin{align}
 \mathcal{P}_n^{\rm NNMHV} = \sum_{2\leq a_1,b_1
\leq n-1} \!\!\!\!\!\!\!\! R_{n;a_1b_1}^{0;0} \Bigl[ &\sum_{a_1+1
\leq a_2,b_2 \leq
  b_1} \!\!\!\!\!\!\!\! R_{n;b_1a_1;a_2b_2}^{0;a_1b_1}
  +  \sum_{b_1 \leq a_2b_2 \leq n-1} \!\!\!\!\!\!\!\! R_{n;a_2b_2}^{a_1b_1;0} \Bigr]\,.
~~\label{PNNMHVnew}\end{align}
For $k=3$,  we have
\begin{align}
&\mathcal{P}^{\rm{N}^3\rm{MHV}}_n = \sum_{2\leq a_1,b_1 \leq n-1}
  \!\!\!\!\!\!\!R_{n;a_1b_1}\biggl[\notag \\
&\sum_{a_1+1 \leq a_2,b_2 \leq b_1} \!\!\!\!\!\!\!\!\!
    R_{n;b_1a_1;a_2b_2}^{0;a_1b_1} \Bigl(\sum_{a_2+1 \leq a_3,b_3 \leq b_2}
    \!\!\!\!\!\!\!\!\!R_{n;b_1a_1;b_2a_2;a_3b_3}^{0;b_1a_1a_2b_2} + \sum_{b_2 \leq
      a_3,b_3 \leq b_1} \!\!\!\!\!\!R_{n;b_1a_1;a_3b_3}^{b_1a_1a_2b_2;a_1b_1}+ \sum_{b_1
      \leq a_3,b_3 \leq n-1} \!\!\!\!\!\!\!\! R_{n;a_3b_3}^{a_1b_1;0}
    \Bigr)\notag \\
& +\sum_{b_1\leq a_2,b_2 \leq n-1} \!\!\!\!\!
    R_{n;a_2b_2}^{a_1b_1;0}\Bigl(\sum_{a_2+1\leq a_3,b_3 \leq b_2} \!\!\!\!\!\!
    R_{n;b_2a_2;a_3b_3}^{0;a_2b_2} + \sum_{b_2 \leq a_3,b_3 \leq n-1}
    \!\!\!\!\!\!R_{n;a_3b_3}^{a_2b_2;0}\Bigr)
\biggr] \,. ~~\label{NNNMHV-2}\end{align}
In these examples, we see a level structure. The rules to write
down general expressions for $\mathcal{P}^{\rm{N}^k\rm{MHV}}$ and
proofs are given in \cite{Drummond:2008cr}.
Here we give a brief summary (one may check following points with Figure
\ref{fig:N4-tree}) :
\begin{itemize}

\item (1) For ${\rm N}^k {\rm MHV}$, we have $k+1$ levels. At the 0-th level,
one has the factor $1$ for MHV. At the first level, one has $(a_1,b_1)$ for NMHV.

\item (2) Each level has several vertices coming from their parent
at the previous level. The vertex specifies the subscript of the $R$
function defined in (\ref{generalR}). If the parent
vertex has $m$-pair indices, there will be $m+1$ children.

\item (3) The general vertex form is $(v_1, u_1; v_2,
u_2;...; v_r, u_r; a_p, b_p)$, where the last pair $(a_p, b_p)$
should be summed over $L\leq a_p< b_p\leq U$ and $a_p+2\leq b_p$
with cyclic ordering. The lower- and upper- bounds $L,U$ are denoted at
the left- and right- handed sides of the line connecting this vertex
with its parent.

\item (4) Given a vertex $(v_1, u_1; v_2,
u_2;...; v_r, u_r; a_p, b_p)$ with $r+1$ pairs as parent, its
$r+2$ children-vertices are as follow. The first vertex is obtained
by reversing the last pair $(a_p, b_p)$ and adding a new pair
$(a_{p+1}, b_{p+1})$, so we have
\bea (v_1, u_1; v_2, u_2;...; v_r, u_r; b_p, a_p;a_{p+1}, b_{p+1}
)\eea
The second vertex is obtained by deleting the pair before $(a_{p+1},
b_{p+1})$. Iterating this way of deleting pairs just before the
$(a_{p+1}, b_{p+1})$ for the vertex obtained in previous step,  we
get the third, fourth and rest vertices until we are left with only
the pair $(a_{p+1}, b_{p+1})$. In total, we get $r+2$ vertices.

\item (5) From this we can count the number of vertices at level $p$, which is given
by the Catalan number
\bea C(p)= {(2p)!\over p! (p+1)!} \eea
The Catalan number has the following recurrence definition
\bea C_0=1,~~~~C_{n+1}= \sum_{i=0}^n C_i C_{n-i} \eea
which is a consequence of on-shell recursion relations.

\item (6) Having determined the subscript (or pairs inside the vertex),  we determine the region of  sum
for the last pair of indices. Assuming these vertices come from the
parent $(v_1, u_1; v_2, u_2;...; v_r, u_r; a_p, b_p)$, we have $r+2$ children.
So there are $r+2$ lines dividing the space into $r+3$ regions.
The left region will be marked by $a_{p}+1$ and
the second region by $b_p$. Starting from  the
third region we will  marked $v_r$ for third region,  $v_{r-1}$ for
fourth-region and so on until $v_1$ for the $(r+2)$-th region.
Finally the last region (the most right one) will be marked by
$n-1$.

\item (7) As shown  in Figure \ref{fig:N4-tree}, these six points have fixed the tree-like diagram completely.
We need to translate these information to expressions like (\ref{PNNMHVnew}) and  (\ref{NNNMHV-2}),
to determine  the left- and right-superscript for the function $R$ in (\ref{superscripts}), coming
from same parent $(v_1, u_1; v_2, u_2;...; v_r, u_r; a_p, b_p)$.

To do this, let us  number these children from left to right by $1,2,...,r+2$.
For the number one child (the most left one) the
left-superscript is zero while for the number $r+2$ child (the
most right one) the right-superscript is zero. The left-superscript
of number $k$ child is the same as the right-superscript of number $k-1$
child. We just need to determine the right-superscript for
number $i=1,...,r+1$.

For the number $k$ child, its right-superscript is given by
following: (1) taking labels of the $k$-th vertex; (2) deleting the
final pair $(a_{p+1}, b_{p+1})$; (3) reversing the order of last
pair after deleting. The sequence after these three steps are the
right-superscript we want.

\item (8) Finally we need to sum over all paths nested from level one to
level $p$.
\end{itemize}
\EPSFIGURE[ht]{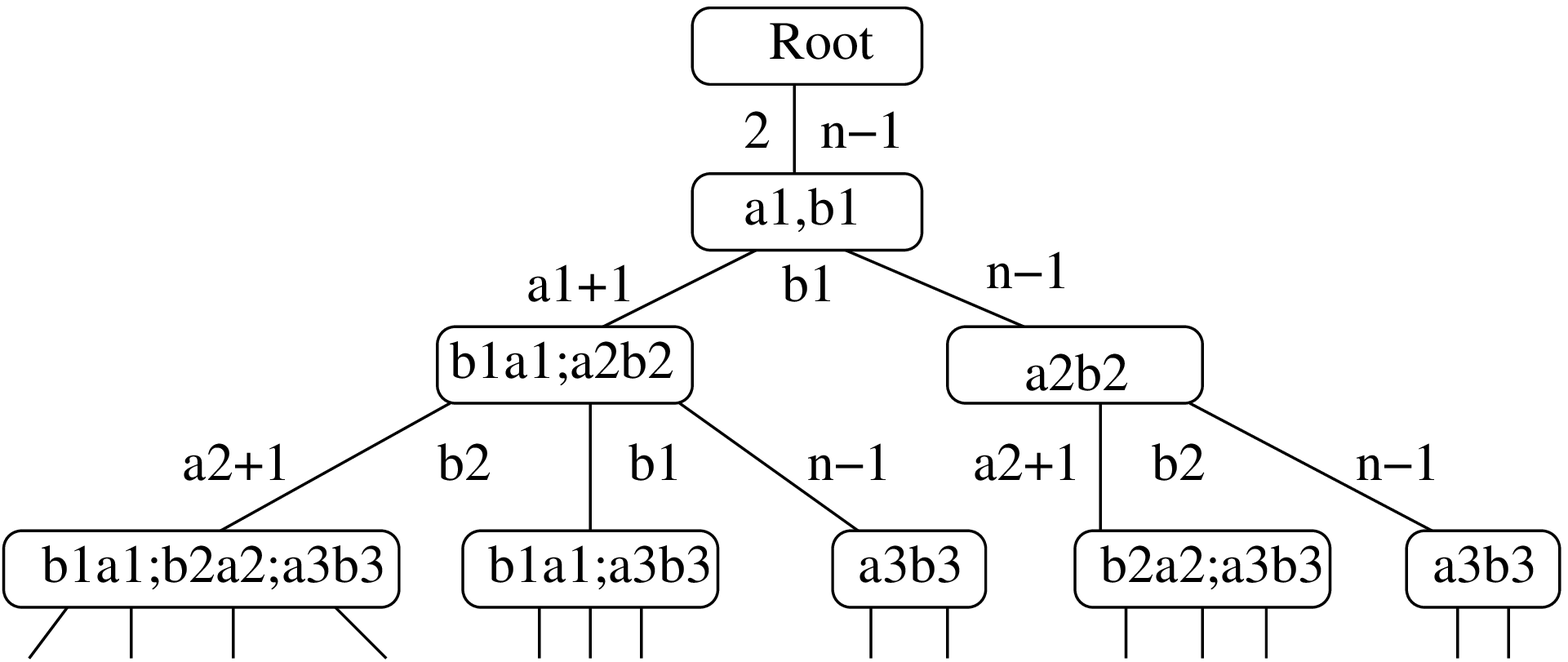,width=14.5cm} {Graphical representation
of tree-level amplitudes of ${\cal N}=4$ SYM.
    \label{fig:N4-tree} }

\subsection{Consistent conditions from on-shell recursion relations}

On-shell recursion relations provide not only powerful tools to
calculate tree-level amplitudes more efficiently and more compactly,
but also new perspectives to understand the theoretical foundation
of field theory \cite{Benincasa:2007xk}.\footnote{Some works along
this line can be found in
\cite{Schuster:2008nh,He:2008nj,Benincasa:2011kn,Benincasa:2011pg}.}

The observation in \cite{Benincasa:2007xk} starts from three-point
amplitudes of massless particles. As  shown in (\ref{Sol-A}) and
(\ref{Sol-B}),  on-shell conditions plus momentum conservation lead
to the splitting of three-point amplitudes into a
``holomorphic"- and an ``anti-holomorphic"-part
\bea M_3 = M_3^H(\Spaa{1|2}, \Spaa{2|3}, \Spaa{3|1})+ M_3^A
(\Spbb{1|2}, \Spbb{2|3}, \Spbb{3|1})~.\label{BS-3-point}\eea
where  $M_3^H, M_3^A$  are not restricted to be rational functions,
to count for full amplitudes (not just tree-level ones).
Furthermore, amplitudes of massless particles satisfy the differential equation (\ref{helicity-counting}).
If we define
\bea F=\Spaa{1|2}^{d_3} \Spaa{2|3}^{d_1} \Spaa{3|1}^{d_2},~~
G=\Spbb{1|2}^{-d_3} \Spbb{2|3}^{-d_1}
\Spbb{3|1}^{-d_2},~~~~d_i=h_i-\sum_{j=1,2,3; j\neq i}
h_j,~~~\label{BS-FG-def} \eea
then  $M_3^H/F$ and $M_3^A/G$ must be ``scalar functions" and have no helicity dependence.
Using $x_i$ to denote either $\Spaa{j|k}$ or
$\Spbb{j|k}$, we have $x_i {\partial {\cal M}(x_1,x_2,x_3)/
\partial x_i}=0$, $\forall i=1,2,3$, where ${\cal M}$ can be either $M_3^H/F$
or $M_3^A/G$. Thus {\sl up to solutions of delta-function support
which we will discard based on analyticity}, the only solution of
${\cal M}$ is a constant and  (\ref{BS-3-point}) is reduced to
\bea M_3=\kappa_H \Spaa{1|2}^{d_3} \Spaa{2|3}^{d_1}
\Spaa{3|1}^{d_2}+\kappa_A\Spbb{1|2}^{-d_3} \Spbb{2|3}^{-d_1}
\Spbb{3|1}^{-d_2}.~~~\label{BS-3-point-1} \eea
To have a well-defined physical amplitude for real momenta, $M_3$
must go to zero when both $\Spaa{i|j}$ and $\Spbb{i|j}$ are taken to
zero, thus we have\footnote{The case $\sum_i h_i=0$ is tricky and
still not fully understood, although one step has been taken in \cite{Benincasa:2011pg}.}\\~\\
%
 {\bf Observation A-1:}
 if $\sum_i h_i>0$, $\kappa_H=0$ and if $\sum_i h_i<0$,
$\kappa_A=0$.
%
\\

One consequence of  (\ref{BS-3-point-1}) is when
all $|h_i|=s$ are the same, we have the following four helicity
configurations:
\bea M_3(1_m^-, 2^-_r, 3^+_s) & = & \kappa_{mrs}\left(
{\Spaa{1|2}\over \Spaa{2|3}\Spaa{3|1}}\right)^s,~~~~M_3(1_m^+,
2^+_r, 3^-_s)  =  \kappa_{mrs}\left( {\Spbb{1|2}\over
\Spbb{2|3}\Spbb{3|1}}\right)^{s},~~~\label{BS-3-h-1}\eea
and
\bea M_3(1_m^-, 2^-_r, 3^-_s) & = & \W\kappa_{mrs}\left(
{\Spaa{1|2}\Spaa{2|3}\Spaa{3|1}}\right)^s,~~~~M_3(1_m^+, 2^+_r,
3^+_s)  =  \kappa_{mrs}\left( {\Spbb{1|2}
\Spbb{2|3}\Spbb{3|1}}\right)^{s},~~~\label{BS-3-h-2} \eea
where coupling constants depend on the type of particles in general.
Using cross symmetries of amplitudes, we have
\\~\\
 {\bf Observation A-2:} If spin $s$ is odd/even,  the coupling constant must be completely antisymmetric/symmetric.
\\~\\
For massless particles of odd spin to have nontrivial three-point
amplitudes, there must be at least three types of them. This is
familiar when we recall that the $U(1)$ gauge theory does not have
self-interaction and the minimal non-Abelian gauge group $SU(2)$ has
three generators.

Having understood three-point amplitudes, one may ask: {\sl assuming
that on-shell recursion relations can be applied to tree-level
amplitudes of a theory,  starting from three-point amplitudes
discussed above, which kind of consequences we can infer ?} To make
the point clear, one defines the concept {\bf constructibility}: we
call a theory constructible if $M_4(z)$ vanishes at $z\to \infty$
and can be computed correctly from $M_3$.

Now let us introduce the notion of consistency conditions. We can
calculate $M_4$ by using arbitrary pairs to do the BCFW-deformation,
for example, the $(1,2)$ or the $(1,4)$ pair. No matter which pair
we have chosen, the final result must be same. That is, we should
have $M_4^{(1,2)}= M_4^{(1,4)}$. By checking this simple
requirement,  we obtain many deep insights.

Under  the $\Spab{1|2}$-deformation, 
one gets the general expression
\begin{equation}
\begin{split}
M_4^{(1,2)}(0) = & \sum_h \left(  \kappa^H_{(1+h_1+h_4+h)} \langle
\hat 1,4\rangle^{h-h_1-h_4}\langle 4,{\hat
p}_{1,4}\rangle^{h_1-h_4-h}
\langle {\hat p}_{1,4} , \hat 1 \rangle^{h_4-h-h_1} + \right. \\
& \left. \kappa^A_{(1-h_1-h_4-h)}[1,4]^{h_1+h_4-h}[4,{\hat
p}_{1,4}]^{h_4+h-h_1}[{\hat p}_{1,4},1]^{h+h_1-h_4}\right)
\times\frac{1}{p_{1,4}^2}\times
\\ & \left( \kappa^H_{(1+h_2+h_3-h)} \langle 3,2\rangle^{-h-h_3-h_2}\langle
2,{\hat p}_{1,4}\rangle^{h_3-h_2+h}\langle {\hat p}_{1,4},3\rangle^{h_2-h_3+h} \right. + \\
 & \left. \kappa^A_{(1-h_2-h_3+h)}[3,\hat 2]^{h+h_3+h_2}[\hat 2,{\hat
p}_{1,4}]^{-h_3+h_2-h}[{\hat p}_{1,4},3]^{-h_2+h_3-h}\right) + \\
& \sum_h (4\leftrightarrow 3).~~\label{geko}
\end{split}
\end{equation}
where subscripts on three-particle couplings denote their
dimensions. The range of the helicity of the internal particle
depends on details of the theory. Even though (\ref{geko}) is
completely general, we choose to exclude theories where $h$ can take
values such that $h+h_1+h_2 =0$ or $-h+h_2+h_3 =0$ to avoid possible
complexity. Although we have kept two pieces of three-particle
amplitudes in (\ref{geko}), we should set to zero either the
holomorphic or the anti-holomorphic coupling constants in
discussions.

With a little algebra, it can be shown that the three-particle
amplitude of coupling constant $\kappa^H_{(1+h_1+h_4+h)}$ in
(\ref{geko}) possesses a factor  of  $\langle 4,4\rangle = 0$ to the power $-h_1-h_4-h$.
Recalling that  if $-h_1-h_4-h$ is less than zero, one must have $\kappa^H_{(1+h_1+h_4+h)}=0$.
No matter which situation it is in, this part gives zero contribution. The
only non-zero contributions to the sum over $h$ come from
the region where $h>-(h_1+h_4)$  of coupling
 $\kappa^A_{(1-h_1-h_4-h)}$.  A similar analysis shows that
only nonzero contributions come from the region where $h>(h_2+h_3)$
of coupling $\kappa^H_{1+h_2+h_3-h}$.

Putting these two conditions together we find that the first term
gives a non-zero contribution only when $h> {\rm
max}(-(h_1+h_4),(h_2+h_3))$. It is
\begin{equation}
\begin{split}
M_4^{(1,2)}(0) = & \sum_{h > {\rm max}(-(h_1+h_4),(h_2+h_3))} \left(
\kappa^A_{1-h_1-h_4-h}\kappa^H_{1+h_2+h_3-h}
\frac{(-p_{3,4}^2)^h}{p_{1,4}^2}\left(\frac{[1,4][3,4]}{[1,3]}\right)^{h_4}
\right. \\
& \left.
\left(\frac{[1,3][1,4]}{[3,4]}\right)^{h_1}\left(\frac{\langle
3,4\rangle}{\langle 2,3\rangle\langle
2,4\rangle}\right)^{h_2}\left(\frac{\langle 2,4\rangle}{\langle
2,3\rangle\langle 3,4\rangle}\right)^{h_3}\right) + \sum_{h > {\rm
max}(-(h_1+h_3),(h_2+h_4))}\!\!\!\!\!\!\!\!(4\leftrightarrow 3).
\end{split}~~~\label{hasi}
\end{equation}
One obtains $M_4^{(1,4)}(0)$ from (\ref{hasi}) by
simply exchanging labels $2$ and $4$.

Now we apply (\ref{hasi}) to various cases. In the first case,
$h_1=s$, $h_2=-s$, $h_3=s$ and $h_4=-s$ with $s$ positive integer
and particles carry new quantum numbers, for example, color. Writing
$\kappa_{a_1 a_2 a_3}=\kappa_{1-s} f_{a_1 a_2 a_3}$,  we have
\begin{equation}
\label{kiko} M^{(1,2)}_4(0) = \kappa_{1-s}^2\sum_{a_I} f_{a_1a_4
a_I}f_{a_I a_3a_2} {\cal A} + \kappa_{1-s}^2\sum_{a_I}f_{a_1a_3
a_I}f_{a_I a_4a_2}{\cal B},
\end{equation}
while
\begin{equation}
\label{kika} M^{(1,4)}_4(0) = \kappa_{1-s}^2\sum_{a_I} f_{a_1a_2
a_I}f_{a_I a_3a_4} {\cal C} + \kappa_{1-s}^2\sum_{a_I} f_{a_1a_3
a_I}f_{a_I a_2a_4}{\cal D}
\end{equation}
with
\begin{equation}
\begin{split}
& {\cal A} = \frac{\langle 2,4 \rangle^4}{\langle 1,2 \rangle\langle
2,3 \rangle\langle 3,4 \rangle\langle 4,1
\rangle}\left(\frac{\langle 2,4 \rangle^3 [1,3]}{\langle
1,2\rangle\langle 3,4\rangle}\right)^{s-1},\quad {\cal B} =
\frac{\langle 2,4 \rangle^3}{\langle 1,2 \rangle\langle 4,3
\rangle\langle 3,1 \rangle}\left(\frac{\langle 2,4 \rangle^3
[1,3]}{\langle
1,2\rangle\langle 3,4\rangle}\right)^{s-1}, \\
& {\cal C} =\frac{\langle 2,4 \rangle^4}{\langle 1,2 \rangle\langle
2,3 \rangle\langle 3,4 \rangle\langle 4,1
\rangle}\left(\frac{\langle 2,4 \rangle^3 [1,3]}{\langle
1,4\rangle\langle 2,3\rangle}\right)^{s-1}, \quad {\cal D} =
\frac{\langle 2,4 \rangle^3}{\langle 1,3 \rangle\langle 3,2
\rangle\langle 4,1 \rangle}\left(\frac{\langle 2,4 \rangle^3
[1,3]}{\langle 1,4\rangle\langle 2,3\rangle}\right)^{s-1}.
\end{split}
\end{equation}

When $s=1$,  ${\cal A}={\cal C}$.  Furthermore we have  (due
to the antisymmetric property of $f$ when $s=1$)
\begin{equation}
\label{jimi} \sum_{a_I}f_{a_1a_3 a_I}f_{a_I a_4a_2}\left({\cal
B}+{\cal D}\right) =-\sum_{a_I}f_{a_1a_3 a_I}f_{a_I a_4a_2}\left(
\frac{\langle 2,4 \rangle^4}{\langle 1,2 \rangle\langle 2,3
\rangle\langle 3,4 \rangle\langle 4,1 \rangle} \right)
\end{equation}
where  the Schouten identity (\ref{Schouten}) has been used. Thus
the requirement $M^{(1,2)}_4(0) - M^{(1,4)}_4(0) = 0$ leads to the
following condition
\begin{equation}
\sum_{a_I} f_{a_1a_4 a_I}f_{a_I a_3a_2} + \sum_{a_I}f_{a_1a_3
a_I}f_{a_I a_4a_2} + \sum_{a_I} f_{a_1a_2 a_I}f_{a_I a_3a_4} = 0.
\end{equation}
which is just the Jacobi identity. We have
\\~\\
 {\bf Observation A-3:}  A theory of
several spin $1$ particles can be non-trivial only if
dimensionless coupling constants $f_{a_1a_2a_3}$ are structure
constants of a Lie algebra.
\\

When $s=2$ and using (\ref{kiko}) and (\ref{kika}), the most general solution requires
\begin{equation}
\label{lasi} \sum_{a_I} f_{a_1a_4 a_I}f_{a_I a_3a_2} =
\sum_{a_I}f_{a_1a_3 a_I}f_{a_I a_4a_2}
\end{equation}
which implies that the algebra defined by
\begin{equation}
{\cal E}_a\star {\cal E}_b = f_{abc}~{\cal E}_c
\end{equation}
must be commutative and associative.
These algebras are reducible and the theory reduces to that of several
non-interacting massless spin 2 particles. Thus we have
\\~\\
{\bf Observation A-4:} it is not possible to define a non-abelian
generalization of a theory of spin 2 particles that is
constructible.
\\~\\
 {\bf Observation A-5:} when $s>2$,   there is no non-trivial way to satisfy the four-particle test.
\\%

In previous cases, we have assumed that four particles have the same spin. Now
we consider a mixed case where there are a spin $s$ particle
$(\Psi)$ and  a spin 2 particle $(G)$. We assume that the
spin 2 particle only has cubic couplings of the form $(++-)$ and
$(--+)$, as the case of graviton.  Let the coupling
constant of three gravitons be $\kappa$ while that of a graviton to
two $\Psi$'s be $\kappa'$.

For configuration $M_4(\Psi_1^{-},\Psi_2^{+},\Psi_3^{-},\Psi_4^{+})$
with $s>1$ and for on-shell recursion relations to be applicable,
$(1^{-}|2^{+})$- and $(1^{-}|4^{+})$-deformations yield
\begin{equation}\label{test1}
\begin{split}
M_4^{(1,2)}\:=&\:(\kappa')^{2}
 \frac{\langle1,4\rangle}{[1,4]}
 \frac{[2,4]^{4s}}{[1,2]^{2s-2}[2,3]^{2}[3,4]^{2s-2}}\\
M_4^{(1,4)}\:=&\:(\kappa')^{2}
 \frac{\langle1,2\rangle}{[1,2]}
 \frac{[2,4]^{4s}}{[1,4]^{2s-2}[3,4]^{2}[2,3]^{2s-2}}.
\end{split}
\end{equation}
The ratio of quantities in (\ref{test1}) is
\begin{equation}\label{ratio1}
\frac{M_4^{(1,2)}}{M_4^{(1,4)}}\:=\:
 \left(\frac{\sf t}{\sf s}\right)^{2s-3},
\end{equation}
where ${\sf s} = P_{12}^{2}$ and ${\sf t}=P_{14}^{2}$. This ratio is
equal to one only if $s=3/2$.

At this point couplings $\kappa$ and $\kappa'$ are independent
and it is not possible to conclude that the theory is the linearized
supergravity. Quite nicely, the next amplitude constrains couplings.

Consider the amplitude $M_4(G_1,G_2,\Psi_3,\Psi_4)$  under  $(1|2)$- and $(1|4)$-deformations
\begin{equation}\label{test2}
\begin{split}
M_{4}^{(1,2)}\:=&\:-(\kappa')^{2}
 \frac{\langle 1,3\rangle^{2}[2,4]^{2s+2}}{[1,2]^{2}[3,4]^{2}[2,3]^{2s-4}}
 \frac{\sf s}{\sf t u}\\
M_{4}^{(1,4)}\:=&\:\kappa' \frac{\langle
1,3\rangle^{2}[2,4]^{2s+2}}{[1,4]^{2}[2,3]^{2s-2}}
 \left(\frac{\kappa}{\sf s}+\frac{\kappa'}{\sf
 u}\right)
\end{split}
\end{equation}
where ${\sf u}=P_{13}^2$. Taking their ratio and setting $s=3/2$, we
get
\begin{equation}\label{ratio2}
1=\frac{M_{4}^{(1,4)}}{M_{4}^{(1,2)}}\:=\: 1 -\frac{\sf u}{\sf t}
 \left(\frac{\kappa}{\kappa'} - 1\right).
\end{equation}
which can be true when and only when $\kappa'=\kappa$. Thus, we have
\\~\\
 {\bf Observation A-6:} The only particle with spin higher than
$1$ which can couple to a graviton, in a constructible theory, has
the same spin as a gravitino in ${\cal N}=1$ supergravity and agree
with linearized ${\cal N}=1$ supergravity.
\\

Thus, combining general principals, such as Lorentz symmetry, with
on-shell recursion relations, one arrives at many important
conclusions without an explicit Lagrangian description\footnote{See
also the work \cite{Antoniadis:2011rr} where tensor gauge bosons in
generalized Yang-Mills theory is considered.} .
In \cite{Benincasa:2007xk}, only tests of four particles have been
done. Tests of general $n$-particles have been carried out in
\cite{Schuster:2008nh, He:2008nj}. These discussions opened a new
way to think about some fundamental properties of quantum field
theory, especially in the language of S-matrix program, where the
use of Lagrangian description is avoided.

\subsubsection{Generalizations}

As we have seen now, using general principles and the assumption of on-shell recursion relations,
important results have been derived.
However, these discussions are based on the validity of  on-shell recursion relations and there are exceptions.
Generalizations are then necessary.
One example is the on-shell recursion relation with nonzero boundary
contribution, as mentioned in the previous section.

Another generalization is to use more complicated deformations by
deforming not only a pair of particles but all particles
\cite{Cohen:2010mi}. The all line anti-holomorphic deformation is
given by
\bea \bket{i}\to \bket{i}+z w_i \bket{X},~~~
i=1,...,n,~~~\label{Cohen-All-anti}\eea
where to ensure momentum conservation, we require $\sum_{i=1}^n w_i\ket{i}=0$.

Under  this deformation, the large $z$ behavior of amplitude is
better. We will have new kinds of on-shell recursion relations which
are helpful in discussions  of general quantum field theory. To see
this, notice that general terms of tree-level amplitudes can be
expressed as
\bea  {\sum \Spaa{...}... \Spbb{..}...\over \sum \Spaa{...}...
\Spbb{..}..}~,\eea
Define $a$ as the number of $\Spaa{..}$ factors in the numerator minus that in the denominator,
and $s$ as the number of $\Spbb{..}$ factors in the numerator minus that in the denominator.
A simple analysis of dimension gives
\bea  a+s+c= 4-n~~~\label{Cohen-Mass}\eea
where $4-n$ is the mass-dimension of $n$-point amplitudes and $c$ the mass-dimension of coupling constants.
The helicity information gives us another identity
\bea a-s= -\sum_i h_i,~~~\label{Cohen-Helicity}\eea
Combining these two equations, we find
\bea 2s= 4-n-c+\sum_i  h_i,~~~2a= 4-n-c-\sum_i
h_i,~~~\label{Cohen-a-s}\eea
thus under the deformation (\ref{Cohen-All-anti}) the large $z$
behavior is
\bea A(z\to \infty) \to z^s 
\eea
For ``power-counting renormalizable" theories, we have $c\geq 0$ and $s+a\leq 4-n$.
Thus when $n>4$, either $s<0$ or
$a<0$, so there is always a deformation\footnote{All line
anti-holomorphic deformation in (\ref{Cohen-All-anti}) for
$s<0$ or all line holomorphic deformation in  the dual of
(\ref{Cohen-All-anti}) for $a<0$.} to write down a on-shell
recursion relation without boundary contribution, so the theory is ``on-shell
constructible".

In non-renormalizable theories  with $c<0$, as long as $n$ is big enough,
this all-line shifting can be used to calculate on-shell amplitudes from lower-point ones.
Using this method, we can deal with many exotic theories.
However, comparing to deformation of a pair of particles, there are many more terms in recursion relations.
In BCFW-deformation, we have freedoms to choose the deformed pair.
For example,  $A(1,2,3,4)$ can be calculated by either
$(1|2)$-deformation with manifest pole $s_{41}$ or $(4|1)$-deformation with manifest pole $s_{12}$.
This provides a consistent requirement when we compare results from these two calculations.
However, there is no such freedom in all-line shifts.
So we cannot get consistent relation directly.

\subsection{KK and BCJ relations of color
ordered gluon partial amplitudes} 

We have seen the power of on-shell recursion relations to understand
properties of quantum field theory in previous subsection. Now we
use on-shell recursion relations \cite{Feng:2010my} to obtain more
properties of color-ordered gluon partial  amplitudes, including the
KK and the BCJ relations.

We start again with on-shell three-point amplitudes presented in the
previous subsection \cite{Benincasa:2007xk}. The key is that three
particle amplitudes are completely fixed by Lorentz symmetry and
satisfy $A(1,2,3)=-A(3,2,1)$ without using any Lie-algebra property
and Lagrangian. In fact, we do not need the explicit form in
(\ref{BS-3-h-1}). Under the assumption that on-shell recursion
relation is applicable, taking  pair $(n,1)$ to do
deformation we get\footnote{As it is familiar now, no matter which helicity
configuration of $n,1$ is, there is always one legitimate
BCFW-deformation available to write down the on-shell recursion
relation. Thus in this subsection, we will not mention explicitly
what deformation should be used. Also, for simplicity we will not
write down the sum over helicities of inner particles.  }
\bean & & A(n,\b_1,.,\b_{n-2},1) \\& = & \sum_{i=1}^{n-3} A(\WH
n,\b_1,.,\b_i,-\WH p_i) {1\over p_i^2} A(\WH
p_i,\b_{i+1},.,\b_{n-2},\WH 1)\nn
& = & \sum_{i=1}^{n-3}(-)^{n-i} A(\WH 1,\b_{n-2},.,\b_{i+1},\WH p_i)
{1\over p_I^2} (-)^{i+2} A(-\WH p_i,\b_i,.,\b_1,\WH n)\nn
& = & (-)^n A(1,\b_{n-2},\b_{n-1},.,\b_1,n)~\eean
where we have expanded the amplitude in the second line, used
induction to reflect both lower-point amplitudes in the third line
and finally recombined them in the fourth line. By this simply
manipulation, we have proved the color-order reversed identity (\ref{color-reverse}).

Although the proof is very simple, it shows the pattern to be
followed. Notice that we do not need to specify details, such as
helicities, the shift $(n|1)$, or  explicit expressions of $A_n$, as
long as on-shell recursion relations without boundary values are
applicable. The conclusion holds for any helicity configuration.

Now we move to the $U(1)$-decoupling equation (\ref{U1-decouple}).
The $n=4$ case is easy to check by using the color-reversed relation
of $A_3$ in on-shell recursion relations. Explicitly,
\bea A(1,2,3,4) & = & \sum_h A_3(4,\WH 1, -\WH p_{14}^{h}) {1\over
s_{14}} A_3( \WH p_{14}^{-h}, \WH 2, 3) =-\sum_h  A_3(4,\WH 1, -\WH
p_{14}^{h}) {1\over s_{14}} A_3( \WH p_{14}^{-h}, 3,\WH 2)\nn
A(1,3,2,4) & = &\sum_h  A_3(4,\WH 1, -\WH p_{14}^{h}) {1\over
s_{14}} A_3( \WH p_{14}^{-h}, 3,\WH 2) + \sum_h A_3(\WH 1,3, -\WH
p_{13}^{h}) {1\over s_{13}} A_3(\WH p_{13}^{-h},  \WH 2,4)\nn
A(1,3,4,2) & = & \sum_h A_3(\WH 1,3, -\WH p_{13}^{h}) {1\over
s_{13}} A_3(\WH p_{13}^{-h}, 4, \WH 2)=-\sum_h A_3(\WH 1,3, -\WH
p_{13}^{h}) {1\over s_{13}} A_3(\WH p_{13}^{-h},  \WH 2,4)\eea
so $A(1,2,3,4)+ A(1,3,2,4)+A(1,3,4,2)=0$.
 To see the strategy of a general proof, we work out the example of $n=5$.
 To make the presentation more clear, we use, for example,
$ A(p_{523},1,4)$ to represent $A(\WH 5,2,3,-\WH
p_{523})A(p_{523},1,4)/s_{523}$. Using this short notation, we
express five-point amplitudes in terms of on-shell  recursion
relations under the $(1|5)$-deformation,
\bea \begin{array}{lllllllllll} A(1,2,3,4,5) & = & A(1, p_{23},
4,5) &
+ & A(1,p_{234},5) & + & 0 & + & 0 & + & 0 \\
A(1,5,2,3,4) & = & A(1,5,p_{23},4) & + & A(1,5, p_{234}) & + & A(1,
p_{52}, 3,4) & + & A(1, p_{523},4) & + & 0\\
A(1,4,5,2,3) & = & A(1,4,5,p_{23}) & + & 0 & + & A(1,4,p_{52},3) & +
& A(1,4,p_{523}) & + & A(1,p_{452}, 3)\\
A(1,3,4,5,2) & = & 0 & + & 0 & + & A(1,3,4,p_{52}) & + & 0 & + &
A(1,3, p_{452})
\end{array} ~~\label{n=5-U1}\eea
Here we have purposely arranged terms such that the sum of each column on the right-handed side is zero,
by using the $U(1)$-decoupling equations for $n=3$ and $n=4$.\footnote{By our short notation,
each column has the same unwritten
factor. For example, the first column has $A(-p_{23},2,3)/s_{23}$.}

The proof for general $n$ is by induction.
Each $n$-point amplitude is first expressed in terms of on-shell recursions, then
regrouped so $U(1)$-identity for lower $m$ can be used. Details can be found in \cite{Feng:2010my}.

Next the KK-relation (\ref{KK-rel}).
For $n=3,4,5$, the KK-relation is equivalent to either color-order reversed relation or
$U(1)$-decoupling relation, depending the set $\a,\b$ in (\ref{KK-rel}).
The first nontrivial KK-relation is  for $n=6$, given in (\ref{KK-6-point}).
We work out this example to demonstrate the idea of a general proof.
Using on-shell recursion relations under the $(1|6)$-deformation,
\bea A(1,2,3,6,4,5) & = & A(5,1,p|p,2,3,6,4)+ A(1,2,p|p, 3,6,4,5)+
A(1,2,3,p|p,6,4,5)+A(5,1,2,p|p,3,6,4)\nn
& & + A(4,5,1,p|p,2,3,6)+ A(5,1,2,3,p|p,6,4)+A(4,5,1,2,p|p,3,6)\eea
Here $A(5,1,p|p,2,3,6,4)$ represents
$\sum_h A(5, \WH 1, -\WH p_{15}^h) (1/ s_{15})A(\WH
p_{15}^{-h}, 2,3,\WH 6,4)$, different from notations in the proof of
$U(1)$-decoupling identity. To match the right-handed side of
(\ref{KK-6-point}), we need to use KK-relations for $n=3,4,5$ to
put $1,6$ at two ends
\bea A(5,1,p|p,2,3,6,4) &= & A(5,1,p) A(p,2,3,6,4)\nn & = &
(-A(1,5,p)) [ -A(p,2,3,4,6)-A(p,4,2,3,6)-A(p,2,4,3,6)]\nn & = &
A(1,5,p|p,2,3,4,6)+A(1,5,p|p,4,2,3,6)+A(1,5,p|p,2,4,3,6)\nn
A(1,2,p|p, 3,6,4,5) & = & A(1,2,p|p,3,5,4,6)+
A(1,2,p|p,5,3,4,6)+A(1,2,p|p,5,4,3,6)\nn
A(1,2,3,p|p,6,4,5) & = & A(1,2,3,p|p,5,4,6)\nn
A(5,1,2,p|p,3,6,4) &= &A(1,2,5,p|p,3,4,6)+ A(1,5,2,p|p,3,4,6)\nn & &
+A(1,2,5,p|p,4,3,6)+ A(1,5,2,p|p,4,3,6) \nn
 A(4,5,1,p|p,2,3,6) & = & A(1,5,4,p|p,2,3,6)\nn
  A(5,1,2,3,p|p,6,4) & = & A(1,2,3,5,p|p,4,6)+A(1,5,2,3,p|p,4,6)+A(1,2,5,3,p|p,4,6)\nn
 A(4,5,1,2,p|p,3,6)& = & A(1,2,5,4,p|p,3,6)+ A(1,5,2,4,p|p,3,6)+A(1,5,4,2,p|p,3,6)\eea
We have $18$ terms in the above and 6 terms on right-handed side of (\ref{KK-6-point}).
One amplitude in (\ref{KK-6-point}) corresponds to three terms here.

The proof of the general case
$A(1,\{\a_1,.,\a_k\}, n,\{\b_1,.,\b_m\})$ is done first by
using on-shell recursion relation under the $(1|n)$-deformation
\bea & & A(1,\{\a_1,,\a_k\}, n,\{\b_1,.,\b_m\}) =
\sum_{i=0}^k \sum_{j=0}^m A(\b_{j+1},.,\b_m,1,\a_1,.,\a_i,
p_{ij}|-p_{ij}, \a_{i+1},.,\a_k, n,\b_1,.,\b_j) 
\label{KK-gen-left}\eea
where two cases $(i=0,j=m)$ and $(i=k,j=0)$ should be excluded from the summation.
Next we use KK-relations for the first factor $A(\b_{j+1},.,\b_m,1,\a_1,.,\a_i, p_{ij})
 = (-)^{m-j}\sum_{\sigma_{ij}} A(1, \sigma_{ij}, p_{ij})$
and the second factor $A(-p_{ij}, \a_{i+1},.,\a_k, n,\b_1,.,\b_j)= (-)^j
\sum_{\W\sigma_{ij}} A(-p_{ij}, \W\sigma_{ij}, n)$.
Similar to the example of $n=6$, for each given set
$\{i,j,\sigma_{ij},\W\sigma_{ij}\}$, (\ref{KK-gen-left}) gives a
term obtained from on-shell recursion relations at the right-handed side of (\ref{KK-rel}).
If we can show that the numbers of terms at  both sides  are the same, the proof is completed.

Now count terms. There are $C^i_{i+m-j}$ and
$C^j_{j+k-i}$ terms for each factor at the right-handed side of (\ref{KK-gen-left}), respectively.
The total number of terms at the right-handed side of (\ref{KK-gen-left}) is
\bea -2 {(m+k)!\over m! k!}+\sum_{i=0}^k \sum_{j=0}^m
{(i+m-j)!\over i! (m-j)!}{(j+k-i)!\over j! (k-i)!}
\eea
where the first term counts the two excluded cases.
The right-handed side of KK-relation (\ref{KK-rel}) has $(k+m-1)C_{k+m}^m$
terms after we use on-shell recursion relations to expand each amplitude into $(k+m-1)$ terms.
These two numbers match up as it should be.
The identity is thus proved.

Now we move to the proof of BCJ relations.
They are much more complicated due to the presence of dynamical factors
$s_{ij}$ \cite{Bern:2008qj}. In its most general form, the sets $\a,\b$ can be arbitrary,
for which an explicit proof exists \cite{Chen:2011jx}.
However, all other equations are redundant except those where the
set $\a$ has only one element, which we call ``fundamental
BCJ-relations". The redundancy is in the following sense.
If fundamental BCJ-relations are true, then combining with KK-relations
we can solve any amplitude by $(n-3)!$ amplitudes of the form
$A(1,2,3,\sigma(4,.n))$, which  is exactly the statement given by
general BCJ-relations.

Fundamental BCJ-relations are given in (\ref{BCJ-funda}).
For an inductive proof, there are two important observations.
The first is the special relation for $n=3$, $A(2,3,1)s_{31}=0$.
The second is the dual form obtained by using momentum
conservation. For example, the case $n=5$ can be rewritten as
\bean 0 & = & A(2,4,3,5,1) s_{24}+A(2,3,4,5,1)(s_{24}+ s_{34})
+A(2,3,5,4,1)(s_{24}+s_{34}+s_{54})\eean

Again, we start with examples to get a sense of the proof.
Take $n=4$ and do the following contour integration under the $(1|2)$-deformation\footnote
{For our proof, there is no need to specify the
details of deformation as long as there exists one deformation to
write down on-shell recursion relations.}
\bea \oint {dz\over z} s_{\WH 23}(z) [ A(\WH 1, \WH 2, 3,4)+A(\WH 1,
3,4,\WH 2)+A(\WH 1, 4,\WH 2, 3)]=0~.~~~\label{BCJ-n=4-contour}\eea
where terms inside the square bracket are added up to zero by $U(1)$-decoupling identity.
Among these three terms,  $s_{\WH 23}(z) A(\WH 1, \WH 2,3,4)/z$ has
only one pole contribution at $z=0$, which yields $T_1= s_{23}
A(1,2,3,4)$. The third term is zero, since $\WH 1, \WH 2$ are not
nearby and the large $z$ limit of amplitude is $z^{-2}$. The second term is $T_2=(s_{23}-s_{23}(z_{13}))
A(1,3,4,2)=-s_{13} A(1,3,4,2)$. putting all results together and
using the color-reserved relation, we get immediately $s_{23}
A(2,3,4,1)+(s_{23}+s_{43}) A(2,4,3,1)=0$.

The  proof for general $n$ is by induction.  To make the argument clear, we demonstrate the case of $n=6$.
The case of arbitrary $n$ follows the same general idea.
Taking the $\Spab{2|1}$-deformation and using on-shell recursion relations to expand each amplitude in $I_6$ given in
(\ref{BCJ-funda}),
 we will get three different splittings for each amplitude, which
 can be grouped as following
{\bea I_6^{[2]}&= & A(\WH 2,4,-\WH p_{24}|\WH p_{24},3,5,6,\WH
1)(s_{43}+s_{45}+s_{46}+s_{41}) + A(\WH 2,3,-\WH p_{23}|\WH
p_{23},4,5,6,\WH 1)(s_{45}+s_{46}+s_{41})\nn
& & +A(\WH 2,3,-\WH p_{23}|\WH p_{23},5,4,6,\WH 1) (s_{46}+s_{41})
+A(\WH 2,3,-\WH p_{23}|\WH p_{23},5,6,4,\WH 1)s_{41}\nn
I_6^{[3]}&= & A(\WH 2,4,3,-\WH p_{243}|\WH p_{243},5,6,\WH
1)(s_{43}+s_{45}+s_{46}+s_{41}) + A(\WH 2,3,4,-\WH p_{234}|\WH
p_{234},5,6,\WH 1)(s_{45}+s_{46}+s_{41})\nn
& & +A(\WH 2,3,5,-\WH p_{235}|\WH p_{235},4,6,\WH 1) (s_{46}+s_{41})
+A(\WH 2,3,5,-\WH p_{235}|\WH p_{235},6,4,\WH 1)s_{41}\nn
I_6^{[4]}&= & A(\WH 2,4,3,5,-\WH p_{2435}|\WH p_{2435},6,\WH
1)(s_{43}+s_{45}+s_{46}+s_{41}) + A(\WH 2,3,4,5,-\WH p_{2345}|\WH
p_{234},6,\WH 1)(s_{45}+s_{46}+s_{41})\nn
& & +A(\WH 2,3,5,4,-\WH p_{2354}|\WH p_{2354},6,\WH 1)
(s_{46}+s_{41}) +A(\WH 2,3,5,6,-\WH p_{2356}|\WH p_{2356},4,\WH
1)s_{41}~~~\label{BCJ-6-split}\eea
Here $A(...p,|p,...)$ is the same as the one used in the proof of
KK-relation. The splitting parameter $[m]$ means that there are $m$
particles on the left handed side of splitting (without counting
$p$).

We analyze $I_{6}^{[2]}$ in (\ref{BCJ-6-split}) first. All terms in
$I_6^{[2]}$ can be divided into two categories: those with particle
$4$ on  the left handed side and those with particle $4$ on the
right handed side. The last three terms with particle $4$ on the
right handed side can be written as
\bean & &  A(\WH 2,3,-\WH p_{23}|\WH p_{23},4,5,6,\WH
1)(s_{45}+s_{46}+s_{4\WH 1})  +A(\WH 2,3,-\WH p_{23}|\WH
p_{23},5,4,6,\WH 1) (s_{46}+s_{4\WH 1})\nn & &  +A(\WH 2,3,-\WH
p_{23}|\WH p_{23},5,6,4,\WH 1)s_{4\WH 1}\nn & & + \left\{A(\WH
2,3,-\WH p_{23}|\WH p_{23},4,5,6,\WH 1) +A(\WH 2,3,-\WH p_{23}|\WH
p_{23},5,4,6,\WH 1) +A(\WH 2,3,-\WH p_{23}|\WH p_{23},5,6,4,\WH
1)\right\}( s_{41}-s_{4\WH 1}(z_{23}))\eean
where we have purposefully written $s_{41}= ( s_{41}-s_{4\WH 1}(z_{23}))+
s_{4\WH 1}(z_{23})$. By induction over the right-handed side of
$A(...,p|p,...)=A_L(...,p) A_R(p,...)$,  the sum of first
two lines is found to be zero. The first term of $I_6^{[2]}$ can be written
in its dual form
\bean & &  -s_{24}A(\WH 2,4,-\WH p_{24}|\WH p_{24},3,5,6,\WH 1)\nn
&= & -s_{\WH 24}(z_{24})A(\WH 2,4,-\WH p_{24}|\WH p_{24},3,5,6,\WH
1) -(s_{24}-s_{\WH 2 4}(z_{24}))A(\WH 2,4,-\WH p_{24}|\WH
p_{24},3,5,6,\WH 1)\eean
where again the first term is zero by induction. Using the fact
$-(s_{24}-s_{\WH 2 4}(z_{24}))= ( s_{41}-s_{4\WH 1}(z_{24}))$, we
obtain
 \bea  I_6^{[2]} &= &  ( s_{41}-s_{4\WH 1}(z_{23}))\left\{ A(\WH
2,3,-\WH p_{23}|\WH p_{23},4,5,6,\WH 1)  +A(\WH 2,3,-\WH p_{23}|\WH
p_{23},5,4,6,\WH 1)\right. \nn & & \left. +A(\WH 2,3,-\WH p_{23}|\WH
p_{23},5,6,4,\WH 1) +A(\WH 2,4,-\WH p_{24}|\WH p_{24},3,5,6,\WH
1)\right\}~~~\label{BCJ-I6-2} \eea
Similar manipulations for $I_6^{[3]}, I_6^{[4]}$ result in
\bea  I_6^{[3]} & = &   ( s_{41}-s_{4\WH 1}(z_{23}))\left\{ A(\WH
2,3,4,-\WH p_{234}|\WH p_{234},5,6,\WH 1)  +A(\WH 2,3,5,-\WH
p_{235}|\WH p_{235},4,6,\WH 1)\right. \nn & & \left. +A(\WH
2,3,5,-\WH p_{235}|\WH p_{235},6,4,\WH 1) +A(\WH 2,4,3,-\WH
p_{243}|\WH p_{243},5,6,\WH 1)\right\} \nn
I_6^{[4]} & = &   ( s_{41}-s_{4\WH 1}(z_{23}))\left\{ A(\WH
2,3,4,5,-\WH p_{2345}|\WH p_{2345},6,\WH 1)  +A(\WH 2,3,5,4,-\WH
p_{2354}|\WH p_{2354},6,\WH 1)\right. \nn & & \left. +A(\WH
2,3,5,6,-\WH p_{2356}|\WH p_{2356},4,\WH 1) +A(\WH 2,4,3,5,-\WH
p_{2435}|\WH p_{2435},6,\WH 1)\right\}\eea
Summing all three together we finally have
{\bea I_6&= & s_{41}\left\{ A(2,4,3,5,6,1) +
A(2,3,4,5,6,1)+A(2,3,5,4,6,1) +A(2,3,5,6,4,1)\right\}\nn & &
+\oint_{z\neq 0} {dz s_{\WH 1 4}\over z}\left\{ A(\WH 2,4,3,5,6,\WH
1) + A(\WH 2,3,4,5,6,\WH 1)+A(\WH 2,3,5,4,6,\WH 1)+A(\WH
2,3,5,6,4,\WH 1)\right\}\nn
& = & \oint {dz s_{\WH 1 4}\over z}\left\{ A(\WH 2,4,3,5,6,\WH 1) +
A(\WH 2,3,4,5,6,\WH 1)+A(\WH 2,3,5,4,6,\WH 1)+A(\WH 2,3,5,6,4,\WH
1)\right\}\eea}
 Using the KK-relation or $U(1)$-decoupling identity, we can rewrite it as
\bea -I_6&= & \oint {dz s_{\WH 1 4}\over z} A(4,\WH 2,3,5,6,\WH
1)~~~~\eea
which is zero since  $(1,2)$ are not nearby, $A(4,\WH
2,3,5,6,\WH 1)\to  z^{-2}$ and the residue at the infinity
pole is zero.

The proof for general $n$ will be exactly the same. First we write down
\bea I_n & = & A(1,2,...,n-2,n,n-1) s_{n,n-1}+ A(1,2,..., n,n-2,n-1)
(s_{n,n-1}+s_{n,n-2})\nn & & +...+ A(1,n,2,3,...,n-1)
\sum_{j=2}^{n-1} s_{n,j}~~~~\label{Fund-BCJ-In}\eea
Then we expand every amplitude $I_{n}^{[m]}$ ($m=2,3,..,n-2$)
in terms of on-shell recursion relations under the $(1|n-1)$-deformation.
Just like the case $n=6$, each $I_{n}^{[m]}$ will be reduced to the form like (\ref{BCJ-I6-2}).
Summing up all $m$, we get
\bea I_n & = & \oint {dz s_{\WH {n-1} n}\over
z}[A(1,2,...,n-2,n,n-1) + A(1,2,..., n,n-2,n-1)
 +...+ A(1,n,2,3,...,n-1)
] \nn
& = & - \oint {dz s_{\WH {n-1} n}\over z}  A(\WH 1,2,3,..,\WH {n-1},
n)=0\eea
where we have used $U(1)$-decoupling identity at the second line.
The integration is zero because $A(z)\to z^{-2}$ under our deformation.

In this proof, it is crucial that when shifted pair $(i,j)$ are
not nearby, there is a deformation rendering the amplitude vanishing as $z^{-2}$.
This is just a bonus relation, as  presented in the previous section and discussed for gravity
theory in \cite{ArkaniHamed:2008gz,Spradlin:2008bu}.
Thus, bonus relations in gauge theories are actually BCJ-relations.

In this proof, we have only assumed:
{\sl (1) Lorentz and Poincare symmetries, which lead
to the antisymmetry $A(1,2,3)=-A(3,2,1)$; (2) the applicability of
on-shell recursion relations, so any tree-level amplitudes can be
obtained from three-point amplitude; (3) $z^{-2}$ vanishing
behavior when $(i,j)$ are not nearby}. Under these
assumptions, we derived all identities without relying on an explicit Lagrangian.
Although assumptions (2) and (3) are proved by using the Lagrangian, but we
can reverse the logic by assuming (2) and (3) as
fundamental to derive all other things. Assumptions (2) and
(3) are statements about  analytic properties of the complex amplitude $A(z)$.
The reversed way of thinking is more or less
along the line of S-matrix program \cite{S-matrix}.

The color-order reversed identity, $U(1)$-decoupling relations,
KK-relations and BCJ relations have been generalized to
the ${\cal N}=4$ theory \cite{Sondergaard:2009za,Jia:2010nz}. It will be
interesting to see if these relations can be found in other
theories.

\subsection{Kawai-Lewellen-Tye (KLT) relations}

Due to the highly non-linear form of the Hilbert-Einstein Lagrangian
and bad divergent behaviors at loop levels,  quantization of gravity
is   one of the most difficult problems in theoretical physics.
Putting the conceptual difficulty aside, we can still use the
standard method to define various amplitudes through Feynman
diagrams. However, calculations of scattering amplitudes directly by
Feynman diagrams from the Lagrangian are difficult even for tree
amplitudes, especially with more than four  external gravitons. We
must look for alternatives. A picture from string theory, where
gravitons are described by closed strings, provides a very good
hint. The celebrated Kawai-Lewellen-Tye relations
\cite{Kawai:1985xq,Bern:1998sv} (see also review \cite{Bern:2002kj,
Sondergaard:2011iv}), which express on-shell graviton amplitudes as
``squares" of on-shell color-ordered gluon amplitudes, were derived
by taking the field theory limit of string theory, based on
relations between open and closed strings.

Although ``squares" like KLT relations are natural in string theory,
they are totally obscure in field theory. The Hilbert-Einstein
Lagrangian and the non-Abelian Yang-Mills Lagrangian are very
different. It is very desirable to have an understanding purely in
field theory. Here we will prove KLT relations with the help of
on-shell  recursion relations.\footnote{See
\cite{BjerrumBohr:2011kc}, for a recent review.}

\subsubsection{KLT relations}

To present KLT relations explicitly, we define several functions. The first one is
\bea {\cal S}[i_1,...,i_k|j_1,j_2,...,j_k]_{p_1} & = & \prod_{t=1}^k
(s_{i_t 1}+\sum_{q>t}^k \theta(i_t,i_q) s_{i_t
i_q})~~~\label{S-def}\eea
where $\theta(i_t, i_q)$ is zero when  the pair $(i_t,i_q)$ has the same
ordering in both sets ${\cal I},{\cal J}$ and one otherwise.
Function ${\cal S}$ is the $f$-function defined in \cite{Bern:1998sv} with improved presentation.
The subscript $p_1$ indicates that there is a term $s_{i_t 1}$ for each $i_t$.
To be familiar with the notation, here are a few examples:
\bean {\cal S}[2,3,4|2,4,3]_{p_1} & = & s_{21} (s_{31}+s_{34})
s_{41},~~~ {\cal
S}[2,3,4|4,3,2]_{p_1}=(s_{21}+s_{23}+s_{24})(s_{31}+s_{34}) s_{41}
\eean
The second is the dual ${\cal \W S}$ defined as (${\cal \W S}$ is the $\O f$-function defined in \cite{Bern:1998sv})
\bea {\cal {\W
S}}[i_2,..,i_{n-1}|j_2,...,j_{n-1}]_{p_n}=\prod_{t=2}^{n-1} (s_{j_t
n}+ \sum_{q<t} \theta(j_t, j_q) s_{j_t j_q})~.~~\label{dual-S-def}
\eea
For example,
\bean {\cal \W S}[2,3,4|4,3,2]_{p_5}= s_{45} (s_{35}+s_{34})
(s_{25}+s_{23}+s_{24}) \eean
${\W {\cal S}}$ and ${\cal S}$ are related as follows:
\bea  {\W {\cal S}}[ {\cal I}|{\cal J}]_{p_n}= {\cal S}[ {\cal
J}^T|{\cal I}^T]_{p_n}~~~~\label{S-tilde-S-rel}\eea
where $T$ means reversing ordering.\footnote{Using
\eqref{S-tilde-S-rel},  we can use only ${\cal S}$ in all formula.
However, to be compatible with  \cite{Bern:1998sv}, we keep
both ${\cal S}$ and ${\cal \W S}$.}

These functions have some interesting properties. For example,
\bea {\cal S}[i_1,...,i_k|j_1,j_2,...,j_k]={\cal
S}[j_k,...,j_1|i_k,..,i_1]~~~\label{S-rev}\eea
where we have exchanged the two sets and reversed orderings in each set.
There is also a vanishing identity
\bea 0 = I \equiv \sum_{\a\in S_k}{\cal
S}[\a(i_1,...,i_k)|j_1,j_2,...,j_k]_{p_1}
A(k+2,\a(i_1,...,i_k),1)~.~~~\label{I-com}\eea
or by reversing the color order
\bea 0 =I \equiv \sum_{\a\in S_k}{\cal S}[j_k,...,j_1|\a]_{p_1} A(1,
\a,k+2)~.~~~\label{I-com-2}\eea
The vanishing identity is very important in the proof of KLT
relations and can be easily  proved by using fundamental  BCJ relations (\ref{Fund-BCJ-In}).
Here we give the example with $n=6$ and ${\cal J}=(2,3,4,5)$.
The permutation $\a(2,3,4,5)$ can be divided into permutation of
$\b(2,3,4)$ plus putting $5$ at all possible positions. Considering
one particular ordering of $\a(2,3,4)$, for example the ordering
$(3,4,2)$, we have in (\ref{I-com})
\bean  I[3,4,2] & \equiv & {\cal S}[ 3,4,2,5 |2,3,4,5]
A(6,3,4,2,5,1) + {\cal S}[ 3,4,5,2 |2,3,4,5] A(6,3,4,5,2,1)\nn & +&
{\cal S}[ 3,5,4,2 |2,3,4,5] A(6,3,5,4,2,1)+ {\cal S}[5, 3,4,2
|2,3,4,5] A(6,5,3,4,2,1)\nn
& = & (s_{31}+s_{32}) (s_{41}+s_{42})s_{21}\Bigg[s_{51}
A(6,3,4,2,5,1) + (s_{51}+s_{52}) A(6,3,4,5,2,1)\nn & +&
(s_{51}+s_{52}+s_{54})A(6,3,5,4,2,1)+
(s_{51}+s_{52}+s_{54}+s_{53})A(6,5,3,4,2,1)\Bigg]\eean
which is zero, by virtue of fundamental BCJ relations.
BCJ relations can also be used to prove the ``moving identity"
\bea & & \sum_{\a\in S_j}\sum_{\b\in S_{n-3-j}}
 {\cal S}[
\a(\sigma_2,..,\sigma_j)|\sigma_2,..,\sigma_j]_{p_1}\nn & & \times
{\cal \W S}[\sigma_{j+1},..,\sigma_{n-2}
|\b(\sigma_{j+1},..,\sigma_{n-2})]_{p_{n-1}} \W
A(\a(\sigma_2,..,\sigma_j),1,n-1,\b(\sigma_{j+1},..,\sigma_{n-2}),n)\nn
& = & \sum_{\a\in S_{j-1}}\sum_{\b\in S_{n-2-j}}
 {\cal S}[
\a(\sigma_2,..,\sigma_{j-1})|\sigma_2,..,\sigma_{j-1}]_{p_1}\nn & &
\times {\cal \W S}[\sigma_{j},..,\sigma_{n-2}
|\b(\sigma_{j},..,\sigma_{n-2})]_{p_{n-1}} \W
A(\a(\sigma_2,..,\sigma_{j-1}),1,n-1,\b(\sigma_{j},..,\sigma_{n-2}),n)~~\label{S-moving}\eea

With these definitions, KLT relations in their original forms can be written as \cite{Bern:1998sv}
\bea M_n & = & (-)^{n+1}\sum_{\sigma\in S_{n-3}}\sum_{\a\in
S_j}\sum_{\b\in S_{n-3-j}}
A(1,\{\sigma_2,..,\sigma_j\},\{\sigma_{j+1},..,\sigma_{n-2}\},
n-1,n) {\cal S}[
\a(\sigma_2,..,\sigma_j)|\sigma_2,..,\sigma_j]_{p_1}\nn & & \times
{\cal \W S}[\sigma_{j+1},..,\sigma_{n-2}
|\b(\sigma_{j+1},..,\sigma_{n-2})]_{p_{n-1}} \W
A(\a(\sigma_2,..,\sigma_j),1,n-1,\b(\sigma_{j+1},..,\sigma_{n-2}),n)~~~\label{KLT-bern}\eea
where $j=[n/2-1]$ is a fixed number, determined by $n$.
However, by using the moving identity (\ref{S-moving}), (\ref{KLT-bern}) will be true for different choices of $j$.
One can actually shift $j$ to make the left or right part to be empty,
thus we have the two following symmetric formulas
\bea M_n  =  (-)^{n+1}\sum_{\sigma,\W\sigma\in S_{n-3}}
A(1,\sigma(2,n-2), n-1,n) {\cal S}[
\W\sigma(2,n-2))|\sigma(2,n-2))]_{p_1} \W
A(n-1,n,\W\sigma(2,n-2),1)~~~~~\label{KLT-Pure-S}\eea
as well as
\bea M_n  =  (-)^{n+1}\sum_{\sigma,\W\sigma\in S_{n-3}}
A(1,\sigma(2,n-2), n-1,n) {\cal \W S}[
\sigma(2,n-2))|\W\sigma(2,n-2))]_{p_{n-1}} \W
A(1,n-1,\W\sigma(2,n-2),n)~~~~~\label{KLT-Dual-S}\eea

The above $(n-3)!$ symmetric form fits well with the picture in
string theory, where three vertices have been inserted at fixed positions dictated by conformal symmetry.
However, scattering amplitudes of gravitons are completely $n!$ symmetric.
It is desirable to have a form manifest of the larger symmetry.
The manifest $(n-2)!$ symmetric KLT form is  \cite{BjerrumBohr:2010ta}
\bea M_n=(-)^n\sum_{\gamma,\b}{\W A(n,\gamma(2,...,n-1),1)  {\cal
S}[ \gamma(2,...,n-1)|\b(2,..,n-1)]_{p_1} A(1,\b(2,...,n-1),n)\over
s_{123..(n-1)}}~~~\label{newKLT}\eea
and its dual form is
\bea M_n & = & (-)^n\sum_{\b,\gamma} {A (1,\b(2,...,n-1),n) {\cal \W
S}[\b(2,...,n-1)|\gamma(2,..,n-1)]_{p_{n}} \W A(n,\gamma(2,...,n-1),
1)\over s_{2...n}}~~~\label{New-KLT-dual} \eea
They can be proved by using on-shell recursion relations. As
discussed in \cite{BjerrumBohr:2010ta}, the numerator in
(\ref{newKLT}) is  zero according to the vanishing identity
(\ref{I-com}) and  the denominator $s_{12..(n-1)} $ is zero when
on-shell. The formula is in fact of the ${0/0}$-type. To make sense
of  the formula, one must give a well defined regularization scheme
to obtain meaningful finite limit. In \cite{Feng:2010hd}, an
explicit regularization was provided and the equivalence between the
new $(n-2)!$ form (\ref{newKLT}) and old $(n-3)!$ form
(\ref{KLT-Pure-S}) is established.

\subsubsection{The proof of KLT relations}

Now we prove KLT relations in (\ref{KLT-Pure-S}) and
(\ref{KLT-Dual-S}) by using on-shell recursion relations with, for
example, the $(1|n)$-deformation. The starting point is the
three-point amplitude, where $M_3(1,2,3)= A(1,2,3)^2$. This result
can be derived directly by using the Lorentz symmetry and spin
information without referring to Lagrangian, as we have done in
(\ref{BS-3-h-1}). In other words, for any  Lagrangian,  we will get
the same conclusion, as long as these general principles are
satisfied.

Consider two contour integrals
\bea I_1  = \oint {dz\over z} M_n(z),~~~~I_2 = \oint {dz\over z}
\sum_{\a,\b} A_n(\a) \W A_n(\b) {\cal S} \eea
where for simplicity, the integrand of $I_2$ is abbreviated.
Both $I_1$ and $I_2$ vanish, due to the boundary behavior of their integrands.
One then has
\bea M_n & = & \sum_i { M_L M_R \over p_i^2},~~~\label{KLT-idea-1} \\
\sum_{\a,\b} A_n(\a) \W A_n(\b) {\cal S} & = & \sum_{\rm Poles} {\rm
Res} \left[ A_n(\a) \W A_n(\b) {\cal S}
\right]~~~\label{KLT-idea-2}\eea
Residues  in  (\ref{KLT-idea-2}) come from poles of both $A_n$ and $\W A_n$.
If the right-handed sides of (\ref{KLT-idea-1})  and (\ref{KLT-idea-2}) are the same, the KLT relation is proved.
To show this, we focus on the pole $s_{12...k}$. Other poles can be
treated similarly by permutations, because both (\ref{KLT-idea-1})
and (\ref{KLT-idea-2}) are total symmetric. The general
pole-structure in (\ref{KLT-idea-2}) can be divided to three
categories:
\begin{itemize}
\item (A-1) Pole is in $A$ and there is no pole in
$\W A$. In this case, the sum over $\sigma$ in (\ref{KLT-bern})
will be factorized as $\sum_{\sigma_1,\sigma_2}
A(1,\sigma_1(2,..,k),\sigma_2(k+1,...,n-2),n-1,n)$.
\item (A-2) Pole is  in part $\W A$ and there is no pole in
$A$. In this case, the sum over $\b$ is factorized as
$\a(\sigma_2,..,\sigma_j)= \a_1(...)  \a_2(2,...,k)$.
\item (B) Both $A$ an $\W A$ have the pole.
Both $\sigma,\a$ will be factorized as $\sum_{\sigma_1,\sigma_2}
A(1,\sigma_1(2,..,k),\sigma_2(k+1,...,n-2),n-1,n)$ and
$\a(\sigma_2,..,\sigma_j)= \a_1(...)  \a_2(2,...,k)$.
Due to the double pole structure of $A\W A$, the dynamical factor ${\cal S} {\cal \W S}$ will
contribute an overall factor $s_{12...k}$  to cancel this naive double pole.
\end{itemize}
We  claim that contributions from (A-1) and (A-2) are zero, while
that from (B) reproduce  the right-handed side of
(\ref{KLT-idea-1}).

For contributions from (A-1), the residue is
\bea &  & (-)^{n+1}\sum_{\sigma_1,\sigma_2}\sum_{\a'}\sum_{\b\in
S_{n-3-j}} {A(\WH 1,\sigma_1(2,..,k),-\WH
 p^{h}) A(\WH p^{-h},,\sigma_2(k+1,...,n-2),\WH {n-1}, n)\over s_{12..k}}\left\{ {\cal S}[
\a(\sigma_2,..,\sigma_j)|\sigma_1,\sigma_2]_{p_1}\right.\nn & &
\left. \times {\cal \W S}[\sigma_{j+1},..,\sigma_{n-2}
|\b(\sigma_{j+1},..,\sigma_{n-2})]_{p_{n-1}} \W
A(\a(\sigma_2,..,\sigma_j),1,n-1,\b(\sigma_{j+1},..,\sigma_{n-2}),n)\right\}_{z_{12..k}}
~~~\label{A1-1}\eea
where $\a'$ denotes the sum over constrained permutations by avoiding the appearance of the same pole in $\W A$.
Since $\W A$ does not have the pole, the general pattern of $\a$ is $(T_1, \a(2),T_2,\a(3),...,T_{k-1}, \a(k),T_k)$,
where $(\a(2),..,\a(k))$ is a permutation of $(2,..,k)$ via $\a$.
In this pattern we consider the factor ${\cal S}$.
For the element $t_1\in T_1$,  we have a factor $s_{1t_1}+\sum_{j=2}^k s_{\a(j) t_1}+{\rm others}$,
by taking $(\sigma_1,\sigma_2)$ as references.
For the element $t_2\in T_2$, the factor will be $s_{1t_2}+\sum_{j=3}^k s_{\a(j) t_1}+{\rm others}$.
These factors are fixed by the ordering of $(\a(2),..,\a(k))$.
These $\cal S$-factors for elements inside the
set $(T_1,...,T_k)$ are the same for all possible permutations $\a_1$.
(\ref{A1-1}) can thus be rewritten as
\bea &  & (-)^{n+1}\sum_{\sigma_2}\sum_{\a'}\sum_{\b\in S_{n-3-j}} {
A(\WH p^{-h},\sigma_2(k+1,...,n-2),\WH {n-1}, n)\over
s_{12..k}}\left\{ {\cal S}_1[ (T_1,..,T_k)|\sigma_2]_{p_1}\right.\nn
& & \left. \times {\cal \W S}[\sigma_{j+1},..,\sigma_{n-2}
|\b(\sigma_{j+1},..,\sigma_{n-2})]_{p_{n-1}} \W
A(\a(\sigma_2,..,\sigma_j),1,n-1,\b(\sigma_{j+1},..,\sigma_{n-2}),n)\right\}_{z_{12..k}}\nn
& & \times (\sum_{\sigma_1} {\cal S}[\a(2),..,\a(k)|
\sigma_1(2,..,k)]_{\WH p_1} A(\WH 1,\sigma_1(2,..,k),-\WH
 p^{h}))\eea
The sum in the third line is just the vanishing form in
(\ref{I-com-2}). So contributions from (A-1) are zero. Contributions
from (A-2) can also be shown to vanish, by similar arguments but
using  the reversing property (\ref{S-rev}) of ${\cal S}$.

For (B),  the residue is
\bea I_B& = &
(-)^{n+1}\sum_{\sigma_1,\sigma_{2}}\sum_{\W\a_1,\W\a_2}\sum_{\b\in
S_{n-2-j}} {A(\WH 1, \sigma_1(2,..,k),-\WH p^{h}) A(\WH
p^{-h},\sigma_2(k+1,..,n-2),\WH {n-1},n)\over s_{12..k}}\nn &
&\left\{ {\cal S}[
\W\a_2(\sigma_2(k+1,..,j)),\W\a_1(2,..,k)|\sigma_1,\sigma_2(k+1,j)]_{\WH
p_1}\times {\cal \W S}[\sigma_2(j+1,...,n-2)|\W
\b(\sigma_2(j+1,...,n-2))]_{\WH p_{n-1}} \right. \nn & & \left.
\times \W
A(\W\a_2(\sigma_2(k+1,..,j),\W\a_1(2,..,k),1,n-1,\W\a_2(\sigma_2(k+1,..,j),\W\a_1(2,..,k)\W
\b(\sigma_2(j+1,...,n-2)),n)\right\}_{z_{12..k}}~~~\label{part-B-1}\eea
To proceed, one can first show the following factorization property
\bea {\cal S}[
\W\a_2(\sigma_2(k+1,..,j)),\W\a_1(2,..,k)|\sigma_1,\sigma_2(k+1,j)]_{p_1}=
{\cal S}[ \W\a_1(2,..,k)|\sigma_1]_{p_1} {\cal S}[
\W\a_2(\sigma_2(k+1,..,j))|\sigma_2(k+1,j)]_{\WH p}~~~\eea
A naive sum over $\sigma_1$ of this factorization form gives
zero by vanishing identity (\ref{I-com}), which indicates
the existence of an $s_{\WH 1 2..k}$ factor.
Due to this factor,  almost
all contributions  inside the curly bracket vanish, except those
having the pole $s_{\WH 1 2..k}$.  Thus
\bea I_B& = &
(-)^{n+1}\sum_{\sigma_1,\sigma_{2}}\sum_{\W\a_1,\W\a_2}\sum_{\b\in
S_{n-2-j}} {A(\WH 1, \sigma_1(2,..,k),-\WH p^{h}) A(\WH
p^{-h},\sigma_2(k+1,..,n-2),\WH {n-1},n)\over s_{12..k}}\nn &
&\left\{ {\cal S}[ \W\a_1(2,..,k)|\sigma_1]_{p_1} {\cal S}[
\W\a_2(\sigma_2(k+1,..,j))|\sigma_2(k+1,j)]_{\WH p}\times {\cal \W
S}[\sigma_2(j+1,...,n-2)|\W \b(\sigma_2(j+1,...,n-2))]_{\WH p_{n-1}}
\right. \nn & & \left. \times \W A(\W\a_2(\sigma_2(k+1,..,j),\WH
p^{-\W h},n-1,\W \b(\sigma_2(j+1,...,n-2)),n){1\over s_{12..k}} \W
A(-\WH p^{\W h},
\W\a_1(2,..,k),1)\right\}_{z_{12..k}}~~~\label{part-B-2}\eea
which can be rewritten as
\bea I_B & = & \sum_{h,\W h} I_L^{h,\W h} {1\over s_{12..k}}
I_R^{h,\W h} \eea
where
\bea I_L^{h,\W h} & = & (-)^{k+1} \sum_{\sigma_1,\W \a_1} {A(\WH 1,
\sigma_1(2,..,k),-\WH p^{h}) {\cal S}[
\W\a_1(2,..,k)|\sigma_1]_{p_{\WH 1}} \W A(-\WH p^{\W h},
\W\a_1(2,..,k),1)\over s_{\WH 1 2 ..k}} ~~~\label{IL}\eea
and
\bea I_R^{h,\W h} & = & (-)^{n-k+1+1} \sum_{\sigma_2,
\W\a_2,\b}A(\WH p^{-h},\sigma_2(k+1,..,n-2),\WH {n-1},n){\cal S}[
\W\a_2(\sigma_2(k+1,..,j))|\sigma_2(k+1,j)]_{\WH p}~~~\label{IR}\\
 & & \times
{\cal \W S}[\sigma_2(j+1,...,n-2)|\W \b(\sigma_2(j+1,...,n-2))]_{\WH
p_{n-1}}\W A(\W\a_2(\sigma_2(k+1,..,j),\WH p^{-\W h},n-1,\W
\b(\sigma_2(j+1,...,n-2)),n)\nonumber \eea
When $h =\W h$, the $I_R$ is just the graviton amplitude of
$(n-k+1)$ particles in (\ref{KLT-bern}), where the helicity of
particle $p$ is $h$. When $h=-\W h$, it is zero,\footnote{The
KLT-like quadratic vanishing identity was discovered and proved in
\cite{BjerrumBohr:2010zb} by on-shell recursion relations, by
following similar ideas in this subsection. Also in
\cite{Feng:2010br}, the KLT relation and the quadratic vanishing
identity are unified in the supersymmetric version.  } as  shown in
\cite{BjerrumBohr:2010zb}. Similarly, when $h=\W h$ the $I_L$ is the
graviton amplitude of $k+1$ particles in (\ref{newKLT}). When $h=-\W
h$, the $I_L$ is zero, as shown in \cite{BjerrumBohr:2010zb}.

Putting all together we finally have
\bea I_B & = & \sum_h {M(\WH 1,2,..,k,-\WH p^h) M(\WH
p^{-h},k+1,...,\WH {n-1}, n)\over s_{12..k}}\eea
which proves the KLT relation.

The reason that Feynman diagram method is very difficult to prove
KLT relation is because it is off-shell method and contains too many
redundant information. On-shell recursion method keeps only on-shell
information, thus we can see the intrinsic connection more
transparently. Some generalizations of above method to other
theories can be found in \cite{Feng:2010hd,Du:2011js, Chen:2010ct}.

\section{Remarks}

Starting with the basics, such as spinor notations and color
decompositions, we have  exposed many aspects of analytic properties
of gauge-boson amplitudes, defined BCFW-deformations, analyzed the
large $z$-behavior of amplitudes, and derived on-shell recursion
relations of gluons. Further developments  have also been discussed,
included the generalization to other quantum field theories, in
particular supersymmetric theories, recursion relations for
off-shell currents, recursion relation with nonzero boundary
contributions, bonus relations, relations for rational parts of
one-loop amplitudes, recursion relations in 3D and a proof of CSW
rules. Sample applications were also provided, including solutions
of split helicity amplitudes  and of ${\cal N} = 4$ SYM theories,
consequences of consistent conditions, the proofs of  KK, BCJ
relations and Kawai-Lewellen-Tye (KLT) relations.

But there are many important topics which deserve the attention, but not covered.
This is partly due to our taste and partly due to our abilities.
We apologize to those whose contributions were not mentioned and papers not cited.
Of course, some topics are more advanced and require more backgrounds.
In particular, we would mention the following topics:
\begin{itemize}
\item {\bf Recursion relations for integrands:} One important
progress in recent years is the establishment of recursion
relations for all-loop integrands in ${\cal N}=4$ SYM theories
\cite{ArkaniHamed:2010kv} (see also
\cite{Bullimore:2010pj,Mason:2010yk, CaronHuot:2010ek,
CaronHuot:2010zt,Brandhuber:2010mi}). This construction used
the language of twistors and momentum twistors.
More background will be needed for this interesting topic.

\item {\bf Recursion relations for AdS/CFT correlators:}
There are generalizations of on-shell recursion relations to
computations of correlation functions of the stress tensor and
conserved currents in the conformal theory with the help of AdS/CFT
correspondence
\cite{Raju:2010by,Raju:2011mp} \cite{Fitzpatrick:2011ia,Paulos:2011ie}.

\item {\bf Recursion relations for tree-amplitudes in string theories:}
On-shell recursion relations have been applied to calculate
tree-amplitudes in string theories
\cite{Boels:2008fc,Cheung:2010vn,Boels:2010bv,Fotopoulos:2010cm,Fotopoulos:2010jz}.
Comparing with field theories, one extra complexity is the infinite
number of fields contributing to inner propagators and we need to
sum their contributions in some ways.

\end{itemize}

Before concluding, we would like to emphasize again the importance
of {\bf analyticity}, which has been central in our discussions.
Analyticity has been known since the early days of quantum field
theory. The S-matrix program tried to build the whole structure
based on it, but had limited success to generate useful results. It
is partly due to the dominance of the conventional Lagrangian
paradigm, besides the lack of suitable tools for studies of the
subject. Progresses in the last few years have put analyticity back
into the center stage again. New perspectives have been provided, by
introducing new concepts and apparatus, such as twistor geometry,
on-shell deformation, Grassmannian geometry, etc. These developments
may just be the beginning of many important discoveries. It is
possible that we are facing a revolution for the study of analytic
properties of quantum field theories. Finally, we would like to
mention several reviews related to our discussions. There has been a
special issue of Journal of Physics A, devoted to ``Scattering
Amplitudes in Gauge Theories"
\cite{Bern:2011qt,arXiv:1104.0816,Drummond:2011ic, Ferro:2011ph,
Dixon:2011xs,Bargheer:2011mm, Ita:2011hi,Britto:2010xq,
Elvang:2010xn, Schabinger:2011kb, Adamo:2011pv, Brandhuber:2011ke,
Carrasco:2011hw, Henn:2011xk}. The $AdS/CFT$ integrability was
discussed in \cite{arXiv:1012.3982}. Relations between twistor
geometry and scattering amplitudes were \cite{Wolf:2010av}.
Relations between Wilson loops and scattering amplitudes were
discussed in \cite{Alday:2008yw}.

\subsection*{Acknowledgements}
We thank Yin Jia, Hui Luo, and Cong-Kao Wen for critical readings of the manuscript.
This work is supported, in part, by fund from Qiu-Shi, the
Fundamental Research Funds for the Central Universities with
contract number 2010QNA3015, as well as Chinese NSF funding under
contract No.10875104, No.11031005,  No.11135006, No. 11125523.


\end{document}